\documentclass[reprint]{revtex4-2}
\usepackage{amsmath,amsfonts,amscd,amssymb,amsthm}
\usepackage{commath}
\usepackage{bm}
\usepackage{dsfont}
\usepackage{graphicx}
\usepackage{epstopdf}
\usepackage{cancel}
\usepackage{rotating}
\usepackage{url}
\usepackage{caption}
\usepackage{subcaption}
\usepackage{rotating}
\usepackage{multirow}
\usepackage{wrapfig}
\usepackage{mathtools}
\usepackage{subeqnarray}
\usepackage{setspace}
\usepackage{hyperref}
\usepackage[group-separator={,}]{siunitx}
\usepackage[titletoc,title]{appendix}

\usepackage[bottom,flushmargin,hang,multiple]{footmisc}
\usepackage{lipsum}

\begin{document} 

\title{\LARGE{Physics-constrained, low-dimensional models for MHD: \\ First-principles and data-driven approaches}}

\author{Alan A. Kaptanoglu}
\affiliation{Department of Physics, University of Washington, Seattle, WA, 98195, USA }
\author{Kyle D. Morgan}
\affiliation{Department of Aeronautics and Astronautics, University of Washington, Seattle, WA, 98195, USA}
\author{Chris J. Hansen}
\affiliation{Department of Aeronautics and Astronautics, University of Washington, Seattle, WA, 98195, USA}
\affiliation{\hspace{-.5in}Department of Applied Physics \& Applied Mathematics, Columbia University, New York, NY, 10027, USA\hspace{-.5in}}
\author{Steven L. Brunton} 
\affiliation{Department of Mechanical Engineering, University of Washington, Seattle, WA, 98195, USA} 

\normalsize
\begin{abstract}{ 
Plasmas are highly nonlinear and multi-scale, motivating a hierarchy of models to understand and describe their behavior. 
However, there is a scarcity of plasma models of lower fidelity than magnetohydrodynamics (MHD), although these reduced models hold promise for understanding key physical mechanisms, efficient computation, and real-time optimization and control.  Galerkin models, obtained by projection of the MHD equations onto a truncated modal basis, and data-driven models, obtained by modern machine learning and system identification, can furnish this gap in the lower levels of the model hierarchy. 
This work develops a reduced-order modeling framework for compressible plasmas, leveraging decades of progress in projection-based and data-driven modeling of fluids.
We begin by formalizing projection-based model reduction for nonlinear MHD systems.
To avoid separate modal decompositions for the magnetic, velocity, and pressure fields, we introduce an energy inner product to synthesize all of the fields into a dimensionally-consistent, reduced-order basis. 
Next, we obtain an analytic model by Galerkin projection of the Hall-MHD equations onto these modes. 
We illustrate how global conservation laws constrain the model parameters, revealing symmetries that can be enforced in data-driven models, directly connecting these models to the underlying physics.  
We demonstrate the effectiveness of this approach on data from high-fidelity numerical simulations of a 3D spheromak experiment.}  
This manuscript builds a bridge to the extensive Galerkin literature in fluid mechanics, and facilitates future principled development of projection-based and data-driven models for plasmas.

\end{abstract}

\flushbottom
\maketitle
\section{Introduction}\label{Sec:introduction}

Plasmas and plasma-enabled technologies are pervasive in everyday life~\cite{roth2001industrial}, 
but their nonlinear, multi-scale behavior poses severe challenges for understanding, modeling, and controlling these systems.
There are a tremendous number of known plasma models of varying model complexity, from magnetohydrodynamics (MHD) to the Klimontovich equations, but a large gap exists in the lower levels of this hierarchy between simple circuit models and the many MHD variants. 
These low-level models are motivated because higher fidelity models typically require computationally intensive and high-dimensional simulations~\cite{Candy2003,Ohia2012,Groselj2018}, obfuscating the dynamics and precluding model-based real-time control. 
Moreover, many high-dimensional nonlinear systems tend to evolve on low-dimensional attractors~\cite{Taira2017aiaa}; plasmas across a large range of parameter regimes, geometry, and degree of nonlinearity exhibit this feature~\cite{jimenez2007analysis,vanMilligen14,pandya,victor2015development,strait2016spatial,byrne2017study,gu2019new,kaptanoglu2020}. 
In these cases, the evolution of only a few coherent structures, obtained from model-reduction techniques~\cite{benner2005dimension, benner2017model}, can closely approximate the evolution of the high-dimensional physical system. Fortunately, recent progress in theoretical, data-driven, and machine learning methods are revolutionizing the analysis, modeling, and control of high-dimensional, nonlinear systems, especially in the field of fluid mechanics~\cite{brunton2020machine}.
Reduced-order modeling is advancing particularly rapidly, enabling the modeling of increasingly complex fluid flows~\cite{Noack2003jfm,Noack2016jfm,Taira2017aiaa,Carlberg2017jcp,Rowley2017arfm,brunton2019data,brunton2020machine},
but many of these advances have not yet been adopted in the plasma physics community.
In this manuscript, we provide a framework for physics-constrained, low-dimensional plasma models which address this important gap in the model hierarchy. 

The applications of reduced-order models include understanding reduced physical mechanisms~\cite{Lorenz1963jas,guan2020sparse}, computationally efficient simulations~\cite{humbird2019transfer}, digital twins (virtual time-dependent models for a dynamic system, constantly updating with sensor measurements)~\cite{kapteyn2020toward}, and real-time control~\cite{ravindran2005real,galperti2017integration,levesque2013multimode}. For example, acceleration of inertial confinement fusion simulations and digital twins can facilitate an exploration of the implosion parameter space~\cite{humbird2019transfer}, surrogate closure models can lead to more accurate and efficient fluid simulations~\cite{wang2020deep}, surrogate gyrokinetic transport models can speed up tokamak simulations by orders of magnitude~\cite{citrin2015real,van2020fast}, and steady-state tokamak operation will require the active avoidance or mitigation of disruptions, which can seriously damage components of the device~\cite{loarte2007power}. For these real-time control challenges, there are a wealth of model-based control techniques such as model predictive control~\cite{allgower1999nonlinear} that can be leveraged for plasma systems. However, existing models can be either too high-dimensional and computationally expensive to operate in real-time, or too low-fidelity to be useful for control. 

In addition to being computational efficient, reduced-order models can help uncover key mechanisms that govern the evolution of the dominant coherent structures.  This aspect of reduced-order modeling has a rich history, from the famous Lorenz model in 1963~\cite{Lorenz1963jas}, through the present era, including the low-order mechanistic model of the cylinder wake in 2003 by Noack et al.~\cite{Noack2003jfm}.  Recently, data-driven algorithms, like the algorithms used in this work, have shown potential to uncover similarly interpretable and useful models. Examples include related fluid systems~\cite{loiseau2018constrained,loiseau2020data} as well as recent work that uncovers a Lorenz-like model of electroconvective chaos by Guan et al.~\cite{guan2020sparse}. Moreover, increasingly reduced order models are used to describe key mechanisms in plasma physics, including ``predator-prey'' dynamics in gyrokinetic simulations~\cite{kobayashi2015direct}, direct data-driven discovery of reduced MHD or kinetic equations from a plasma dataset~\cite{alves2020data}, and data-driven fluid models for the L-H mode transition in tokamaks~\cite{Dam2017pf}. These models are often critical for providing insight into the physical system, including energy transfers and other nonlinear interactions.

Reduced-order models traditionally fall into two categories: projection-based model reduction and data-driven system identification. 
Projection-based model reduction is achieved by first computing the evolution of a governing partial differential equation (PDE) model, often by spatially discretizing the domain, resulting in a high-dimensional system of ordinary differential equations (ODEs). In our case, we consider a 3D MHD simulation. 
Then a low-dimensional orthogonal basis is computed, often via the proper orthogonal decomposition (POD)~\cite{holmes2012turbulence,Taira2017aiaa}. 
Finally, the high-dimensional model is  ``Galerkin-projected'' onto this basis~\cite{willcox2002balanced, benner2015survey}, resulting in an efficient reduced system that describes how the amplitudes of the POD modes evolve in time. 
However, this projection is intrusive since it requires knowledge of the governing physics and a working high-fidelity solver.

In contrast, system identification techniques attempt to identify data-driven models directly from measurement data, often without knowledge of the governing equations. 
Increasingly, data-driven methods are producing effective bases beyond POD for different experimental or computational tasks~\cite{Taira2017aiaa}; modern methods include balanced POD~\cite{willcox2002balanced,rowley2005model}, spectral POD~\cite{Towne2018jfm}, DMD~\cite{schmid_dynamic_2010,Rowley2009jfm,Tu2014jcd}, the Koopman decomposition~\cite{koopman_hamiltonian_1931,mezic_analysis_2013,brunton2021modern}, resolvent analysis~\cite{mckeon2010critical,luhar2014opposition}, and neural-network-based autoencoders~\cite{lusch2018deep,champion2019data, lee2020model}. 
Data-driven techniques, including modern machine learning, are also being widely applied to discover dynamical systems models of complex physical systems~\cite{bongard_automated_2007,schmidt_distilling_2009,Brunton2016pnas,raissi2017hidden,raissi2017physics1,Yair2017pnas,klus2017data,Wehmeyer2018jcp,Mardt2018natcomm,Duraisamy2019arfm,pathak2018model,Noe2019science,bar2019learning,kochkov2021machine}, with a particular emphasis on hybrid physics-inspired or physics-informed machine learning~\cite{raissi2017physics1,raissi2017physics2,battaglia2018relational,loiseau2018constrained,cranmer2019learning,mohebujjaman2019physically,cranmer2020lagrangian}. 
In fluid mechanics, sparse model discovery has been used to develop interpretable nonlinear models that enforce known physics by construction~\citep{Brunton2016pnas,loiseau2018constrained,loiseau2018sparse}. 
Particular emphasis is put on understanding how these models alter or retain the ``direct energy cascade'' coming from the interaction of terms in the Navier-Stokes equations; in Hall-MHD there are also inverse cascades and bidirectional cascades~\cite{pouquet2019helicity}, complicating the analysis of model stability and generalizability.

In this work we develop theoretical foundations for principled projection-based and data-driven plasma models. In fluid mechanics, careful development of a dimensionalized inner product  enabled the extension of POD from incompressible to compressible fluid flows~\citep{rowley2004model}. 
It is also common in fluid mechanics to obtain nonlinear reduced-order models by Galerkin projection of the Navier-Stokes equations onto POD modes, making it possible to enforce known symmetries and conservation laws, such as conservation of energy~\citep{noack2011galerkin,balajewicz2013low,Schlegel2015jfm,Carlberg2017jcp}. These symmetries have recently been utilized to constrain the identification of data-driven fluid models~\cite{loiseau2018constrained}.
The present work extends and unifies these three innovations for compressible plasmas, enabling a wealth of advanced modeling and control machinery. The culmination of this work is illustrated by accurately forecasting the evolution of a 3D isothermal Hall-MHD simulation of the HIT-SI experiment~\cite{jarboe2006spheromak}, described in detail in Appendix~\ref{sec:HITSI}. 

\begin{figure*}
\begin{subfigure}[t]{0.3\textwidth}
    \centering
     \includegraphics[width=\linewidth]{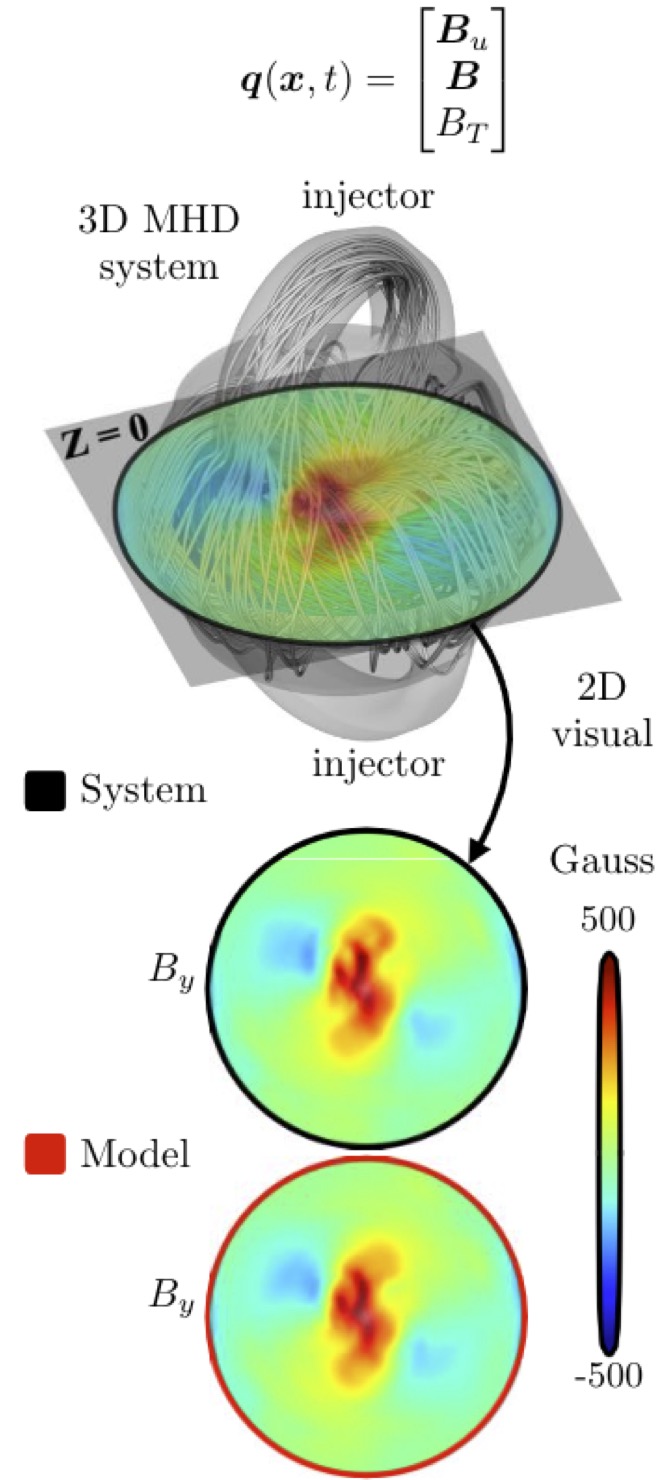}
    \caption{}
\end{subfigure}
\begin{subfigure}[t]{0.3\textwidth}
    \centering
     \includegraphics[width=\linewidth]{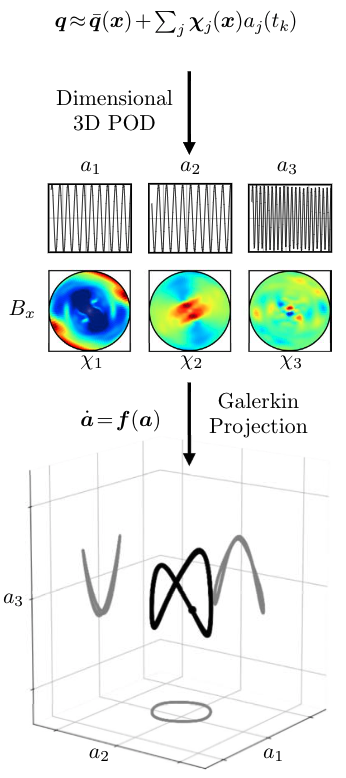}
    \caption{}
\end{subfigure}
\begin{subfigure}[t]{0.3\textwidth}
    \centering
     \includegraphics[width=\linewidth]{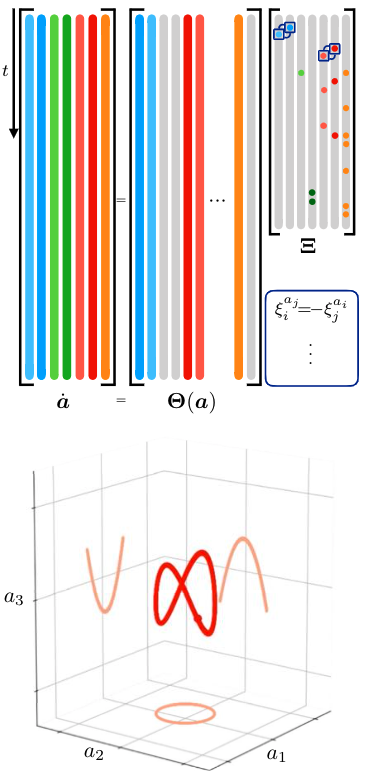}
    \caption{}
\end{subfigure}
    \caption{Proposed approach for filling in lower rungs of the plasma model hierarchy: (a) Collect data, (b) perform projection-based model reduction (c) discover data-driven models using physics-constrained system identification. Throughout this work, see the online version of this manuscript for the color bars and other color-coding.}
    \label{fig:overview}
\end{figure*}

In Section~\ref{sec:pod}, we propose a formalism for reduced plasma models, and in Section~\ref{sec:conservation_laws} we derive how global conservation laws in MHD manifest in the subsequent reduced-order models. After establishing a framework for projection-based model reduction in Hall-MHD, in Section~\ref{sec:sindy} we utilize physics-constrained machine learning to discover low-dimensional plasma models directly from data.
Figure~\ref{fig:overview} summarizes the steps for using plasma data with projection-based model reduction and data-driven system identification. As described in detail in Sections~\ref{sec:pod} and~\ref{sec:sindy}, data is collected, projected onto a low-dimensional basis, constrained by any known physical laws, and analyzed to find a descriptive model for its evolution. This last task may be achieved through projection-based, data-driven, or hybrid methods.

\section{Projection-based \\reduced-order models in MHD}\label{sec:pod}
Despite the many ways to obtain low-dimensional models, projection-based model reduction such as Galerkin methods have seen significant use and remarkable success in fluid mechanics. This success stems from the direct ties with the first-principles physics, lending interpretability to Galerkin models so that physical characteristics such as linear stability can be investigated. 
When the low-dimensional basis is obtained from the POD, the resulting models are referred to as POD-Galerkin models. Since the 1990's, it has been common in fluid mechanics to obtain these nonlinear reduced-order models by Galerkin projection of the Navier-Stokes equations onto POD modes. Further developments increased the utility of these methods, including a dimensionalized inner product which enabled the extension of POD-Galerkin models from incompressible to compressible fluids~\cite{rowley2004model}. Towards the goal of extending these developments into plasma physics, we first need to define a new inner product for MHD plasmas in Section~\ref{sec:energy_inner_prod} and then illustrate how the subsequent proper orthogonal decomposition is performed for plasma datasets in Section~\ref{sec:pod_review}. This formalism facilitates our the derivation of a POD-Galerkin model for Hall-MHD in Section~\ref{sec:generic_pod_galerkin}.

The present work adapts and extends these innovations for plasmas, enabling a wealth of advanced modeling and control machinery. For clarity and robust connection with the Galerkin literature in fluid mechanics, we primarily consider models which are quadratic in nonlinearity. This includes ideal MHD, incompressible Hall-MHD, and variants such as compressible Hall-MHD with a slowly time-varying density, which together describe a broad class of space and laboratory plasmas~\cite{Schnack2006,ma2001hall,krishan2004magnetic,Ebrahimi2011,Ferraro2012,kaptanoglu2020two}. 

\subsection{An MHD energy inner product\label{sec:energy_inner_prod}}
Traditional use of the POD on the MHD fields (velocity, magnetic, and temperature) would either require separate decompositions for $\bm{u}$, $\bm{B}$, and $T$, or an arbitrary choice of dimensionalization. However, separate decompositions of the fields obfuscates the interpretability and increases the complexity of a low-dimensional model, and choosing the units of the combined matrix of measurement data can have a significant impact on the performance and energy spectrum of the resulting POD basis. %
Inspired by the inner product defined for compressible fluids~\cite{rowley2004model}, we define an inner product for MHD through
\medmuskip=0mu
\thinmuskip=2mu
\thickmuskip=2mu
\begin{align}
    \label{eq:q_def}
\bm{q}(\bm{x},t) = \begin{bmatrix}\bm{B}_u \\ \bm{B} \\ B_T\end{bmatrix},\,\,\,
\bm{B}_u = \sqrt{\rho\mu_0}\bm{u}, \,\,\, B_T = \sqrt{\frac{4\rho \mu_0k_bT}{m_i(\gamma-1)}}.
\end{align}
Here $\rho$ is the mass density, $k_b$ is Boltzmann's constant, $\mu_0$ is the permeability, $m_i$ is the ion mass, $\gamma$ is the adiabatic index, ${p = 2\rho T/m_i}$ is the plasma pressure, and the total plasma energy is
\begin{equation}
    \label{eq:inner_prod}
W = \frac{1}{2\mu_0}\langle\bm{q},\bm{q}\rangle = \int \left(\frac{1}{2}\rho u^2+\frac{B^2}{2\mu_0}+\frac{p}{\gamma-1}\right) d^3\bm{x}.
\end{equation}
Normalizing the MHD fields to magnetic field units produces a natural interpretation of inner products of the vector $\bm{q}$ as the total plasma energy. This formulation is also useful because reduced order models built for $\bm{q}$ can be constrained by conservation of energy via Eq.~\eqref{eq:inner_prod}, as we illustrate in detail in Section~\ref{sec:conservation_laws}.

\subsection{Proper orthogonal decomposition \\ for plasma datasets\label{sec:pod_review}}
The POD is already used extensively for interpreting plasma physics data across a range of parameter regimes~\cite{dudok1994biorthogonal,levesque2013multimode,galperti2014development,van2014use,hansen2015numerical}, but some formalism is required to effectively use it for modeling and forecasting.
For POD, a set of point measurements at time $t_k$ are arranged in a vector $\bm{q}_k\in\mathds{R}^D$, called a snapshot, where the dimension $D$ is the product of the number of spatial locations and the number of variables measured at each point. For instance, we could have obtained the magnetic field data from $D/3$ magnetic probes that measure the magnetic field components at a fixed location and sampling rate.  
Now we assume that the data is sampled at some times $t_1,\hspace{0.05in}t_2,\hspace{0.05in}...,\hspace{0.05in}t_M$, arranged in a matrix  $\bm{X}\in\mathds{R}^{D\times M}$, and the average in time $\bar{\bm{q}}$ is subtracted off. The singular value decomposition (SVD) provides a low-rank approximation
 \medmuskip=2mu
 \thinmuskip=2mu
 \thickmuskip=2mu
\begin{eqnarray}
\bm{X}
= \overset{\text{\normalsize time}}{\left.\overrightarrow{\overset{~~}{\begin{bmatrix}
q_1(t_1) & q_1(t_2) & \cdots & q_1(t_M)\\
q_2(t_1) & q_2(t_2) & \cdots & q_2(t_M)\\
\vdots & \vdots & \ddots & \vdots \\
q_D(t_1) & q_D(t_2) & \cdots & q_D(t_M)
\end{bmatrix}}}\right\downarrow}\begin{rotate}{270}\hspace{-.125in}state~~\end{rotate} \hspace{.125in}= \bm{U}\bm{\Sigma}\bm{V}^*,
\label{Eq:DataMatrix}
\end{eqnarray}
where $\bm{U}\in\mathds{R}^{D\times D}$ and $\bm{V}\in\mathds{R}^{M\times M}$ are unitary matrices, and $\bm{\Sigma} \in \mathds{R}^{D\times M}$ is a diagonal matrix containing non-negative and decreasing entries $s_{jj}$ called the singular values of $\bm{X}$. $\bm{V}^*$ denotes the complex-conjugate transpose of $\bm{V}$.  
The singular values indicate the relative importance of the corresponding columns of $\bm{U}$ and $\bm{V}$ for describing the spatio-temporal structure of $\bm{X}$.

It is often possible to discard small values of $\bm{\Sigma}$, resulting in a truncated matrix $\bm{\Sigma}_r\in\mathds{R}^{r\times r}$. With the first $r\ll \min(D,M)$ columns of $\bm{U}$ and $\bm{V}$, denoted $\bm{U}_r$ and $\bm{V}_r$, we have
\begin{equation}
\bm{X}\approx \bm{U}_r\bm{\Sigma}_r\bm{V}_r^*.
\label{eq:svd_trunc}
\end{equation}
The truncation rank $r$ is typically chosen to balance accuracy and complexity~\citep{brunton2019data}. 
The computational complexity of the SVD is $\mathcal{O}(DM^2+M^3)$~\cite{golub1996cf}, although there are randomized singular value decompositions~\cite{frieze2004fast,liberty2007randomized,woolfe2008fast} for very large problems that can be as fast as $\mathcal{O}(DM\log(r))$. Therefore, even for $r \gg 1$, the SVD typically produces significant computational speedup over codes which evolve the full spatio-temporal dynamics.
The computational speed~\cite{golub1996cf,woolfe2008fast} of the SVD also enables online computations to update a model for real-time control. 

To proceed, a well-defined SVD requires that the measurements in $\bm{X}$ have the same physical dimensions.
With a dimensionalized measurement vector $\bm{q}$, the matrix $\bm{X}^*\bm{X}$ satisfies
\medmuskip = 0.5mu 
\begin{align}
    \label{eq:snapshots1}
    \bm{X}^*\bm{X} \approx 
    \langle\bm{q}(t_k),\bm{q}(t_m)\rangle, \qquad k,m\in \{1,2,...,M\}. 
\end{align}
The equality is not exact because the inner product (an integral) is approximated by the discrete sum from the matrix product of $\bm{X}^*\bm{X}$, but it is important that we can relate the matrix $\bm{X}$ to the total plasma energy through Eq.~\eqref{eq:inner_prod}.
The temporal SVD modes, or chronos, $\bm{v}_{j}$ are the columns of $\bm{V}_r$. The spatial modes, or topos, $\bm{\chi}$ form the columns of $\bm{U}_r$. 
We scale 
$    a_{j}(t_k) = 
    {v_{j}(t_k)}/{\sum_{j=1}^r\max_k|v_{j}(t_k)|}$.
Finally, 
\begin{equation}
    \label{eq:q_expansion}
    \bm{q}(\bm{x}_i,t_k) \approx \bar{\bm{q}}(\bm{x}_i) + 
    \sum_{j=1}^r \bm{\chi}_j(\bm{x}_i)a_j(t_k).
\end{equation}
We have absorbed the normalization of $a_{j}(t_k)$ and the singular values into the definition of $\bm{\chi}_j(\bm{x}_i)$. By construction $\langle \bm{\chi}_i,\bm{\chi}_j \rangle \propto \delta_{ij}$. 
Note that, in principle, we could have expanded $\bm{q}$ in any set of modes, although orthonormal modes are preferred because this property facilitates the analysis in Section~\ref{sec:conservation_laws}. Non-orthogonal modes are also suitable, but introduce a complication in the form of a mass matrix~\cite{rempfer1994dynamics}.
The advantage of the POD basis is that the modes are ordered by energy content; a truncation of the system still captures a majority of the dynamics. 
A separate POD of each of the MHD fields would lead to three sets of POD modes with independent time dynamics and mixed orthogonality properties.
In contrast, our approach captures all the fields simultaneously, resulting in a single set of modes $a_i(t)$ in Eq.~\eqref{eq:q_expansion}.

\begin{figure*}
\vspace{.3in}
    \centering
 \begin{subfigure}[t]{0.5\textwidth}
    \includegraphics[width=0.99\linewidth]{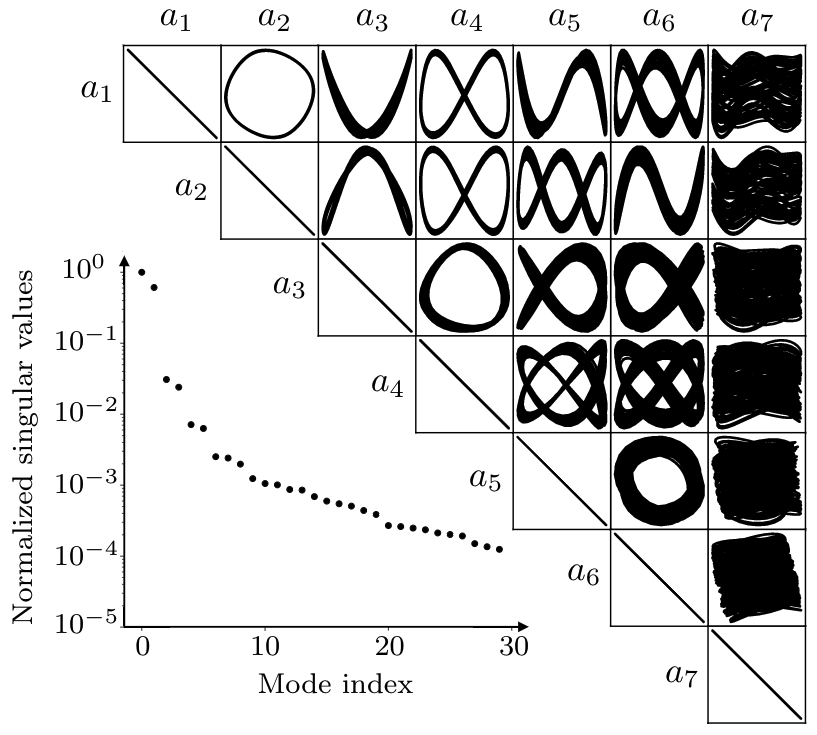}
    \caption{}
\end{subfigure}
 \begin{subfigure}[t]{0.4675\textwidth}
    \includegraphics[width=0.85\linewidth]{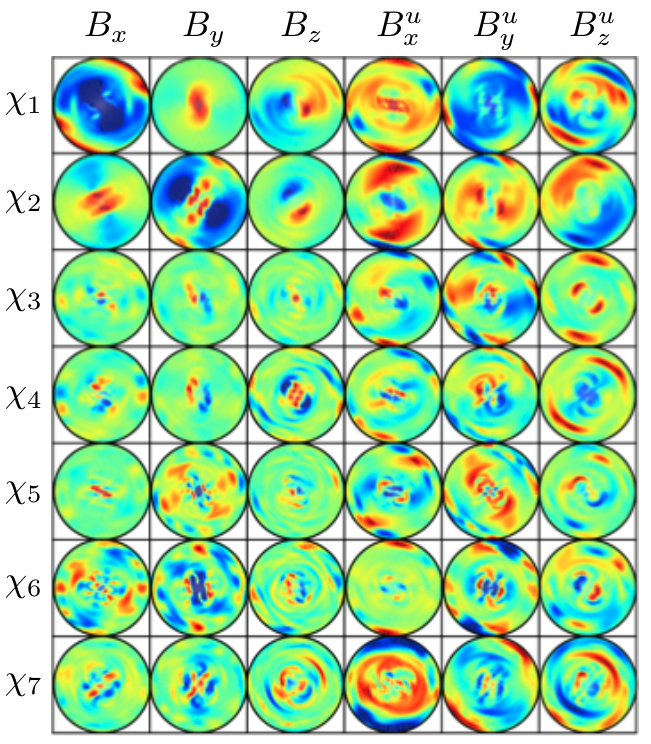}
    \caption{}
\end{subfigure}
 \begin{subfigure}[t]{0.95\textwidth}
    \includegraphics[width=0.99\linewidth]{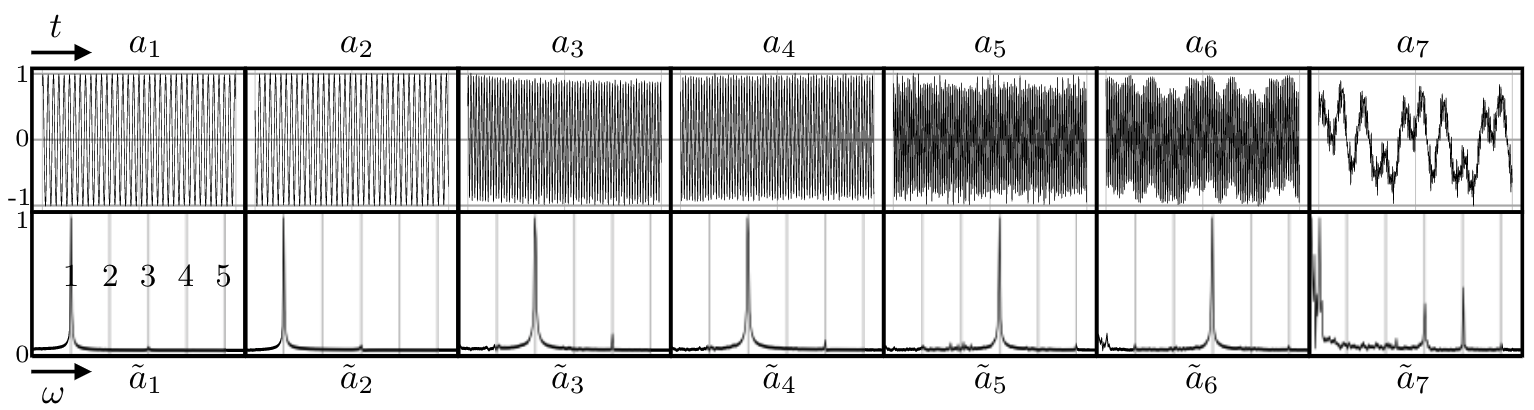}
    \caption{}
\end{subfigure}
    \caption{The first seven POD modes for a 3D isothermal Hall-MHD simulation of the HIT-SI device detailed in Appendix~\ref{sec:HITSI}. The mean-flow-subtracted chronos indicate that the primary dynamics are forcing at the driving injector frequency and its harmonics; (a) Mode pair trajectories evolved in time and the corresponding singular values; (b) 3D spatial modes in the $Z = 0$ midplane illustrate a complicated mix of length scales; (c) Normalized temporal modes and corresponding Fourier transforms exhibit harmonics of the driving frequency, labeled 1-5.}
    \label{fig:pod_modes}
   \vspace{-0.2in}
\end{figure*}

An example of this decomposition is illustrated in Fig.~\ref{fig:pod_modes} for a 3D isothermal Hall-MHD simulation, described in detail in Appendix~\ref{sec:HITSI} and modeled in Section~\ref{sec:sindy_results}; the dominant dynamics are harmonics and sub-harmonics of a driving frequency imposed by two actuating injectors on the top and bottom of the device. In general, examining the structure and symmetry in the spatial and temporal POD modes can inform physical understanding. For instance, in Fig.~\ref{fig:pod_modes}, the short-wavelength structures exhibited in the 3D spatial modes derive both from dispersive whistler waves via the Hall term and the small characteristic scale associated with the injectors (actuators). 
The steep fall-off in the singular values also indicates that models of only the first few modes would be enough to accurately forecast and control the dominant dynamics. While we show stable and accurate $r=16$ models in Section~\ref{sec:sindy_results} to illustrate the strengths of our methods for more complicated dynamics, in our test case it is true that more modest models of three or four modes can already achieve reasonable forecasting accuracies. 

\subsection{\label{sec:generic_pod_galerkin} POD-Galerkin models}
Now that we have an expansion of the fields in a low-dimensional basis in Eq.~\eqref{eq:q_expansion}, we can project the Hall-MHD equations onto these POD modes in order to obtain a POD-Galerkin model.
Hall-MHD, using the definitions of the electromagnetic current $\mu_0\bm{J} = \nabla\times\bm{B}$, electron fluid velocity $\bm{u}_e = \bm{u} - \bm{J}/ne$, electron and ion temperature $T_e=T_i=T$, and the definitions in Eq.~\eqref{eq:q_def}, can be written:
\medmuskip = 0.0mu 
\begin{widetext}\small
\begin{align} \label{eq:continuity_equation}
    \dot{\rho} = &-\nabla\cdot\left(\sqrt{\frac{\rho}{\mu_0}}\bm{B}_u\right),
    \\  \notag
    \dot{\bm{B}} =& 
 \nabla \times \left[ \frac{1}{\sqrt{\rho\mu_0}}\left(\bm{B}_u \times \bm{B}- d_i  ((\nabla\times\bm{B}) \times \bm{B}) \right)  \right] + \eta_0\rho^{\frac{3}{2}}B_T^{-3} \nabla^2\bm{B} + \frac{d_i}{\sqrt{\rho\mu_0}}(\gamma-1)B_T\nabla B_T\times\frac{\nabla \rho}{2\rho},
    \\ \notag
\small
    \dot{\bm{B}}_u = &-\frac{1}{\sqrt{\rho\mu_0}}\left[\frac{1}{2}\bm{B}_u\nabla\cdot\bm{B}_u + \bm{B}_u\cdot\nabla\bm{B}_u-\frac{1}{4\rho}\bm{B}_u(\nabla\rho\cdot\bm{B}_u) -(\nabla\times\bm{B})\times\bm{B}+\frac{(\gamma-1)B_T^2}{2}\frac{\nabla\rho}{\rho} - (\gamma-1)B_T\nabla B_T\right] \\ \notag &+\nu\left[\nabla^2\bm{B}_u-\frac{\nabla^2\rho}{2\rho}\bm{B}_u +\frac{3\bm{B}_u}{4\rho^2}(\nabla\rho)^2+ \frac{1}{\rho}(\nabla\rho\cdot\nabla)\bm{B}_u-\frac{1}{6\rho}\nabla(\nabla\rho\cdot\bm{B}_u) +\frac{1}{4\rho^2}(\nabla\rho\cdot\bm{B}_u)\nabla\rho + \frac{1}{3}\nabla(\nabla\cdot\bm{B}_u) -
    \frac{1}{6\rho}(\nabla\cdot\bm{B}_u)\nabla\rho\vphantom{\nabla^2\bm{B}_u-\frac{\nabla^2\rho}{2\rho}\bm{B}_u +\frac{3\bm{B}_u}{4\rho^2}\nabla\rho\cdot\nabla\rho+ \frac{1}{\rho}(\nabla\rho\cdot\nabla)\bm{B}_u)-\frac{1}{6\rho}\nabla(\nabla\rho\cdot\bm{B}_u)}\right],
    \\ \notag
    \dot{B}_T =&-\frac{1}{\sqrt{\rho\mu_0}} \left[\bm{B}_u\cdot\nabla B_T - \gamma B_T (\nabla\cdot\bm{B}_u - \frac{\nabla\rho}{2\rho}\cdot\bm{B}_u) \right] - \frac{2}{B_T}\left[\nabla\cdot\bm{h} + Q_\text{visc} \right] + 4\eta_0\rho^{\frac{3}{2}}B_T^{-4}(\nabla\times\bm{B})^2,
\normalsize
\end{align}
where we have used that $\nabla\cdot\bm{B} = 0$ and the definitions of the heat flux $\bm{h}$ and viscous heating $Q_\text{visc}$,
\begin{align}
    \bm{h} &= -\frac{ (\gamma-1)B_T}{4} \left[ \chi_{\parallel} \hat{\bm{b}} \hat{\bm{b}} + \chi_{\perp} \left( \bm{I} - \hat{\bm{b}} \hat{\bm{b}} \right) \right] \cdot \left(\nabla B_T - B_T\frac{\nabla\rho}{\rho}\right), 
    \\ \notag
    Q_\text{visc} &= -\tilde{\nu}( \nabla \bm{B}_u - \bm{B}_u\frac{\nabla\rho}{2\rho})^T \bm{:} \left[( \nabla \bm{B}_u - \bm{B}_u\frac{\nabla\rho}{2\rho}) + ( \nabla \bm{B}_u - \bm{B}_u\frac{\nabla\rho}{2\rho})^T - \frac{2}{3} \bm{I} (\nabla \cdot \bm{B}_u - \bm{B}_u\cdot\frac{\nabla\rho}{2\rho})\right].
\end{align}
\end{widetext}
Here $\nu = \tilde{\nu}$/$\rho$ is the dynamic viscosity, $\eta = \eta_0T^{-\frac{3}{2}} \propto \rho^{\frac{3}{2}}B_T^{-3}$ is the Spitzer resistivity~\cite{spitzer2006physics}, $d_i = m_i$/$(e\sqrt{\rho\mu_0})$ is the ion inertial length, and $\chi_{\perp}$ and $\chi_{\parallel}$ are the anisotropic Braginskii thermal diffusivities with temperature and magnetic field dependencies~\cite{braginskii1965reviews}. Although many of the nonlinear terms are only quadratic in $\bm{q}$, 
we consider the isothermal limit and limit of time-independent density to restrict ourselves to the pure quadratic nonlinear case:
\begin{align}
\label{eq:qpde}
\dot{\bm{q}} &= \bm{C} +  \bm{L}(\bm{q})+\bm{Q}(\bm{q},\bm{q}), \\ \notag
\bm{C} &= \begin{bmatrix}
-\frac{(\gamma-1)B_T^2}{2}\sqrt{\frac{1}{\mu_0\rho}}\frac{\nabla\rho}{\rho} \\ 
0 \\
0
\end{bmatrix}, \\ \notag 
\bm{Q}(\bm{q},\bm{q}) &= \begin{bmatrix}
-\frac{1}{\sqrt{\rho\mu_0}}\left(\bm{B}_u(\nabla\cdot\bm{B}_u)+\bm{B}_u\cdot\nabla\bm{B}_u- (\nabla\times\bm{B})\times\bm{B}\right) \\ 
\nabla \times \left( \frac{1}{\sqrt{\rho\mu_0}}\left(\bm{B}_u \times \bm{B}- d_i(\nabla\times\bm{B}) \times \bm{B}\right)   \right) \\
0
\end{bmatrix}, \\ \notag
\bm{L}(\bm{q}) &= \begin{bmatrix}
\nu\left(\nabla^2\bm{B}_u-
\cdots - \frac{1}{3\rho}(\nabla\cdot\bm{B}_u)\nabla\rho\right) \\ 
\frac{\eta}{\mu_0}\nabla^2\bm{B}  \\
0
\end{bmatrix}. 
\end{align}
Increasingly sophisticated models may be tractable in future work, since the data-driven approach that we adopt in Section~\ref{sec:sindy} is not limited to quadratic nonlinearities. Other models, such as those assuming incompressibility and finite temperature evolution, can also be derived straightforwardly from the results here.
Substituting Eq.~\eqref{eq:q_expansion} into Eq.~\eqref{eq:qpde} and utilizing the orthonormality of the $\bm{\chi}_j$ produces: 
\begin{align}
\label{eq:Galerkin_model}
\dot{a}_i(t) &= C_i^0 + \sum_{j=1}^rL^0_{ij}a_j + \sum_{j,k=1}^r Q^0_{ijk}a_ja_k, \\ \notag
C_i^0 &= \langle \bm{C} + \bm{L}(\bar{\bm{q}}) + \bm{Q}(\bar{\bm{q}},\bar{\bm{q}}),\bm{\chi}_i\rangle,\\ \notag
L^0_{ij} &= \langle \bm{L}(\bm{\chi}_j) + \bm{Q}(\bar{\bm{q}},\bm{\chi}_j) + \bm{Q}(\bm{\chi}_j,\bar{\bm{q}}),\bm{\chi}_i\rangle, \\ \notag 
Q^0_{ijk} &= \langle \bm{Q}(\bm{\chi}_j,\bm{\chi}_k),\bm{\chi}_i\rangle.
\end{align}
The model is quadratic in the temporal POD modes $a_i(t)$. 
The zero superscript is meant to distinguish the coefficient tensors $C_i^0$, $L_{ij}^0$, and $Q_{ijk}^0$ from the spatio-temporal operators $\bm{C}$, $\bm{L}$, and $\bm{Q}$.
If $\bar{\bm{q}}$ satisfies the steady-state MHD equations, then $C_i^0 = 0$ for all $i$. This is a reasonable assumption for any approximately steady-state device, such as a tokamak, which can be sustained for many characteristic timescales. 
In contrast to Eq.~\eqref{eq:Galerkin_model}, a Galerkin model based on separate POD expansions for each field would involve significant mixing and a lack of orthonormality $\langle \bm{\chi}^{\bm{u}}_i,\bm{\chi}^{\bm{B}}_j \rangle \neq \delta_{ij}$ between the POD modes. 
Although Eq.~\eqref{eq:Galerkin_model} contains only quadratic nonlinearities, 
the influence of truncated low-energy modes can sometimes be modeled with cubic nonlinearities in the Galerkin model~\cite{Noack2003jfm,loiseau2018constrained}.

\subsection{Relation to Fourier-Galerkin methods}
Similar analytic Fourier-Galerkin models (also called MHD shell models) have been used for modeling incompressible MHD turbulence~\cite{plunian2013shell}. Shell models in MHD have primarily been used to describe the statistics of homogeneous and isotropic turbulence in spectral space, rather than as reduced order models~\cite{biskamp1994cascade}. The differences in application likely stem from shell models preserving the MHD invariants within each triad of wave vectors but POD models providing a dataset-tailored and energy-optimal basis. However, in various homogeneous and symmetric limits the POD reduces to the Fourier basis~\cite{couplet2003intermodal, holmes2012turbulence}. In both Fourier-Galerkin and POD-Galerkin models, truncation of the model at some rank $r$ can lead to under-resolving the dissipation rate or approximately breaking the global conservation laws, and a closure scheme may be required to re-introduce the full dissipation. Additionally, if energy is not conserved, as in some dissipative MHD models, the stability of the truncated system is no longer guaranteed. Two advantages of the data-driven approach in Section~\ref{sec:sindy} over either POD-Galerkin or Fourier-Galerkin is that 1) we need not laboriously compute the coefficients in Eq.~\eqref{eq:Galerkin_model} from full state knowledge, and 2) we can enforce global energy or cross-helicity conservation directly into the truncated model (even without energy conservation, we may be able to enforce other generic stability properties~\cite{kaptanoglu2021promoting}). Lastly, preserving the features of the direct energy cascade in truncated Galerkin and data-driven models for incompressible fluid flows is a current field of research. Since even this ``simple case'' is unsettled, there is much research to be done regarding the preservation of direct, inverse, and even bidirectional cascades~\cite{pouquet2019helicity} in truncated Galerkin models for Hall-MHD beyond highly simplified cases such as isotropic, incompressible, isothermal Hall-MHD turbulence on simple geometries.

\vspace{0.2in}
\section{Deriving constraints on projection-based models\label{sec:conservation_laws}}
We have successfully obtained a POD-Galerkin model for the dynamic fields in Hall-MHD. However, there is substantial additional structure in the coefficients in Eq.~\eqref{eq:Galerkin_model} because local and global MHD conservation laws are in principle retained in this low-dimensional basis.
Vanishing $\nabla\cdot\bm{B}$ and the linear independence of the temporal POD modes produce
\begin{align}
    \nabla\cdot\bm{\chi}_i^B = 0,\,\,\,\,\,\,\forall i.
\end{align}
In other words, there is a local divergence constraint for each of the $\bm{\chi}_i^B$, but this does not produce insight into the coefficients defined in Eq.~\eqref{eq:Galerkin_model}
In contrast, global energy conservation produces substantial constraints on the structure of the Galerkin model coefficients.
\subsection{\label{sec:global_energy}Global conservation of energy}
For an examination of the global conservation laws, we consider isothermal Hall-MHD with a very slowly time-varying density. This model reduces to ideal MHD and incompressible resistive or Hall MHD in the appropriate limits, and produces (Galtier~\cite{galtier2016introduction} Eq. 3.22)
\medmuskip=-2mu
\thickmuskip=-1mu
\thinmuskip=-1mu
\begin{align}
\label{eq:galtier_deriv}
    \frac{\partial W}{\partial t} = &- \int \left[\tilde{\nu}(\nabla\times\bm{u})^2+\frac{\eta}{\mu_0}(\nabla\times\bm{B})^2 + \frac{4}{3}\tilde{\nu}(\nabla\cdot\bm{u})^2\right] d^3\bm{x}  \\ \notag
    &-\oint \left[\left(\frac{1}{2}\rho u^2+p\right)\bm{u} + \bm{P} - \frac{4}{3}\tilde{\nu}(\nabla\cdot\bm{u})\bm{u} - \tilde{\nu}\bm{u}\times(\nabla\times\bm{u})\right]\cdot\hat{\bm{n}}dS.
\end{align}
Here $\hat{\bm{n}}$ is a unit normal vector to the boundary, and 
\medmuskip=1mu
\thickmuskip=1mu
\thinmuskip=1mu
\begin{align}
\label{eq:poyntingvec}
\bm{P} = \frac{1}{\mu_0}\bm{E}\times\bm{B} = \frac{\bm{u}_e}{\mu_0}\cdot(B^2\bm{I} - \bm{B}\bm{B})  - \frac{\eta}{\mu_0^2}(\nabla\times\bm{B})\times\bm{B},
\end{align}
is the Poynting vector ($\bm{E}$ is the electric field), which is often an imposed and experimentally-known function of space and time. 
Omission of the Hall term changes $\bm{u}_e$ to $\bm{u}$ in Eq.~\eqref{eq:poyntingvec}. Even with temperature evolution, the electron diamagnetic term in $\bm{P}$ does not alter the energy balance if Dirichlet conditions are used for $\rho$ and $T$. To simplify, we assume that $\bm{u}\cdot\hat{\bm{n}} = \bm{u}\times\hat{\bm{n}} = 0$, $\bm{J}\cdot\hat{\bm{n}} = 0$, and $\bm{B}\cdot\hat{\bm{n}} = 0$ at the wall, consistent with the Hall-MHD HIT-SI simulation described in Appendix~\ref{sec:HITSI} and modeled in Section~\ref{sec:sindy_results}.
Now assume steady-state, define $a_0(t) = 1$, and substitute Eq.~\eqref{eq:q_expansion} into Eq.~\eqref{eq:galtier_deriv}, 
\medmuskip=1mu
\begin{widetext}
\begin{align}
\label{eq:W_decomp_reduced}
    0 \approx \frac{\partial W}{\partial t} &= \oint\frac{\eta}{\mu_0^2} ((\nabla\times\bm{B})\times\bm{B})\cdot\hat{\bm{n}}dS - \int \left[\frac{\nu}{\mu_0}(\nabla\times\bm{B}_u-\frac{\nabla\rho}{2\rho}\times\bm{B}_u)^2+\frac{\eta}{\mu_0^2}(\nabla\times\bm{B})^2 + \frac{4}{3}\frac{\nu}{\mu_0}(\nabla\cdot\bm{B}_u-\frac{\nabla\rho}{2\rho}\cdot\bm{B}_u)^2\right] d^3\bm{x}, \\
\notag
&= W^C + \sum_{i=1}^rW_i^L a_i + \sum_{i,j=1}^rW_{ij}^\text{Q}a_ia_j 
= \sum_{i, j=0}^rW_{ij}^Qa_ia_j,
\end{align}
We have padded the matrix in the last step so that $W_{0i}^Q = 0$, $W_{i0}^Q = W_i^L$ for $i \in \{1,...,r\}$, and $W^Q_{00} = W^C$. Eq.~\eqref{eq:W_decomp_reduced} is generally satisfied for anti-symmetric $W_{ij}^Q$, from which it follows that
\medmuskip=-2mu
\thinmuskip=-2mu
\thickmuskip=-2mu
\begin{align}
\label{eq:antisymmetry_energy}
0 = W^\text{Q}_{00} = &\frac{\eta}{\mu_0}\oint \left[(\nabla\times\bar{\bm{B}})\times\bar{\bm{B}}\right]\cdot\hat{\bm{n}}dS -\int\left[\nu(\nabla\times\bar{\bm{B}}_u-\frac{\nabla\rho}{2\rho}\times\bar{\bm{B}}_u)^2+ \frac{\eta}{\mu_0}(\nabla\times\bar{\bm{B}})^2+\frac{4}{3}\nu(\nabla\cdot\bar{\bm{B}}_u-\frac{\nabla\rho}{2\rho}\cdot\bar{\bm{B}}_u)^2\right]d^3\bm{x}, \\ \notag
0 = W_{i0}^\text{Q} = &\frac{\eta}{\mu_0}\oint \left[(\nabla\times\bar{\bm{B}})\times\bm{\chi}_i^{B} +(\nabla\times\bm{\chi}_i^B)\times\bar{\bm{B}} \right]\cdot\hat{\bm{n}}dS \\ \notag &-2\int\left[\nu(\nabla\times\bar{\bm{B}}_u-\frac{\nabla\rho}{2\rho}\times\bar{\bm{B}}_u)\cdot(\nabla\times\bm{\chi}^{B_u}_i-\frac{\nabla\rho}{2\rho}\times\bm{\chi}^{B_u}_i)+\frac{\eta}{\mu_0}(\nabla\times\bar{\bm{B}})\cdot(\nabla\times\bm{\chi}^{B}_i)+ \frac{4}{3}\nu(\nabla\cdot\bar{\bm{B}}_u-\frac{\nabla\rho}{2\rho}\cdot\bar{\bm{B}}_u)\cdot(\nabla\cdot\bm{\chi}^{B_u}_i-\frac{\nabla\rho}{2\rho}\cdot\bm{\chi}^{B_u}_i) \vphantom{\nu(\nabla\times\bm{\chi}^{B_u}_i-\frac{\nabla\rho}{2\rho}\times\bm{\chi}^{B_u}_i)\cdot(\nabla\times\bm{\chi}^{B_u}_j}\right] d^3\bm{x}, \\ \notag
W_{ij}^\text{Q} = &-W_{ji}^\text{Q} = \frac{\eta}{\mu_0}\oint \left[(\nabla\times\bm{\chi}_i^B)\times\bm{\chi}_j^B\right]\cdot\hat{\bm{n}}dS \\ \notag &-\int\left[\nu(\nabla\times\bm{\chi}^{B_u}_i-\frac{\nabla\rho}{2\rho}\times\bm{\chi}^{B_u}_i)\cdot(\nabla\times\bm{\chi}^{B_u}_j-\frac{\nabla\rho}{2\rho}\times\bm{\chi}^{B_u}_j)+\frac{\eta}{\mu_0}(\nabla\times\bm{\chi}^{B}_i)\cdot(\nabla\times\bm{\chi}^{B}_j)+  \frac{4}{3}\nu(\nabla\cdot\bm{\chi}^{B_u}_i-\frac{\nabla\rho}{2\rho}\cdot\bm{\chi}^{B_u}_i)\cdot(\nabla\cdot\bm{\chi}^{B_u}_j-\frac{\nabla\rho}{2\rho}\cdot\bm{\chi}^{B_u}_j)\vphantom{\nu(\nabla\times\bm{\chi}^{B_u}_i-\frac{\nabla\rho}{2\rho}\times\bm{\chi}^{B_u}_i)\cdot(\nabla\times\bm{\chi}^{B_u}_j}\right]d^3\bm{x}. 
\end{align}
\medmuskip=-1mu
\thickmuskip=2mu
\thinmuskip=-1mu
\end{widetext}
Evaluating Eq.~\eqref{eq:antisymmetry_energy} and the Galerkin coefficients in Eq.~\eqref{eq:Galerkin_model} relies on the existence of all of the $\nabla\times\bm{\chi}_i$. These spatial POD modes are evaluated on a discrete set of spatial locations, but in practice we can always choose an interpolation such that the curl operator is well-defined. In such a case, $\nabla\times\chi_i^B$ and $\nabla\times\chi_i^{B_u}$ have natural interpretations as the spatial POD modes of the electromagnetic current and vorticity fields. However, in the present work these computations only serve as formal manipulations so we need not evaluate these curls; our data-driven method in Section~\ref{sec:sindy} uses sparse regression to determine these coefficients from data. Continuing on with our analysis, we can compute $a_i\dot{a}_i$ for $i \in \{1,...,r\}$,
\begin{align}
\label{eq:q2evo}
    a_i\dot{a}_i &= \sum_{i,j=1}^ra_i\frac{\partial a_j}{\partial t}\int \chi_i \chi_j d^3\bm{x} = \int \frac{1}{2}\frac{\partial q^2}{\partial t} d^3\bm{x} = \frac{\partial W}{\partial t},
\\
\label{eq:energy_conservation}
    a_i\dot{a}_i &= a_iC^0_i +  a_iL^0_{ij}a_j + a_i Q^0_{ijk}a_ja_k,\,\,\,\,\,\,\,\, i,j,k \in \{1,...,r\}.
\end{align}
First, note that $W^\text{Q}_{i0}=0$ produces $C^0_i = 0$ for all $i \in \{1,...,r\}$. There are no constant terms in the Galerkin model.
This is a physical consequence of our assumption that $\bar{q}$ is steady-state; nonzero constant terms would imply the possibility of unbounded growth in the energy norm. The anti-symmetry of $W_{ij}^Q$ for $i,j \in \{1,...,r\}$ constrains the quadratic structure of the energy $\bm{a}^T\bm{a}$, 
\begin{align}
\label{eq:L_constraint}
    \bm{a}^T\bm{L}^0\bm{a} \approx 0.
\end{align}
This physical interpretation is also clear; if the plasma is steady-state but has finite dissipation, the input power, here manifested through a purely quadratic Poynting flux $\bm{P} \propto \eta\bm{J}\times\bm{B}$, must be balancing these losses.  Finally, there are no cubic terms in the energy, implying 
\begin{align}
\label{eq:Q_constraint}
    \bm{a}^T\bm{Q}^0\bm{a}\bm{a} = 0,
\end{align}
or equivalently,
\begin{align}
    Q_{ijk}^0 + Q_{jik}^0 + Q_{kij}^0 = 0.
\end{align}
In other words, the quadratic nonlinearities in the Galerkin model of Eq.~\eqref{eq:Galerkin_model} are energy-preserving; this conclusion did not rely on any assumption of steady-state and energy-preserving structure in other quadratic nonlinearities is well-studied in fluid mechanics~\cite{Schlegel2015jfm,loiseau2018sparse,kaptanoglu2021promoting}.
The lack of nonlinear energy losses is a physical consequence coming from the boundary conditions $\bm{B}\cdot\hat{\bm{n}} = 0$, $\bm{J}\cdot\hat{\bm{n}} = 0$, $\bm{u}\cdot\hat{\bm{n}} = \bm{u}\times\hat{\bm{n}} = 0$ (and constant temperature).

\subsection{\label{sec:global_helicity}Global conservation of cross-helicity}
An analogous derivation can be done to further constrain the model-building for systems which conserve cross-helicity, although this is inappropriate for the Hall-MHD HIT-SI simulation in Section~\ref{sec:sindy_results}. 
Consider the local form of cross-helicity $H_c = \bm{u}\cdot\bm{B}$. Using Galtier~\cite{galtier2016introduction} Eq. (3.36), 
\begin{widetext}
\begin{align}
    \frac{\partial H_c}{\partial t} = &-\nabla\cdot\left[\left(\frac{u^2}{2}+ \frac{\gamma p}{(\gamma-1)\rho}\right)\bm{B} + \bm{u}\times(\bm{u}\times\bm{B})-\frac{d_i}{\sqrt{\rho\mu_0}}\bm{u}\times\left((\nabla\times\bm{B})\times\bm{B}\right) - \eta\bm{u}\times(\nabla\times\bm{B})  \right] \\ \notag &+ \nu \nabla\cdot\left(\bm{B}\times\bm(\nabla\times\bm{u}) + \frac{4}{3}(\nabla\cdot\bm{u})\bm{B}\right)-\frac{d_i}{\sqrt{\rho\mu_0}}(\nabla\times\bm{u})\cdot\left((\nabla\times\bm{B})\times\bm{B}\right) - (\eta+\nu)(\nabla\times\bm{B})\cdot(\nabla\times\bm{u}).
\end{align}
Consider again the simplifying case $\bm{J}\cdot\hat{\bm{n}} = 0$, $\bm{B}\cdot\hat{\bm{n}} = 0$, and $\bm{u}\cdot\hat{\bm{n}} = \bm{u}\times\hat{\bm{n}} = 0$. If global cross-helicity is conserved, the integral form is
\medmuskip=0mu
\begin{align}
    0 \approx \int \frac{\partial H_c}{\partial t}d^3\bm{x} = \int \left[\nu \frac{\nabla\rho}{\rho}\cdot\left(\bm{B}\times(\nabla\times\bm{u})+\frac{4}{3}(\nabla\cdot\bm{u})\bm{B}\right)-\frac{d_i}{\sqrt{\rho\mu_0}}(\nabla\times\bm{u})\cdot\left((\nabla\times\bm{B})\times\bm{B}\right) - (\eta+\nu)(\nabla\times\bm{B})\cdot(\nabla\times\bm{u}) \right]d^3\bm{x}.
\end{align}
\end{widetext}
Substituting in Eq.~\eqref{eq:q_expansion} produces terms up to cubic in the temporal POD modes,
\begin{align}
\label{eq:Hc_constraints}
    0 \approx \int\frac{\partial H_c}{\partial t}d^3\bm{x} &= \frac{\partial}{\partial t}(a_ia_j)\int \frac{1}{\sqrt{\rho\mu_0}}\bm{\chi}_i^{B_u}\cdot\bm{\chi}_j^Bd^3\bm{x}  \\ \notag 
    &= A_{ij}\frac{\partial}{\partial t}(a_ia_j) \to 
    \begin{bmatrix}
    A_{ij}C^0_ja_i \\
    A_{ij}L^0_{jk}a_ia_k \\
    A_{ij}Q^0_{jkl}a_ia_ka_l 
    \end{bmatrix} \approx \begin{bmatrix}
    0 \\
    0 \\
    0
    \end{bmatrix}
\end{align}
Note that if the system is energy-preserving, $C^0_j = 0$ for all j, so the first equality is already satisfied. The second equality determines that $A_{ij}L^0_{jk}$ is anti-symmetric under swapping $i$ and $k$, and energy-preservation in Eq.~\eqref{eq:L_constraint} produces anti-symmetry under swapping $j$ and $k$. The most straightforward solution is $L^0_{jk} = 0$ for all $j$,$k$;
this solution is precisely the ideal limit corresponding to $\eta = \nu = 0$. Since $A_{ij}$ is not symmetric, this constraint can also apply to systems which conserve cross-helicity despite finite dissipation.

Lastly, $A_{ij}Q^0_{jkl}$, containing only the contribution from the Hall-term, exhibits the same structure as (and is compatible with) our constraint on the energy-preserving nonlinearities in Eq.~\eqref{eq:Q_constraint}. The simplest solution is $A_{ij}Q^0_{jkl}=0$ for all $i,k,l$, since this corresponds to standard MHD without the Hall term. Like the analysis of the linear terms, this constraint indicates that it is possible that there are indices for which $A_{ij}Q^0_{jkl} \neq 0$ but overall satisfy $A_{ij}Q^0_{jkl}a_ia_ka_l = 0$, so that nonzero Hall contributions can still conserve cross-helicity. 
Lastly, although inviscid Hall-MHD has two other time-invariants, enforcing the remaining invariants may require alternative formulations to the one presented here, since derived fields like the vector potential are involved.
\subsection{\label{sec:vel_units}Conservation laws with velocity units}
The previous sections have illustrated that our choice of magnetic field units in Eq.~\eqref{eq:q_def} allowed us to relate global MHD conservation laws to the structure of the coefficients in the POD-Galerkin model. It is worth exploring any alterations in velocity units (in closer analogy to fluid dynamics) $\bm{q} = [\bm{u},\bm{u}_A, u_s]$,
\begin{align}
    u_s^2 &= \frac{4 T}{m_i(\gamma-1)}, \quad \bm{u}_A = \frac{\bm{B}}{\sqrt{\mu_0\rho}},
\\ 
\frac{1}{2}\langle\bm{q},\bm{q}\rangle
&= \frac{1}{2}\int \left( u^2+u_A^2+u_s^2\right) d^3\bm{x}.
\end{align}
We have defined a scaled plasma sound speed, $u_s$.  
If $\rho$ is uniform 
    $\rho\langle\bm{q},\bm{q}\rangle/2 = W$.
The isothermal and time-independent density assumptions allow us to derive another quadratic model in $\bm{q}$, for which a POD-Galerkin model is readily available (the form is identical to Eq.~\eqref{eq:Galerkin_model} but the POD modes and coefficients have changed). Once again, assume $\bm{u}\cdot\hat{\bm{n}}= \bm{u}\times\hat{\bm{n}}=0$, $\bm{J}\cdot\hat{\bm{n}} = 0$, and $\bm{B}\cdot\hat{\bm{n}}=0$ on the boundary, so that
\begin{align}
     \int \frac{\rho}{2}\frac{dq^2}{dt} d^3\bm{x} = \frac{\partial W}{\partial t}.
\end{align}
This is equivalent to Eq.~\eqref{eq:q2evo} in the particular case of time-independent density. Without this assumption, an extra term appears, proportional to $\int\bm{u}\cdot\nabla(u^2+u_A^2)d^3\bm{x}$. Although from dimensional analysis this term is potentially very large, this may not be the case for many laboratory devices with strong anisotropy introduced by a large external magnetic field. For instance, steady-state toroidal plasmas with large closed flux surfaces would expect $\bm{u}\cdot\nabla u_A^2$ and $\bm{u}\cdot\nabla u^2$ to be small, as the fluid velocity is primarily along field lines and gradients in both the magnetic and velocity fields are primarily across field lines. For this reason, in certain devices the use of $\bm{q} = [\bm{u},\bm{u}_A,u_s]$ could be a useful alternative to the formulation used in the main body of this work. It is possible that, in these units, the structure of the nonlinearities in the associated POD-Galerkin model may prove more amenable to analysis or computation. 

\subsection{Hyper-reduction techniques}
Now that we have illustrated how global conservation laws manifest as structure in Galerkin models, we could compute the coefficients in Eq.~\eqref{eq:Galerkin_model} and evolve the subsequent model. However, in order to calculate the model coefficients, spatial derivatives for $\rho$, $\bm{B}_u$, and $\bm{B}$ (and $B_T$ if temperature is evolved) must be well-approximated in the region of experimental interest. 
In some cases, high-resolution diagnostics can resolve these quantities in a particular plasma region. 
Even if the high-quality data is available, for instance through simulations, computing these inner products and evaluating the nonlinear terms is expensive, because the fields have the original spatial dimension $D$. This somewhat reduces the usefulness of projection-based model reduction.
Fortunately, there are hyper-reduction techniques from fluid dynamics~\cite{benner2015survey}, such as the discrete empirical interpolation method (DEIM)~\cite{chaturantabut2009discrete}, QDEIM~\cite{drmac2016new}, missing-point estimation (MPE)~\cite{astrid2008missing} and gappy POD~\cite{willcox2006unsteady,carlberg2013gnat}, which can enable efficient computations. Instead of using hyper-reduction, we will turn to emerging and increasingly sophisticated machine learning methods in Section~\ref{sec:sindy} to discover Galerkin models from data. There are two primary reasons we have derived the POD-Galerkin model structure here anyways: 1) it indicates that we can search plasma datasets for systems of ODEs consisting only up to quadratic polynomials, 2) it provides a theoretical basis for projection-based model reduction and hyper-reduction techniques in future MHD work. 
\section{\label{sec:sindy} Constrained identification of data-driven models}
In projection-based modeling, the expansion in the POD basis is data-driven, but the projection step is intrusive, requiring access to a numerical solver of the known governing equations. Purely data-driven techniques are useful because they are non-intrusive and in principle do not require high-resolution simulations or knowledge of the governing equations. From the projection-based analysis in Sections~\ref{sec:pod}$-$\ref{sec:conservation_laws}, we are now able to define physical constraints for improved data-driven models.

This is an opportune time to discover data-driven models;
throughout the scientific community, emerging techniques in system identification and optimization are increasingly facilitating the discovery of physical models directly from data~\cite{brunton2019data,schmidt_distilling_2009}. 
We use the sparse identification of nonlinear dynamics (SINDy) algorithm~\cite{Brunton2016pnas} to identify nonlinear reduced-order models for plasmas because
SINDy models are \emph{parsimonious}, having as few terms as are required to explain the dynamics. This feature of the SINDy algorithm promotes models that are interpretable and generalizable.

\subsection{\label{sec:sindy_review}The constrained SINDy method}
In our case, we compute a set of POD modes from a plasma dataset, and then use SINDy to search for low-dimensional models for $\bm{a}(t)$ as a sparse linear combination of elements from a library of candidate terms $\boldsymbol{\Theta}$, 
\begin{align}
       \label{eq:theta_matrix}
\bm{\Theta}(\bm{a}) = 
\begin{bmatrix} 
~~\vline&\vline & \vline & ~ \\
~~\bm{1}&\hspace{-.025in}\bm{a} & \hspace{-.025in}\bm{a}\otimes\bm{a} & \cdots\\
~~\vline &\vline & \vline & ~ 
\end{bmatrix}. 
\end{align}
Here $\bm{a}\otimes\bm{a}$ is all combinations of $a_ia_j$ without duplicates, and similarly for the other candidate terms. 
We now assume that the evolution of $\bm{a}$ can be approximated as
\begin{equation}\label{Eq:SINDyExpansion}
     \dot{\bm{a}} = \bm{f}(\bm{a}) \approx \boldsymbol{\Theta}(\bm{a})\boldsymbol{\Xi}.
\end{equation}
The optimization problem solves for a sparse matrix of coefficients  $\bm{\Xi}$, which represents the coefficients (strengths) of the candidate terms in $\bm{\Theta}$.
To address this combinatorically hard problem, it leverages sparse regression techniques, optimizing for the sparsest set of equations that produces an accurate fit of the data. 
To incorporate known physical laws, a constrained SINDy formulation was first introduced to conserve energy in incompressible fluids~\cite{loiseau2018constrained}.
The constrained SINDy optimization problem can be written 
\begin{align}
\label{eq:constrained_sindy}
    \text{min}_{\bm{\Xi}}&||\dot{\bm{a}}-\bm{\Theta}(\bm{a})\bm{\Xi}||_2^2 + \lambda R(\bm{\Xi}), \\ \notag
    &\text{subject to} \,\,\,\,\, \bm{D}\bm{\Xi}[:] = \bm{d},
\end{align}
where $R(\bm{\Xi})$ is a regularizer such as the $L^0$ or $L^1$ norm, which promotes sparsity in the coefficients $\bm{\Xi}$. $\bm{D}$ is a constraint matrix that allows us to impose that affine combinations of the coefficients in $\bm{\Xi}$ have the fixed values in $\bm{d}$. The original unconstrained SINDy algorithm   solves Eq.~\eqref{eq:constrained_sindy} without using the constraint $\bm{D}\bm{\Xi}[:] = \bm{d}$. Here $\bm{a},\dot{\bm{a}} \in \mathds{R}^{M\times r}$, $\bm{\Theta}(\bm{a}) \in \mathds{R}^{M\times N}$, $\bm{\Xi} \in \mathds{R}^{N\times r}$, $\bm{D} \in \mathds{R}^{N_c\times rN}$, $\bm{\Xi}[:] \in \mathds{R}^{rN}$, $\bm{d} \in \mathds{R}^{N_c}$, where $N$ is the number of candidate terms, $N_c$ is the number of constraints, and $
\bm{\Xi}[:] = 
\begin{bmatrix}
    \xi_1^{a_1} &
    \cdots &
    \xi_1^{a_r} &
    \cdots &
    \xi_N^{a_1} &
    \cdots &
    \xi_N^{a_r}
\end{bmatrix}$ denotes the flattened or ``vectorized'' set of model coefficients.
Motivated by the Galerkin model we have derived, we restrict the library of candidate terms $\bm{\Theta}$ to first and second order polynomials in $\bm{a}(t)$,  although this is not a requirement of the SINDy algorithm; more complicated nonlinear terms may also be included for modeling the effect of truncated POD modes~\cite{loiseau2018constrained} or capturing POD-Galerkin models which exhibit higher order nonlinearities. 
$\bm{\Xi}$ is typically identified via sparse regression, for example by sequentially thresholded least-squares~\cite{Brunton2016pnas,silva2020pysindy}, LASSO~\cite{Tibshirani1996lasso}, or sparse regularized relaxed regression (SR3)~\cite{zheng2019unified}. In Appendix~\ref{sec:sindy_coeffs}, we explicitly derive the SINDy constraints required for the identified models to satisfy the global conservation laws discussed in Section~\ref{sec:conservation_laws}. 

To summarize, we use a physics-informed sparse regression method that requires only $\dot{\bm{a}}$ to discover data-driven models for the evolution of $\bm{a}$. In the next section, we compute the POD for an example 3D MHD simulation and feed the chronos into the SINDy algorithm to identify data-driven models that we can use for forecasting future data.
\begin{figure}
\begin{subfigure}[t]{0.5\textwidth}
    \centering
    \includegraphics[width=0.95\linewidth]{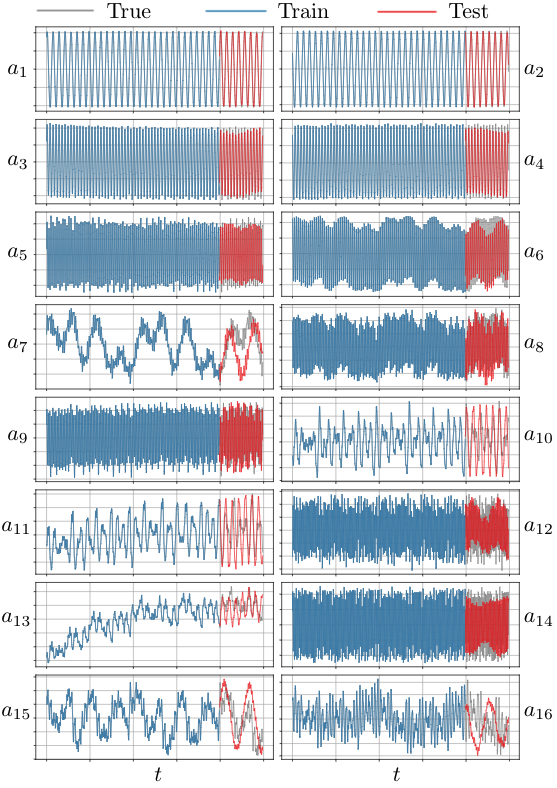}
    \caption{}
    \vspace{0.2in}
    \label{fig:sindy_evo_pred}
\end{subfigure}
\begin{subfigure}[t]{0.5\textwidth}
\centering
    \includegraphics[width=0.8\linewidth]{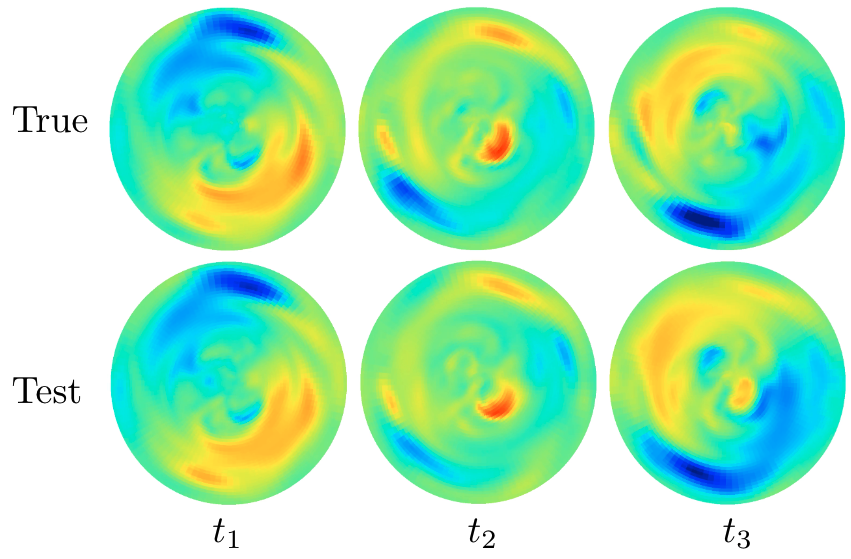}
    \caption{}
    \label{fig:sindy_recons}
\end{subfigure}
    \caption{Summary of the constrained SINDy performance on a 3D Hall-MHD simulation of the HIT-SI device described in Appendix~\ref{sec:HITSI}. (a) Constrained SINDy prediction of $a_1,...,a_{16}$. The true evolution is in gray, the training data used for the model-finding is in blue, and the SINDy prediction is in red (see online version for color). (b) Constrained SINDy predictions of $u_z$ (Test) in the $Z = 0$ midplane are compared with the true $u_z$ evolution at three snapshots in time, indicating strong algorithm performance.}
    \label{fig:sindy_results}
\end{figure}
\subsection{Initial results \label{sec:sindy_results}}
The theoretical structure of this reduced-order modeling framework is appealing, but its value to the community ultimately depends on the quality of the analysis when applied to plasma systems. Guided by the theory, we construct a nonlinear, physics-constrained SINDy model for an isothermal Hall-MHD simulation of this device, described in detail in Appendix~\ref{sec:HITSI}. The density, velocity, and magnetic field are sampled at a set of equally-spaced 3D points in the volume and sampling intervals $\Delta \phi = \pi$/$16$, $\Delta R \approx \Delta Z \approx 2$ cm. The result is that each component of $\bm{u}$ and $\bm{B}$ has 47712 samples. This high-resolution is ideal for visualization but substantial size reduction can be done with little or no change to the spatial or temporal POD modes. For instance, in Figures~\ref{fig:pod_modes} and \ref{fig:sindy_evo_pred}, the $Z=0$ visualizations of the 3D spatial POD modes are constructed from the 1440 sample locations at $Z=0$; with a non-uniform set of 50 points in the midplane, the only change to the visualization is a smoothing out of the shortest wavelengths. The temporal resolution of the measurements is $\Delta t_k = 1$ $\mu$s.
The analysis is essentially unchanged for time steps as large as $10$ $\mu$s, but smaller time steps are required in HIT-SI to resolve harmonics of the injector frequency that appear in the temporal POD modes. For instance, at $\Delta t_k = 10$ $\mu$s, the fourth injector harmonic is sampled, on average, less than twice per period. 

From these measurements of the density, velocity, and magnetic field, we compute the topos and chronos via the SVD in Eq.~\eqref{eq:svd_trunc}, obtaining a Galerkin expansion for the velocity and magnetic fields in magnetic field units, as in Eq.~\eqref{eq:q_expansion}. Now a constrained SINDy model is identified for the first 16 chronos $a_i(t)$ and the forecasting is illustrated in Fig.~\ref{fig:sindy_evo_pred}. 
The SINDy model accurately captures most of the $a_j(t)$ dynamics, with larger errors for the  less energetic modes. Some of the low-frequency content in the $a_j(t)$ is not captured by this particular constrained SINDy model, but this is largely because the low frequencies are not well-resolved in the time range used for training. Despite this deficiency in the data, the SINDy model illustrates strong prediction performance in the $Z=0$ midplane reconstructions of the simulation data in Fig.~\ref{fig:sindy_recons} and forecasts much of the time evolution for a high-dimensional simulation that used $\num{589824}$ grid points, a tremendous efficiency gain of $\mathcal{O}(10^5)$. Furthermore, this model was obtained by training on a dataset representing a single discharge. Further improvements are likely accessible by training on a dataset of many discharges of varying trajectories.

\begin{figure*}
    \centering
    \includegraphics[width=0.8\linewidth]{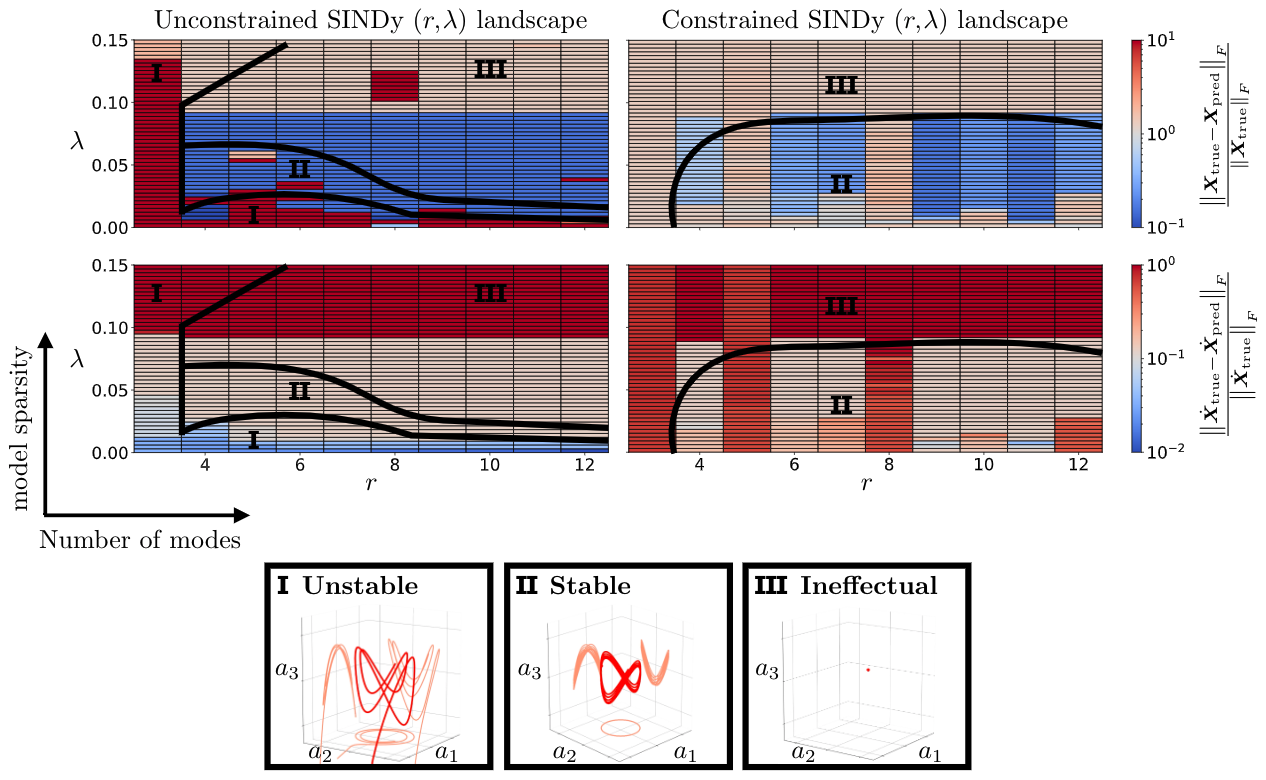}
    \caption{Summary of the $(r,\lambda)$ space of unconstrained and constrained SINDy models from the HIT-SI simulation. The unconstrained models approximately separate into three distinct classes. Class I illustrates nonsparse and typically unstable models. Class II consists of sparse and accurate solutions. Class III denotes solutions which are too sparse to accurately capture the dynamics. Computed errors are for the testing part of the dataset; the colorbar (see online version for color) is limited to $10^1$ as unstable model errors grow arbitrarily large. Constrained SINDy guarantees the energy norm is preserved and thus class I vanishes. Algorithmic advances~\cite{ gelss2019multidimensional,callaham2020nonlinear,bramburger2020sparse,kaheman2020sindy,cortiella2021sparse,bruckner2020inferring,beetham2020formulating,kaptanoglu2021promoting} may help further expand the size of class II.
    }
    \label{fig:pareto}
\end{figure*}

We have found a quality forecasting model from the SINDy system identification method, but it is interesting to see how the model quality varies with the algorithm hyperparameters like the model sparsity $\lambda$ and model rank $r$. In Figure~\ref{fig:pareto}, we illustrate how the normalized reconstruction errors of $\bm{X}$ and $\dot{\bm{X}}$ vary in the ``Pareto-space" of $(r,\lambda)$ for both the unconstrained and constrained SINDy algorithms, with the goal to explore the space of possible models obtained from this system identification technique. Although the exact reconstruction error values are unique to the simulation examined here, there are some interesting qualitative features that we expect to be quite general. The unconstrained SINDy algorithm indicates a significant region of $(r,\lambda)$ where numerically unstable models are found. For $r \gtrapprox 10$, the models are typically either unstable or too sparse to be effective for forecasting. In contrast, by construction the constrained SINDy algorithm conserves the energy and therefore exhibits no unstable models. This is promising for discovering models on historically challenging systems for machine learning methods $-$ multi-scale or turbulent systems that require $r \gg 1$ to properly capture the dynamics. At first glance, it may appear that the constrained SINDy errors in $\dot{\bm{X}}$ are worse than the unconstrained errors, but the low-error values in the unconstrained case are precisely the unstable models. These nonsparse models are overfitting, leading to instability in the numerical integration. Finally, we can see that at $\lambda \approx 0.091$, all the SINDy models are rendered ineffective. This value is precisely at the driving frequency of the HIT-SI injectors in this simulation; if $\lambda$ is larger than this frequency, SINDy thresholds off the primary dynamics in the system.  

\section{Conclusions}
A hierarchy of models with varying fidelity is essential for understanding and controlling plasmas, and our work provides a principled lower level on this hierarchy $-$ low-dimensional and interpretable plasma models which can be used for physical discovery, forecasting, stability analysis, and real-time control. We have discussed how these models are obtained from either projection-based or data-driven methods. Furthermore, we illustrated how Galerkin plasma models retain the global conservation laws of MHD, and machine learning or system identification methods like SINDy can use these constraints directly in an optimization procedure for discovering such models from data. We demonstrated the effectiveness of this approach for a 3D isothermal Hall-MHD simulation of a self-organized plasma.
This framework may be used more broadly for discovering low-dimensional models, forecasting, or real-time control of complex plasmas. This principled enforcement of global conservation laws is critical for the stability and success of future low-dimensional plasma models. 

There are a number of potential numerical limitations to the methodology presented here, including stability issues, the curse of dimensionality, addressing turbulent or stochastic systems, and extrapolation beyond the training dataset. 
Fortunately, all of these potential caveats are currently the subjects of intense ongoing research efforts. 
Generally, the systems of nonlinear ODEs identified by the unconstrained SINDy algorithm tend to have depreciating stability properties as the number of modes increases; unless steps to constrain the model structure are taken, as in the present work and other recent work in provably stable data-driven models~\cite{kolter2019learning,pan2020physics,kaptanoglu2021promoting}, this may prove a difficult obstacle.  
Fortunately, there are also several alternative formulations of the SINDy algorithm that may be more robust in some circumstances and are worth exploring on plasma systems~\cite{boninsegna2018sparse,gelss2019multidimensional,callaham2020nonlinear,bramburger2020sparse,kaheman2020sindy,cortiella2021sparse,bruckner2020inferring,beetham2020formulating}. 

In regards to the curse of dimensionality, the SINDy library grows combinatorially with the number of state variables
and the optimization problem can become very ill-conditioned and computationally intensive. 
Even for the relatively modest candidate library used in the present work, limited to quadratic polynomials, the size scales as $\mathcal{O}(r^3)$ because $Q^0_{ijk}$ is a three index tensor. However, there is recent work utilizing low-rank tensor decompositions to significantly reduce memory usage and computational latency~\cite{gelss2019multidimensional}. 
There has also been considerable recent progress in the modeling of turbulent systems that exhibit broadband turbulence, which generally require a prohibitive number of modes to faithfully reconstruct the field. New approaches bypass this requirement by using stochastic techniques~\cite{boninsegna2018sparse,callaham2020nonlinear,bruckner2020inferring} or finding new data-driven closures for the Navier-Stokes equations~\cite{beetham2020formulating,zucatti2021data}. 
Finally, extrapolation beyond the training set is a central challenge for all machine learning techniques and this issue is primarily addressed in system identification methods by additional steps to mitigate overfitting, such as the additional of a sparsity-promoting regularizer in the optimization problem. 

Lastly, to promote reproducible research, the python code used for this analysis can be found at \url{https://github.com/akaptano/POD-Galerkin_MHD}. The results presented below have also been incorporated into an advanced example of the PySINDy software package~\cite{silva2020pysindy}.

\section{Acknowledgements}
The authors would like to extend their gratitude to Dr. Uri Shumlak for his input on this work. This work was supported by the Army Research Office ({ARO W}911{NF}-19-1-0045) and the Air Force Office of Scientific Research (AFOSR {FA}9550-18-1-0200). Simulations were supported by the U.S. Department of Energy under award numbers {DE-SC}0016256 and DE-AR0001098 and used resources of the National Energy Research Scientific Computing Center, supported by the Office of Science of the U.S. Department of Energy under Contract No. DE-AC02–05CH11231.

\begin{appendices}
\subsection{\label{sec:HITSI} The HIT-SI experiment and simulations}
HIT-SI was a laboratory plasma device that formed and sustained spheromak plasmas for the study of plasma self-organization and steady inductive helicity injection (SIHI)~\cite{jarboe2006spheromak}. 
It consisted of an axisymmetric flux conserver and two inductive injectors (actuators) mounted on each end as illustrated in the top left panel of Fig. \ref{fig:overview}. Magnetic coils on each injector, generating helical fields linking through the flux conserver, were oscillated in phase at a frequency with values between $10-70$ kHz. The magnetic fields generated by the two injectors were spatially and temporally 90$^\circ$ out of phase, resulting in approximately constant power and helicity injection. The fields from these injectors provided the power and magnetic helicity to both form and sustain a spheromak during experimental discharges, with a quasi-steady-state period of roughly constant spheromak amplitude lasting $< 1$~ms. Additional details of the experiment and its operation can be found in references ~\cite{jarboe2006spheromak,wrobel2011study,victor2014sustained}.

Simulations of HIT-SI were performed using the Hall-MHD equations and solved by the NIMROD code~\cite{sovinec2004nonlinear}. NIMROD discretizes equations in cylindrical coordinates $(R,Z,\phi)$; the R-Z plane is composed of finite elements and the $\phi$ component is expanded in a finite Fourier series. Mesh convergence was obtained previously by Akcay~\cite{akcay2013extended} on a grid with $28\times 28$ finite elements of polynomial degree 4 and 22 Fourier components, so we use the same grid for the simulation here. 
Due to the 2D grid, the HIT-SI injectors
cannot be directly modeled in the simulation. Rather, the injectors are implemented as $B_\perp$ and $E_\parallel$ boundary conditions at the top and bottom device surfaces to match the experimental waveforms. 
A detailed description of the
implementation of these boundary conditions can be found
in Akcay~\cite{akcay2013}. Dirichlet boundary conditions are used for all other variables; the plasma density satisfies $n_e = 2 \times 10^{19}$ $\text{m}^{-3}$ and the temperatures satisfy $T_i = T_e = 14$ eV. Isotropic viscosity $\nu = 550$ $\text{m}^2/$s and Spitzer resistivity~\cite{spitzer2006physics} is used. 
The remaining boundary conditions are $\bm{u}\times\hat{\bm{n}} = \bm{u}\cdot\hat{\bm{n}} = 0$, $\bm{J}\cdot\hat{\bm{n}} = 0$, and $\bm{B}\cdot\hat{\bm{n}} = 0$. For more information on the numerical model used in this simulation, see Morgan et al.~\cite{morgan2017validation}. The data for training and testing are obtained during the approximately steady-state phase of the simulation so that the energy constraints derived in Section~\ref{sec:global_energy} are applicable. 

\subsection{\label{sec:sindy_coeffs}Derivation of the SINDy constraints}
In Sec.~\ref{sec:conservation_laws}, we derived constraints for the POD-Galerkin model coefficients from global conservation laws; our goal here is to rewrite these constraints to be compatible with the formulation of the SINDy system identification method. The conclusions for the global conservation of energy were: 1) no constant terms, 2) an anti-symmetry constraint on the linear part of the coefficient matrix $\bm{\Xi}$, and 3) a more complicated energy-preserving structure in the quadratic coefficients.
Consider a quadratic library in a set of $r$ modes, ordered as $\bm{\Theta}(\bm{a}) = [a_1, ..., a_r, a_1a_2,...,a_{r-1}a_r,a_1^2,...,a_r^2]$. Note that this arrangement of the polynomials in $\bm{\Theta}$ differs from Loiseau et al.~\cite{loiseau2018sparse}, so the indexing and subscripts are also different here. First we will consider the constraint on the linear part of the Galerkin model in Eq.~\eqref{eq:Galerkin_model}, $\bm{a}^T\bm{L}_0\bm{a} \approx 0$. We can rewrite this in the SINDy notation as
\begin{align}
    \label{eq:linear_constraint_deriv}
    0 = \begin{bmatrix}
    a_1 & \cdots & a_r
    \end{bmatrix}
    \begin{bmatrix}
    \xi_1^{a_1} & \cdots & \xi_r^{a_1} \\
    \vdots & \ddots & \vdots \\
    \xi_r^{a_r} & \cdots & \xi_r^{a_r} \\
    \end{bmatrix}
    \begin{bmatrix}
    a_1 \\
    \vdots \\
    a_r
    \end{bmatrix}.
\end{align}
We conclude $\xi_i^{a_j} = -\xi_j^{a_i}$ for $i,j \in \{1,...,r\}$ and identify $\xi_i^{a_j}$ by accessing the $(i-1)r+j$ index in the vector of model coefficients $\bm{\Xi}[:]$. Note we are only accessing the first $r^2$ elements of $\bm{\Xi}[:]$. For models of linear and quadratic polynomials, $N = (r^2+3r)/2$ and the number of constraints from anti-symmetry of the linear coefficients is $N_L = (r^2+r)/2$. Thus there are now only $rN - N_L = r(r^2+2r-1)/2$ free parameters. Since the constrained SINDy algorithm solves linear equality constraints of the form $\bm{D}\bm{\Xi}[:] = \bm{d}$, we can write this out explicitly for $r=3$,
\setcounter{MaxMatrixCols}{30}
\setlength{\arraycolsep}{3.0pt}
\medmuskip = 4.0mu 
\begin{align}
    \begin{bmatrix}
    \bm{1} & 0 & 0 & 0 & 0 & 0 & 0 & 0 & 0 & 0 & \cdots \\
    0 & 0 & 0 & 0 & \bm{1} & 0 & 0 & 0 & 0 & 0 & \cdots \\
    0 & 0 & 0 & 0 & 0 & 0 & 0 & 0 & \bm{1} & 0 & \cdots \\
    0 & \bm{1} & 0 & \bm{1} & 0 & 0 & 0 & 0 & 0 & 0 & \cdots \\
    0 & 0 & \bm{1} & 0 & 0 & 0 & \bm{1} & 0 & 0 & 0 & \cdots \\
    0 & 0 & 0 & 0 & 0 & \bm{1} & 0 & \bm{1} & 0 & 0 & \cdots \\
    \end{bmatrix}\begin{bmatrix}
    \xi_1^{a_1} \\
    \xi_1^{a_2} \\
    \xi_1^{a_3} \\
    \xi_2^{a_1} \\
    \vdots \\
\end{bmatrix}
= \begin{bmatrix}
    0 \\ 
    0 \\
    0 \\
    0 \\
    0 \\
    0 \\
\end{bmatrix}.
\end{align}
\medmuskip = 0mu 
The boundary conditions $\bm{u}\cdot\hat{\bm{n}} = 0$,  $\bm{J}\cdot\hat{\bm{n}} = 0$, $\bm{B}\cdot\hat{\bm{n}} = 0$ guaranteed that the quadratic nonlinearities were energy-preserving, and thus that cubic terms in Eq.~\eqref{eq:galtier_deriv} vanish,
\begin{align}
    \sum_{i,j,k=0}^rQ_{ijk}^0a_ia_ja_k \approx 0.
\end{align}
This constraint is significantly more involved to reformat. Written in SINDy notation, this is equivalent to
\begin{align}
\label{eq:quad_constraint_formulation}
    0 =\begin{bmatrix}
        a_1 & \cdots & a_r
    \end{bmatrix}
    \begin{bmatrix}
        \xi_{r+1}^{a_1} & \xi_{r+2}^{a_1} & \cdots & \xi_{N}^{a_1} \\
        \vdots & \vdots & \vdots & \vdots \\
        \xi_{r+1}^{a_r} & \xi_{r+2}^{a_r} & \cdots & \xi_{N}^{a_r}
    \end{bmatrix}
    \begin{bmatrix}
        a_1a_2 \\
        \vdots \\
        a_{r-1}a_r \\
        a_1^2 \\
        \vdots \\
        a_r^2
    \end{bmatrix}.
\end{align}
Expand this all out and group the like terms, i.e. terms which look like $a_i^3$, $a_ia_j^2$ or $a_ia_ja_k$, $i,j,k \in \{1,...,r\}$, $i\neq j \neq k$. All of the like terms can be straightforwardly shown to be linearly independent, so we can consider three constraints separately for the three types of terms. The number of each of these respective terms is ${r \choose 1} = r$, 2${r \choose 2} = r(r-1)$, and ${r \choose 3} = r(r-1)(r-2)/6$, for a total of $r(r+1)(r+2)/6 = N_Q$ constraints. With both constraints, we have $rN-N_L-N_Q = r(r-1)(2r+5)/6$ free parameters, and $N_c = N_L + N_Q$ constraints. 
Further considering the quadratic case, we find that coefficients which adorn $a_i^3$ must vanish, $
\xi_{N-r+i}^{a_{i}} = 0$. 
Now define
\begin{equation}
    \tilde{\xi}_{ijk} = \xi_{r+\frac{j}{2}(2r-j-3)+k-1}^{a_i}.
\end{equation}
The second type of constraint, with $i\neq j$, produces 
\begin{equation}
\xi_{N-r+j}^{a_i} =\begin{cases}
 \tilde{\xi}_{jij} & i < j \\
\tilde{\xi}_{jji} & i > j,
\end{cases}
\end{equation}
while the third type of constraint produces 
\begin{equation}
    \label{eq:ai_aj_ak}
\tilde{\xi}_{ijk} + \tilde{\xi}_{jik} + \tilde{\xi}_{kij} = 0.
\end{equation}
This relation is equivalent to the energy-preserving conditions in Schlegel et al.~\cite{Schlegel2015jfm}, but the indexing is not straightforward, even after fully expanding Eq.~\eqref{eq:quad_constraint_formulation}. 
This equation is an arbitrary $r$ generalization to the $r=3$ constraint used in Loiseau et al.~\cite{loiseau2018constrained}. 
For the specific case where the plasma system is Hamiltonian (for instance in ideal~\cite{morrison1980noncanonical}, Hall~\cite{yoshida2013canonical}, and extended~\citep{abdelhamid2015hamiltonian} MHD without dissipation) and the measurements are assumed to be sufficient to represent the Hamiltonian, one could alternatively use formulations of SINDy to directly discover the Hamiltonian~\cite{chu2020discovering} and subsequently derive the equations of motion.
Lastly, if the global energy conservation constraint on the quadratic terms in the SINDy coefficient matrix $\bm{\Xi}$ is written $D_{jk}\Xi_k = 0$, then the quadratic cross-helicity constraint can be written $D_{jk}A_{kl}\Xi_l = 0$. 
\end{appendices}

 \bibliography{Galerkin}

\begin{thebibliography}{131}%
\makeatletter
\providecommand \@ifxundefined [1]{%
 \@ifx{#1\undefined}
}%
\providecommand \@ifnum [1]{%
 \ifnum #1\expandafter \@firstoftwo
 \else \expandafter \@secondoftwo
 \fi
}%
\providecommand \@ifx [1]{%
 \ifx #1\expandafter \@firstoftwo
 \else \expandafter \@secondoftwo
 \fi
}%
\providecommand \natexlab [1]{#1}%
\providecommand \enquote  [1]{``#1''}%
\providecommand \bibnamefont  [1]{#1}%
\providecommand \bibfnamefont [1]{#1}%
\providecommand \citenamefont [1]{#1}%
\providecommand \href@noop [0]{\@secondoftwo}%
\providecommand \href [0]{\begingroup \@sanitize@url \@href}%
\providecommand \@href[1]{\@@startlink{#1}\@@href}%
\providecommand \@@href[1]{\endgroup#1\@@endlink}%
\providecommand \@sanitize@url [0]{\catcode `\\12\catcode `\$12\catcode
  `\&12\catcode `\#12\catcode `\^12\catcode `\_12\catcode `\%12\relax}%
\providecommand \@@startlink[1]{}%
\providecommand \@@endlink[0]{}%
\providecommand \url  [0]{\begingroup\@sanitize@url \@url }%
\providecommand \@url [1]{\endgroup\@href {#1}{\urlprefix }}%
\providecommand \urlprefix  [0]{URL }%
\providecommand \Eprint [0]{\href }%
\providecommand \doibase [0]{https://doi.org/}%
\providecommand \selectlanguage [0]{\@gobble}%
\providecommand \bibinfo  [0]{\@secondoftwo}%
\providecommand \bibfield  [0]{\@secondoftwo}%
\providecommand \translation [1]{[#1]}%
\providecommand \BibitemOpen [0]{}%
\providecommand \bibitemStop [0]{}%
\providecommand \bibitemNoStop [0]{.\EOS\space}%
\providecommand \EOS [0]{\spacefactor3000\relax}%
\providecommand \BibitemShut  [1]{\csname bibitem#1\endcsname}%
\let\auto@bib@innerbib\@empty
\bibitem [{\citenamefont {Roth}(2001)}]{roth2001industrial}%
  \BibitemOpen
  \bibfield  {author} {\bibinfo {author} {\bibfnamefont {J.~R.}\ \bibnamefont
  {Roth}},\ }\href@noop {} {\emph {\bibinfo {title} {Industrial plasma
  engineering: Volume 2: Applications to nonthermal plasma processing}}},\
  Vol.~\bibinfo {volume} {2}\ (\bibinfo  {publisher} {CRC press},\ \bibinfo
  {year} {2001})\BibitemShut {NoStop}%
\bibitem [{\citenamefont {Candy}\ and\ \citenamefont
  {Waltz}(2003)}]{Candy2003}%
  \BibitemOpen
  \bibfield  {author} {\bibinfo {author} {\bibfnamefont {J.}~\bibnamefont
  {Candy}}\ and\ \bibinfo {author} {\bibfnamefont {R.~E.}\ \bibnamefont
  {Waltz}},\ }\bibfield  {title} {\bibinfo {title} {Anomalous transport scaling
  in the {DIII-D} tokamak matched by supercomputer simulation},\ }\href
  {https://doi.org/10.1103/PhysRevLett.91.045001} {\bibfield  {journal}
  {\bibinfo  {journal} {Phys. Rev. Lett.}\ }\textbf {\bibinfo {volume} {91}},\
  \bibinfo {pages} {045001} (\bibinfo {year} {2003})}\BibitemShut {NoStop}%
\bibitem [{\citenamefont {Ohia}\ \emph {et~al.}(2012)\citenamefont {Ohia},
  \citenamefont {Egedal}, \citenamefont {Lukin}, \citenamefont {Daughton},\
  and\ \citenamefont {Le}}]{Ohia2012}%
  \BibitemOpen
  \bibfield  {author} {\bibinfo {author} {\bibfnamefont {O.}~\bibnamefont
  {Ohia}}, \bibinfo {author} {\bibfnamefont {J.}~\bibnamefont {Egedal}},
  \bibinfo {author} {\bibfnamefont {V.~S.}\ \bibnamefont {Lukin}}, \bibinfo
  {author} {\bibfnamefont {W.}~\bibnamefont {Daughton}},\ and\ \bibinfo
  {author} {\bibfnamefont {A.}~\bibnamefont {Le}},\ }\bibfield  {title}
  {\bibinfo {title} {Demonstration of anisotropic fluid closure capturing the
  kinetic structure of magnetic reconnection},\ }\href
  {https://doi.org/10.1103/PhysRevLett.109.115004} {\bibfield  {journal}
  {\bibinfo  {journal} {Phys. Rev. Lett.}\ }\textbf {\bibinfo {volume} {109}},\
  \bibinfo {pages} {115004} (\bibinfo {year} {2012})}\BibitemShut {NoStop}%
\bibitem [{\citenamefont {Gro\ifmmode~\check{s}\else \v{s}\fi{}elj}\ \emph
  {et~al.}(2018)\citenamefont {Gro\ifmmode~\check{s}\else \v{s}\fi{}elj},
  \citenamefont {Mallet}, \citenamefont {Loureiro},\ and\ \citenamefont
  {Jenko}}]{Groselj2018}%
  \BibitemOpen
  \bibfield  {author} {\bibinfo {author} {\bibfnamefont {D.}~\bibnamefont
  {Gro\ifmmode~\check{s}\else \v{s}\fi{}elj}}, \bibinfo {author} {\bibfnamefont
  {A.}~\bibnamefont {Mallet}}, \bibinfo {author} {\bibfnamefont {N.~F.}\
  \bibnamefont {Loureiro}},\ and\ \bibinfo {author} {\bibfnamefont
  {F.}~\bibnamefont {Jenko}},\ }\bibfield  {title} {\bibinfo {title} {Fully
  kinetic simulation of {3D} kinetic {A}lfv\'en turbulence},\ }\href
  {https://doi.org/10.1103/PhysRevLett.120.105101} {\bibfield  {journal}
  {\bibinfo  {journal} {Phys. Rev. Lett.}\ }\textbf {\bibinfo {volume} {120}},\
  \bibinfo {pages} {105101} (\bibinfo {year} {2018})}\BibitemShut {NoStop}%
\bibitem [{\citenamefont {Taira}\ \emph {et~al.}(2017)\citenamefont {Taira},
  \citenamefont {Brunton}, \citenamefont {Dawson}, \citenamefont {Rowley},
  \citenamefont {Colonius}, \citenamefont {McKeon}, \citenamefont {Schmidt},
  \citenamefont {Gordeyev}, \citenamefont {Theofilis},\ and\ \citenamefont
  {Ukeiley}}]{Taira2017aiaa}%
  \BibitemOpen
  \bibfield  {author} {\bibinfo {author} {\bibfnamefont {K.}~\bibnamefont
  {Taira}}, \bibinfo {author} {\bibfnamefont {S.~L.}\ \bibnamefont {Brunton}},
  \bibinfo {author} {\bibfnamefont {S.}~\bibnamefont {Dawson}}, \bibinfo
  {author} {\bibfnamefont {C.~W.}\ \bibnamefont {Rowley}}, \bibinfo {author}
  {\bibfnamefont {T.}~\bibnamefont {Colonius}}, \bibinfo {author}
  {\bibfnamefont {B.~J.}\ \bibnamefont {McKeon}}, \bibinfo {author}
  {\bibfnamefont {O.~T.}\ \bibnamefont {Schmidt}}, \bibinfo {author}
  {\bibfnamefont {S.}~\bibnamefont {Gordeyev}}, \bibinfo {author}
  {\bibfnamefont {V.}~\bibnamefont {Theofilis}},\ and\ \bibinfo {author}
  {\bibfnamefont {L.~S.}\ \bibnamefont {Ukeiley}},\ }\bibfield  {title}
  {\bibinfo {title} {Modal analysis of fluid flows: An overview},\ }\href@noop
  {} {\bibfield  {journal} {\bibinfo  {journal} {AIAA Journal}\ }\textbf
  {\bibinfo {volume} {55}},\ \bibinfo {pages} {4013} (\bibinfo {year}
  {2017})}\BibitemShut {NoStop}%
\bibitem [{\citenamefont {Jim{\'e}nez-G{\'o}mez}\ \emph
  {et~al.}(2007)\citenamefont {Jim{\'e}nez-G{\'o}mez}, \citenamefont
  {Ascas{\'\i}bar}, \citenamefont {Estrada}, \citenamefont
  {Garc{\'\i}a-Cort{\'e}s}, \citenamefont {Van~Milligen}, \citenamefont
  {L{\'o}pez-Fraguas}, \citenamefont {Pastor},\ and\ \citenamefont
  {L{\'o}pez-Bruna}}]{jimenez2007analysis}%
  \BibitemOpen
  \bibfield  {author} {\bibinfo {author} {\bibfnamefont {R.}~\bibnamefont
  {Jim{\'e}nez-G{\'o}mez}}, \bibinfo {author} {\bibfnamefont {E.}~\bibnamefont
  {Ascas{\'\i}bar}}, \bibinfo {author} {\bibfnamefont {T.}~\bibnamefont
  {Estrada}}, \bibinfo {author} {\bibfnamefont {I.}~\bibnamefont
  {Garc{\'\i}a-Cort{\'e}s}}, \bibinfo {author} {\bibfnamefont {B.}~\bibnamefont
  {Van~Milligen}}, \bibinfo {author} {\bibfnamefont {A.}~\bibnamefont
  {L{\'o}pez-Fraguas}}, \bibinfo {author} {\bibfnamefont {I.}~\bibnamefont
  {Pastor}},\ and\ \bibinfo {author} {\bibfnamefont {D.}~\bibnamefont
  {L{\'o}pez-Bruna}},\ }\bibfield  {title} {\bibinfo {title} {Analysis of
  magnetohydrodynamic instabilities in {TJ-II} plasmas},\ }\href@noop {}
  {\bibfield  {journal} {\bibinfo  {journal} {Fusion science and technology}\
  }\textbf {\bibinfo {volume} {51}},\ \bibinfo {pages} {20} (\bibinfo {year}
  {2007})}\BibitemShut {NoStop}%
\bibitem [{\citenamefont {van Milligen}\ \emph {et~al.}(2014)\citenamefont {van
  Milligen}, \citenamefont {S{\'{a}}nchez}, \citenamefont {Alonso},
  \citenamefont {Pedrosa}, \citenamefont {Hidalgo}, \citenamefont
  {de~Aguilera},\ and\ \citenamefont {Fraguas}}]{vanMilligen14}%
  \BibitemOpen
  \bibfield  {author} {\bibinfo {author} {\bibfnamefont {B.~P.}\ \bibnamefont
  {van Milligen}}, \bibinfo {author} {\bibfnamefont {E.}~\bibnamefont
  {S{\'{a}}nchez}}, \bibinfo {author} {\bibfnamefont {A.}~\bibnamefont
  {Alonso}}, \bibinfo {author} {\bibfnamefont {M.~A.}\ \bibnamefont {Pedrosa}},
  \bibinfo {author} {\bibfnamefont {C.}~\bibnamefont {Hidalgo}}, \bibinfo
  {author} {\bibfnamefont {A.~M.}\ \bibnamefont {de~Aguilera}},\ and\ \bibinfo
  {author} {\bibfnamefont {A.~L.}\ \bibnamefont {Fraguas}},\ }\bibfield
  {title} {\bibinfo {title} {The use of the biorthogonal decomposition for the
  identification of zonal flows at {TJ}-{II}},\ }\href@noop {} {\bibfield
  {journal} {\bibinfo  {journal} {Plasma Physics and Controlled Fusion}\
  }\textbf {\bibinfo {volume} {57}},\ \bibinfo {pages} {025005} (\bibinfo
  {year} {2014})}\BibitemShut {NoStop}%
\bibitem [{\citenamefont {Pandya}(2016)}]{pandya}%
  \BibitemOpen
  \bibfield  {author} {\bibinfo {author} {\bibfnamefont {M.}~\bibnamefont
  {Pandya}},\ }\emph {\bibinfo {title} {Low edge safety factor disruptions in
  the Compact Toroidal Hybrid: Operation in the low-q regime, passive
  disruption avoidance and the nature of {MHD} precursors}},\ \href@noop {}
  {Ph.D. thesis},\ \bibinfo  {school} {{A}uburn {U}niversity} (\bibinfo {year}
  {2016})\BibitemShut {NoStop}%
\bibitem [{\citenamefont {Victor}\ \emph {et~al.}(2015)\citenamefont {Victor},
  \citenamefont {Akcay}, \citenamefont {Hansen}, \citenamefont {Jarboe},
  \citenamefont {Nelson},\ and\ \citenamefont
  {Morgan}}]{victor2015development}%
  \BibitemOpen
  \bibfield  {author} {\bibinfo {author} {\bibfnamefont {B.}~\bibnamefont
  {Victor}}, \bibinfo {author} {\bibfnamefont {C.}~\bibnamefont {Akcay}},
  \bibinfo {author} {\bibfnamefont {C.}~\bibnamefont {Hansen}}, \bibinfo
  {author} {\bibfnamefont {T.}~\bibnamefont {Jarboe}}, \bibinfo {author}
  {\bibfnamefont {B.}~\bibnamefont {Nelson}},\ and\ \bibinfo {author}
  {\bibfnamefont {K.}~\bibnamefont {Morgan}},\ }\bibfield  {title} {\bibinfo
  {title} {Development of validation metrics using biorthogonal decomposition
  for the comparison of magnetic field measurements},\ }\href@noop {}
  {\bibfield  {journal} {\bibinfo  {journal} {Plasma Physics and Controlled
  Fusion}\ }\textbf {\bibinfo {volume} {57}},\ \bibinfo {pages} {045010}
  (\bibinfo {year} {2015})}\BibitemShut {NoStop}%
\bibitem [{\citenamefont {Strait}\ \emph {et~al.}(2016)\citenamefont {Strait},
  \citenamefont {King}, \citenamefont {Hanson},\ and\ \citenamefont
  {Logan}}]{strait2016spatial}%
  \BibitemOpen
  \bibfield  {author} {\bibinfo {author} {\bibfnamefont {E.}~\bibnamefont
  {Strait}}, \bibinfo {author} {\bibfnamefont {J.}~\bibnamefont {King}},
  \bibinfo {author} {\bibfnamefont {J.}~\bibnamefont {Hanson}},\ and\ \bibinfo
  {author} {\bibfnamefont {N.}~\bibnamefont {Logan}},\ }\bibfield  {title}
  {\bibinfo {title} {Spatial and temporal analysis of {DIII-D 3D} magnetic
  diagnostic data},\ }\href@noop {} {\bibfield  {journal} {\bibinfo  {journal}
  {Review of Scientific Instruments}\ }\textbf {\bibinfo {volume} {87}},\
  \bibinfo {pages} {11D423} (\bibinfo {year} {2016})}\BibitemShut {NoStop}%
\bibitem [{\citenamefont {Byrne}(2017)}]{byrne2017study}%
  \BibitemOpen
  \bibfield  {author} {\bibinfo {author} {\bibfnamefont {P.~J.}\ \bibnamefont
  {Byrne}},\ }\emph {\bibinfo {title} {Study of External Kink Modes in Shaped
  HBT-EP Plasmas}},\ \href@noop {} {Ph.D. thesis},\ \bibinfo  {school}
  {Columbia University} (\bibinfo {year} {2017})\BibitemShut {NoStop}%
\bibitem [{\citenamefont {Gu}\ \emph {et~al.}(2019)\citenamefont {Gu},
  \citenamefont {Wan}, \citenamefont {Sun}, \citenamefont {Chu}, \citenamefont
  {Liu}, \citenamefont {Shi}, \citenamefont {Wang}, \citenamefont {Jia},\ and\
  \citenamefont {He}}]{gu2019new}%
  \BibitemOpen
  \bibfield  {author} {\bibinfo {author} {\bibfnamefont {S.}~\bibnamefont
  {Gu}}, \bibinfo {author} {\bibfnamefont {B.}~\bibnamefont {Wan}}, \bibinfo
  {author} {\bibfnamefont {Y.}~\bibnamefont {Sun}}, \bibinfo {author}
  {\bibfnamefont {N.}~\bibnamefont {Chu}}, \bibinfo {author} {\bibfnamefont
  {Y.}~\bibnamefont {Liu}}, \bibinfo {author} {\bibfnamefont {T.}~\bibnamefont
  {Shi}}, \bibinfo {author} {\bibfnamefont {H.}~\bibnamefont {Wang}}, \bibinfo
  {author} {\bibfnamefont {M.}~\bibnamefont {Jia}},\ and\ \bibinfo {author}
  {\bibfnamefont {K.}~\bibnamefont {He}},\ }\bibfield  {title} {\bibinfo
  {title} {A new criterion for controlling edge localized modes based on a
  multi-mode plasma response},\ }\href@noop {} {\bibfield  {journal} {\bibinfo
  {journal} {Nuclear Fusion}\ }\textbf {\bibinfo {volume} {59}},\ \bibinfo
  {pages} {126042} (\bibinfo {year} {2019})}\BibitemShut {NoStop}%
\bibitem [{\citenamefont {Kaptanoglu}\ \emph
  {et~al.}(2020{\natexlab{a}})\citenamefont {Kaptanoglu}, \citenamefont
  {Morgan}, \citenamefont {Hansen},\ and\ \citenamefont
  {Brunton}}]{kaptanoglu2020}%
  \BibitemOpen
  \bibfield  {author} {\bibinfo {author} {\bibfnamefont {A.~A.}\ \bibnamefont
  {Kaptanoglu}}, \bibinfo {author} {\bibfnamefont {K.~D.}\ \bibnamefont
  {Morgan}}, \bibinfo {author} {\bibfnamefont {C.~J.}\ \bibnamefont {Hansen}},\
  and\ \bibinfo {author} {\bibfnamefont {S.~L.}\ \bibnamefont {Brunton}},\
  }\bibfield  {title} {\bibinfo {title} {Characterizing magnetized plasmas with
  dynamic mode decomposition},\ }\href@noop {} {\bibfield  {journal} {\bibinfo
  {journal} {Physics of Plasmas}\ }\textbf {\bibinfo {volume} {27}},\ \bibinfo
  {pages} {032108} (\bibinfo {year} {2020}{\natexlab{a}})}\BibitemShut
  {NoStop}%
\bibitem [{\citenamefont {Benner}\ \emph {et~al.}(2005)\citenamefont {Benner},
  \citenamefont {Mehrmann},\ and\ \citenamefont
  {Sorensen}}]{benner2005dimension}%
  \BibitemOpen
  \bibfield  {author} {\bibinfo {author} {\bibfnamefont {P.}~\bibnamefont
  {Benner}}, \bibinfo {author} {\bibfnamefont {V.}~\bibnamefont {Mehrmann}},\
  and\ \bibinfo {author} {\bibfnamefont {D.~C.}\ \bibnamefont {Sorensen}},\
  }\href@noop {} {\emph {\bibinfo {title} {Dimension reduction of large-scale
  systems}}},\ Vol.~\bibinfo {volume} {45}\ (\bibinfo  {publisher} {Springer},\
  \bibinfo {year} {2005})\BibitemShut {NoStop}%
\bibitem [{\citenamefont {Benner}\ \emph {et~al.}(2017)\citenamefont {Benner},
  \citenamefont {Ohlberger}, \citenamefont {Cohen},\ and\ \citenamefont
  {Willcox}}]{benner2017model}%
  \BibitemOpen
  \bibfield  {author} {\bibinfo {author} {\bibfnamefont {P.}~\bibnamefont
  {Benner}}, \bibinfo {author} {\bibfnamefont {M.}~\bibnamefont {Ohlberger}},
  \bibinfo {author} {\bibfnamefont {A.}~\bibnamefont {Cohen}},\ and\ \bibinfo
  {author} {\bibfnamefont {K.}~\bibnamefont {Willcox}},\ }\href@noop {} {\emph
  {\bibinfo {title} {Model reduction and approximation: theory and
  algorithms}}}\ (\bibinfo  {publisher} {SIAM},\ \bibinfo {year}
  {2017})\BibitemShut {NoStop}%
\bibitem [{\citenamefont {Brunton}\ \emph {et~al.}(2020)\citenamefont
  {Brunton}, \citenamefont {Noack},\ and\ \citenamefont
  {Koumoutsakos}}]{brunton2020machine}%
  \BibitemOpen
  \bibfield  {author} {\bibinfo {author} {\bibfnamefont {S.~L.}\ \bibnamefont
  {Brunton}}, \bibinfo {author} {\bibfnamefont {B.~R.}\ \bibnamefont {Noack}},\
  and\ \bibinfo {author} {\bibfnamefont {P.}~\bibnamefont {Koumoutsakos}},\
  }\bibfield  {title} {\bibinfo {title} {Machine learning for fluid
  mechanics},\ }\href@noop {} {\bibfield  {journal} {\bibinfo  {journal}
  {Annual Review of Fluid Mechanics}\ }\textbf {\bibinfo {volume} {52}},\
  \bibinfo {pages} {477} (\bibinfo {year} {2020})}\BibitemShut {NoStop}%
\bibitem [{\citenamefont {Noack}\ \emph {et~al.}(2003)\citenamefont {Noack},
  \citenamefont {Afanasiev}, \citenamefont {Morzynski}, \citenamefont
  {Tadmor},\ and\ \citenamefont {Thiele}}]{Noack2003jfm}%
  \BibitemOpen
  \bibfield  {author} {\bibinfo {author} {\bibfnamefont {B.~R.}\ \bibnamefont
  {Noack}}, \bibinfo {author} {\bibfnamefont {K.}~\bibnamefont {Afanasiev}},
  \bibinfo {author} {\bibfnamefont {M.}~\bibnamefont {Morzynski}}, \bibinfo
  {author} {\bibfnamefont {G.}~\bibnamefont {Tadmor}},\ and\ \bibinfo {author}
  {\bibfnamefont {F.}~\bibnamefont {Thiele}},\ }\bibfield  {title} {\bibinfo
  {title} {A hierarchy of low-dimensional models for the transient and
  post-transient cylinder wake},\ }\href@noop {} {\bibfield  {journal}
  {\bibinfo  {journal} {Journal of Fluid Mechanics}\ }\textbf {\bibinfo
  {volume} {497}},\ \bibinfo {pages} {335} (\bibinfo {year}
  {2003})}\BibitemShut {NoStop}%
\bibitem [{\citenamefont {Noack}\ \emph {et~al.}(2016)\citenamefont {Noack},
  \citenamefont {Stankiewicz}, \citenamefont {Morzynski},\ and\ \citenamefont
  {Schmid}}]{Noack2016jfm}%
  \BibitemOpen
  \bibfield  {author} {\bibinfo {author} {\bibfnamefont {B.~R.}\ \bibnamefont
  {Noack}}, \bibinfo {author} {\bibfnamefont {W.}~\bibnamefont {Stankiewicz}},
  \bibinfo {author} {\bibfnamefont {M.}~\bibnamefont {Morzynski}},\ and\
  \bibinfo {author} {\bibfnamefont {P.~J.}\ \bibnamefont {Schmid}},\ }\bibfield
   {title} {\bibinfo {title} {Recursive dynamic mode decomposition of a
  transient cylinder wake},\ }\href@noop {} {\bibfield  {journal} {\bibinfo
  {journal} {Journal of Fluid Mechanics}\ }\textbf {\bibinfo {volume} {809}},\
  \bibinfo {pages} {843} (\bibinfo {year} {2016})}\BibitemShut {NoStop}%
\bibitem [{\citenamefont {Carlberg}\ \emph {et~al.}(2017)\citenamefont
  {Carlberg}, \citenamefont {Barone},\ and\ \citenamefont
  {Antil}}]{Carlberg2017jcp}%
  \BibitemOpen
  \bibfield  {author} {\bibinfo {author} {\bibfnamefont {K.}~\bibnamefont
  {Carlberg}}, \bibinfo {author} {\bibfnamefont {M.}~\bibnamefont {Barone}},\
  and\ \bibinfo {author} {\bibfnamefont {H.}~\bibnamefont {Antil}},\ }\bibfield
   {title} {\bibinfo {title} {Galerkin v. least-squares {P}etrov--{G}alerkin
  projection in nonlinear model reduction},\ }\href@noop {} {\bibfield
  {journal} {\bibinfo  {journal} {Journal of Computational Physics}\ }\textbf
  {\bibinfo {volume} {330}},\ \bibinfo {pages} {693} (\bibinfo {year}
  {2017})}\BibitemShut {NoStop}%
\bibitem [{\citenamefont {Rowley}\ and\ \citenamefont
  {Dawson}(2017)}]{Rowley2017arfm}%
  \BibitemOpen
  \bibfield  {author} {\bibinfo {author} {\bibfnamefont {C.~W.}\ \bibnamefont
  {Rowley}}\ and\ \bibinfo {author} {\bibfnamefont {S.~T.}\ \bibnamefont
  {Dawson}},\ }\bibfield  {title} {\bibinfo {title} {Model reduction for flow
  analysis and control},\ }\href@noop {} {\bibfield  {journal} {\bibinfo
  {journal} {Annual Review of Fluid Mechanics}\ }\textbf {\bibinfo {volume}
  {49}},\ \bibinfo {pages} {387} (\bibinfo {year} {2017})}\BibitemShut
  {NoStop}%
\bibitem [{\citenamefont {Brunton}\ and\ \citenamefont
  {Kutz}(2019)}]{brunton2019data}%
  \BibitemOpen
  \bibfield  {author} {\bibinfo {author} {\bibfnamefont {S.~L.}\ \bibnamefont
  {Brunton}}\ and\ \bibinfo {author} {\bibfnamefont {J.~N.}\ \bibnamefont
  {Kutz}},\ }\href@noop {} {\emph {\bibinfo {title} {Data-driven science and
  engineering: Machine learning, dynamical systems, and control}}}\ (\bibinfo
  {publisher} {Cambridge University Press},\ \bibinfo {year}
  {2019})\BibitemShut {NoStop}%
\bibitem [{\citenamefont {Lorenz}(1963)}]{Lorenz1963jas}%
  \BibitemOpen
  \bibfield  {author} {\bibinfo {author} {\bibfnamefont {E.~N.}\ \bibnamefont
  {Lorenz}},\ }\bibfield  {title} {\bibinfo {title} {Deterministic nonperiodic
  flow},\ }\href@noop {} {\bibfield  {journal} {\bibinfo  {journal} {Journal of
  the atmospheric sciences}\ }\textbf {\bibinfo {volume} {20}},\ \bibinfo
  {pages} {130} (\bibinfo {year} {1963})}\BibitemShut {NoStop}%
\bibitem [{\citenamefont {Guan}\ \emph {et~al.}(2020)\citenamefont {Guan},
  \citenamefont {Brunton},\ and\ \citenamefont {Novosselov}}]{guan2020sparse}%
  \BibitemOpen
  \bibfield  {author} {\bibinfo {author} {\bibfnamefont {Y.}~\bibnamefont
  {Guan}}, \bibinfo {author} {\bibfnamefont {S.~L.}\ \bibnamefont {Brunton}},\
  and\ \bibinfo {author} {\bibfnamefont {I.}~\bibnamefont {Novosselov}},\
  }\bibfield  {title} {\bibinfo {title} {Sparse nonlinear models of chaotic
  electroconvection},\ }\href@noop {} {\bibfield  {journal} {\bibinfo
  {journal} {arXiv preprint arXiv:2009.11862}\ } (\bibinfo {year}
  {2020})}\BibitemShut {NoStop}%
\bibitem [{\citenamefont {Humbird}\ \emph {et~al.}(2019)\citenamefont
  {Humbird}, \citenamefont {Peterson}, \citenamefont {Spears},\ and\
  \citenamefont {McClarren}}]{humbird2019transfer}%
  \BibitemOpen
  \bibfield  {author} {\bibinfo {author} {\bibfnamefont {K.~D.}\ \bibnamefont
  {Humbird}}, \bibinfo {author} {\bibfnamefont {J.~L.}\ \bibnamefont
  {Peterson}}, \bibinfo {author} {\bibfnamefont {B.}~\bibnamefont {Spears}},\
  and\ \bibinfo {author} {\bibfnamefont {R.}~\bibnamefont {McClarren}},\
  }\bibfield  {title} {\bibinfo {title} {Transfer learning to model inertial
  confinement fusion experiments},\ }\href@noop {} {\bibfield  {journal}
  {\bibinfo  {journal} {IEEE Transactions on Plasma Science}\ }\textbf
  {\bibinfo {volume} {48}},\ \bibinfo {pages} {61} (\bibinfo {year}
  {2019})}\BibitemShut {NoStop}%
\bibitem [{\citenamefont {Kapteyn}\ \emph {et~al.}(2020)\citenamefont
  {Kapteyn}, \citenamefont {Knezevic},\ and\ \citenamefont
  {Willcox}}]{kapteyn2020toward}%
  \BibitemOpen
  \bibfield  {author} {\bibinfo {author} {\bibfnamefont {M.~G.}\ \bibnamefont
  {Kapteyn}}, \bibinfo {author} {\bibfnamefont {D.~J.}\ \bibnamefont
  {Knezevic}},\ and\ \bibinfo {author} {\bibfnamefont {K.}~\bibnamefont
  {Willcox}},\ }\bibfield  {title} {\bibinfo {title} {Toward predictive digital
  twins via component-based reduced-order models and interpretable machine
  learning},\ }in\ \href@noop {} {\emph {\bibinfo {booktitle} {AIAA Scitech
  2020 Forum}}}\ (\bibinfo {year} {2020})\ p.\ \bibinfo {pages}
  {0418}\BibitemShut {NoStop}%
\bibitem [{\citenamefont {Ravindran}(2005)}]{ravindran2005real}%
  \BibitemOpen
  \bibfield  {author} {\bibinfo {author} {\bibfnamefont {S.}~\bibnamefont
  {Ravindran}},\ }\bibfield  {title} {\bibinfo {title} {Real-time computational
  algorithm for optimal control of an {MHD} flow system},\ }\href@noop {}
  {\bibfield  {journal} {\bibinfo  {journal} {SIAM Journal on Scientific
  Computing}\ }\textbf {\bibinfo {volume} {26}},\ \bibinfo {pages} {1369}
  (\bibinfo {year} {2005})}\BibitemShut {NoStop}%
\bibitem [{\citenamefont {Galperti}\ \emph {et~al.}(2017)\citenamefont
  {Galperti}, \citenamefont {Coda}, \citenamefont {Duval}, \citenamefont
  {Llobet}, \citenamefont {Milne}, \citenamefont {Sauter}, \citenamefont
  {Moret},\ and\ \citenamefont {Testa}}]{galperti2017integration}%
  \BibitemOpen
  \bibfield  {author} {\bibinfo {author} {\bibfnamefont {C.}~\bibnamefont
  {Galperti}}, \bibinfo {author} {\bibfnamefont {S.}~\bibnamefont {Coda}},
  \bibinfo {author} {\bibfnamefont {B.}~\bibnamefont {Duval}}, \bibinfo
  {author} {\bibfnamefont {X.}~\bibnamefont {Llobet}}, \bibinfo {author}
  {\bibfnamefont {P.}~\bibnamefont {Milne}}, \bibinfo {author} {\bibfnamefont
  {O.}~\bibnamefont {Sauter}}, \bibinfo {author} {\bibfnamefont
  {J.}~\bibnamefont {Moret}},\ and\ \bibinfo {author} {\bibfnamefont
  {D.}~\bibnamefont {Testa}},\ }\bibfield  {title} {\bibinfo {title}
  {Integration of a real-time node for magnetic perturbations signal analysis
  in the distributed digital control system of the {TCV} tokamak},\ }\href@noop
  {} {\bibfield  {journal} {\bibinfo  {journal} {IEEE Transactions on Nuclear
  Science}\ }\textbf {\bibinfo {volume} {64}},\ \bibinfo {pages} {1446}
  (\bibinfo {year} {2017})}\BibitemShut {NoStop}%
\bibitem [{\citenamefont {Levesque}\ \emph {et~al.}(2013)\citenamefont
  {Levesque}, \citenamefont {Rath}, \citenamefont {Shiraki}, \citenamefont
  {Angelini}, \citenamefont {Bialek}, \citenamefont {Byrne}, \citenamefont
  {DeBono}, \citenamefont {Hughes}, \citenamefont {Mauel}, \citenamefont
  {Navratil} \emph {et~al.}}]{levesque2013multimode}%
  \BibitemOpen
  \bibfield  {author} {\bibinfo {author} {\bibfnamefont {J.}~\bibnamefont
  {Levesque}}, \bibinfo {author} {\bibfnamefont {N.}~\bibnamefont {Rath}},
  \bibinfo {author} {\bibfnamefont {D.}~\bibnamefont {Shiraki}}, \bibinfo
  {author} {\bibfnamefont {S.}~\bibnamefont {Angelini}}, \bibinfo {author}
  {\bibfnamefont {J.}~\bibnamefont {Bialek}}, \bibinfo {author} {\bibfnamefont
  {P.}~\bibnamefont {Byrne}}, \bibinfo {author} {\bibfnamefont
  {B.}~\bibnamefont {DeBono}}, \bibinfo {author} {\bibfnamefont
  {P.}~\bibnamefont {Hughes}}, \bibinfo {author} {\bibfnamefont
  {M.}~\bibnamefont {Mauel}}, \bibinfo {author} {\bibfnamefont
  {G.}~\bibnamefont {Navratil}}, \emph {et~al.},\ }\bibfield  {title} {\bibinfo
  {title} {Multimode observations and 3{D} magnetic control of the boundary of
  a tokamak plasma},\ }\href@noop {} {\bibfield  {journal} {\bibinfo  {journal}
  {Nuclear Fusion}\ }\textbf {\bibinfo {volume} {53}},\ \bibinfo {pages}
  {073037} (\bibinfo {year} {2013})}\BibitemShut {NoStop}%
\bibitem [{\citenamefont {Wang}\ \emph {et~al.}(2020)\citenamefont {Wang},
  \citenamefont {Xu}, \citenamefont {Zhu}, \citenamefont {Ma},\ and\
  \citenamefont {Lei}}]{wang2020deep}%
  \BibitemOpen
  \bibfield  {author} {\bibinfo {author} {\bibfnamefont {L.}~\bibnamefont
  {Wang}}, \bibinfo {author} {\bibfnamefont {X.}~\bibnamefont {Xu}}, \bibinfo
  {author} {\bibfnamefont {B.}~\bibnamefont {Zhu}}, \bibinfo {author}
  {\bibfnamefont {C.}~\bibnamefont {Ma}},\ and\ \bibinfo {author}
  {\bibfnamefont {Y.-a.}\ \bibnamefont {Lei}},\ }\bibfield  {title} {\bibinfo
  {title} {Deep learning surrogate model for kinetic {L}andau-fluid closure
  with collision},\ }\href@noop {} {\bibfield  {journal} {\bibinfo  {journal}
  {AIP Advances}\ }\textbf {\bibinfo {volume} {10}},\ \bibinfo {pages} {075108}
  (\bibinfo {year} {2020})}\BibitemShut {NoStop}%
\bibitem [{\citenamefont {Citrin}\ \emph {et~al.}(2015)\citenamefont {Citrin},
  \citenamefont {Breton}, \citenamefont {Felici}, \citenamefont {Imbeaux},
  \citenamefont {Aniel}, \citenamefont {Artaud}, \citenamefont {Baiocchi},
  \citenamefont {Bourdelle}, \citenamefont {Camenen},\ and\ \citenamefont
  {Garcia}}]{citrin2015real}%
  \BibitemOpen
  \bibfield  {author} {\bibinfo {author} {\bibfnamefont {J.}~\bibnamefont
  {Citrin}}, \bibinfo {author} {\bibfnamefont {S.}~\bibnamefont {Breton}},
  \bibinfo {author} {\bibfnamefont {F.}~\bibnamefont {Felici}}, \bibinfo
  {author} {\bibfnamefont {F.}~\bibnamefont {Imbeaux}}, \bibinfo {author}
  {\bibfnamefont {T.}~\bibnamefont {Aniel}}, \bibinfo {author} {\bibfnamefont
  {J.}~\bibnamefont {Artaud}}, \bibinfo {author} {\bibfnamefont
  {B.}~\bibnamefont {Baiocchi}}, \bibinfo {author} {\bibfnamefont
  {C.}~\bibnamefont {Bourdelle}}, \bibinfo {author} {\bibfnamefont
  {Y.}~\bibnamefont {Camenen}},\ and\ \bibinfo {author} {\bibfnamefont
  {J.}~\bibnamefont {Garcia}},\ }\bibfield  {title} {\bibinfo {title}
  {Real-time capable first principle based modelling of tokamak turbulent
  transport},\ }\href@noop {} {\bibfield  {journal} {\bibinfo  {journal}
  {Nuclear Fusion}\ }\textbf {\bibinfo {volume} {55}},\ \bibinfo {pages}
  {092001} (\bibinfo {year} {2015})}\BibitemShut {NoStop}%
\bibitem [{\citenamefont {van~de Plassche}\ \emph {et~al.}(2020)\citenamefont
  {van~de Plassche}, \citenamefont {Citrin}, \citenamefont {Bourdelle},
  \citenamefont {Camenen}, \citenamefont {Casson}, \citenamefont {Dagnelie},
  \citenamefont {Felici}, \citenamefont {Ho}, \citenamefont {Van~Mulders},\
  and\ \citenamefont {Contributors}}]{van2020fast}%
  \BibitemOpen
  \bibfield  {author} {\bibinfo {author} {\bibfnamefont {K.~L.}\ \bibnamefont
  {van~de Plassche}}, \bibinfo {author} {\bibfnamefont {J.}~\bibnamefont
  {Citrin}}, \bibinfo {author} {\bibfnamefont {C.}~\bibnamefont {Bourdelle}},
  \bibinfo {author} {\bibfnamefont {Y.}~\bibnamefont {Camenen}}, \bibinfo
  {author} {\bibfnamefont {F.~J.}\ \bibnamefont {Casson}}, \bibinfo {author}
  {\bibfnamefont {V.~I.}\ \bibnamefont {Dagnelie}}, \bibinfo {author}
  {\bibfnamefont {F.}~\bibnamefont {Felici}}, \bibinfo {author} {\bibfnamefont
  {A.}~\bibnamefont {Ho}}, \bibinfo {author} {\bibfnamefont {S.}~\bibnamefont
  {Van~Mulders}},\ and\ \bibinfo {author} {\bibfnamefont {J.}~\bibnamefont
  {Contributors}},\ }\bibfield  {title} {\bibinfo {title} {Fast modeling of
  turbulent transport in fusion plasmas using neural networks},\ }\href@noop {}
  {\bibfield  {journal} {\bibinfo  {journal} {Physics of Plasmas}\ }\textbf
  {\bibinfo {volume} {27}},\ \bibinfo {pages} {022310} (\bibinfo {year}
  {2020})}\BibitemShut {NoStop}%
\bibitem [{\citenamefont {Loarte}\ \emph {et~al.}(2007)\citenamefont {Loarte},
  \citenamefont {Lipschultz}, \citenamefont {Kukushkin}, \citenamefont
  {Matthews}, \citenamefont {Stangeby}, \citenamefont {Asakura}, \citenamefont
  {Counsell}, \citenamefont {Federici}, \citenamefont {Kallenbach},
  \citenamefont {Krieger} \emph {et~al.}}]{loarte2007power}%
  \BibitemOpen
  \bibfield  {author} {\bibinfo {author} {\bibfnamefont {A.}~\bibnamefont
  {Loarte}}, \bibinfo {author} {\bibfnamefont {B.}~\bibnamefont {Lipschultz}},
  \bibinfo {author} {\bibfnamefont {A.}~\bibnamefont {Kukushkin}}, \bibinfo
  {author} {\bibfnamefont {G.}~\bibnamefont {Matthews}}, \bibinfo {author}
  {\bibfnamefont {P.}~\bibnamefont {Stangeby}}, \bibinfo {author}
  {\bibfnamefont {N.}~\bibnamefont {Asakura}}, \bibinfo {author} {\bibfnamefont
  {G.}~\bibnamefont {Counsell}}, \bibinfo {author} {\bibfnamefont
  {G.}~\bibnamefont {Federici}}, \bibinfo {author} {\bibfnamefont
  {A.}~\bibnamefont {Kallenbach}}, \bibinfo {author} {\bibfnamefont
  {K.}~\bibnamefont {Krieger}}, \emph {et~al.},\ }\bibfield  {title} {\bibinfo
  {title} {Power and particle control},\ }\href@noop {} {\bibfield  {journal}
  {\bibinfo  {journal} {Nuclear Fusion}\ }\textbf {\bibinfo {volume} {47}},\
  \bibinfo {pages} {S203} (\bibinfo {year} {2007})}\BibitemShut {NoStop}%
\bibitem [{\citenamefont {Allg{\"o}wer}\ \emph {et~al.}(1999)\citenamefont
  {Allg{\"o}wer}, \citenamefont {Badgwell}, \citenamefont {Qin}, \citenamefont
  {Rawlings},\ and\ \citenamefont {Wright}}]{allgower1999nonlinear}%
  \BibitemOpen
  \bibfield  {author} {\bibinfo {author} {\bibfnamefont {F.}~\bibnamefont
  {Allg{\"o}wer}}, \bibinfo {author} {\bibfnamefont {T.~A.}\ \bibnamefont
  {Badgwell}}, \bibinfo {author} {\bibfnamefont {J.~S.}\ \bibnamefont {Qin}},
  \bibinfo {author} {\bibfnamefont {J.~B.}\ \bibnamefont {Rawlings}},\ and\
  \bibinfo {author} {\bibfnamefont {S.~J.}\ \bibnamefont {Wright}},\ }\bibfield
   {title} {\bibinfo {title} {Nonlinear predictive control and moving horizon
  estimation—an introductory overview},\ }in\ \href@noop {} {\emph {\bibinfo
  {booktitle} {Advances in control}}}\ (\bibinfo  {publisher} {Springer},\
  \bibinfo {year} {1999})\ pp.\ \bibinfo {pages} {391--449}\BibitemShut
  {NoStop}%
\bibitem [{\citenamefont {Loiseau}\ and\ \citenamefont
  {Brunton}(2018)}]{loiseau2018constrained}%
  \BibitemOpen
  \bibfield  {author} {\bibinfo {author} {\bibfnamefont {J.-C.}\ \bibnamefont
  {Loiseau}}\ and\ \bibinfo {author} {\bibfnamefont {S.~L.}\ \bibnamefont
  {Brunton}},\ }\bibfield  {title} {\bibinfo {title} {Constrained sparse
  {G}alerkin regression},\ }\href@noop {} {\bibfield  {journal} {\bibinfo
  {journal} {Journal of Fluid Mechanics}\ }\textbf {\bibinfo {volume} {838}},\
  \bibinfo {pages} {42} (\bibinfo {year} {2018})}\BibitemShut {NoStop}%
\bibitem [{\citenamefont {Loiseau}(2020)}]{loiseau2020data}%
  \BibitemOpen
  \bibfield  {author} {\bibinfo {author} {\bibfnamefont {J.-C.}\ \bibnamefont
  {Loiseau}},\ }\bibfield  {title} {\bibinfo {title} {Data-driven modeling of
  the chaotic thermal convection in an annular thermosyphon},\ }\href@noop {}
  {\bibfield  {journal} {\bibinfo  {journal} {Theoretical and Computational
  Fluid Dynamics}\ }\textbf {\bibinfo {volume} {34}},\ \bibinfo {pages} {339}
  (\bibinfo {year} {2020})}\BibitemShut {NoStop}%
\bibitem [{\citenamefont {Kobayashi}\ \emph {et~al.}(2015)\citenamefont
  {Kobayashi}, \citenamefont {G{\"u}rcan},\ and\ \citenamefont
  {Diamond}}]{kobayashi2015direct}%
  \BibitemOpen
  \bibfield  {author} {\bibinfo {author} {\bibfnamefont {S.}~\bibnamefont
  {Kobayashi}}, \bibinfo {author} {\bibfnamefont {{\"O}.~D.}\ \bibnamefont
  {G{\"u}rcan}},\ and\ \bibinfo {author} {\bibfnamefont {P.~H.}\ \bibnamefont
  {Diamond}},\ }\bibfield  {title} {\bibinfo {title} {Direct identification of
  predator-prey dynamics in gyrokinetic simulations},\ }\href@noop {}
  {\bibfield  {journal} {\bibinfo  {journal} {Physics of Plasmas}\ }\textbf
  {\bibinfo {volume} {22}},\ \bibinfo {pages} {090702} (\bibinfo {year}
  {2015})}\BibitemShut {NoStop}%
\bibitem [{\citenamefont {Alves}\ and\ \citenamefont
  {Fiuza}(2020)}]{alves2020data}%
  \BibitemOpen
  \bibfield  {author} {\bibinfo {author} {\bibfnamefont {E.~P.}\ \bibnamefont
  {Alves}}\ and\ \bibinfo {author} {\bibfnamefont {F.}~\bibnamefont {Fiuza}},\
  }\bibfield  {title} {\bibinfo {title} {Data-driven discovery of reduced
  plasma physics models from fully-kinetic simulations},\ }\href@noop {}
  {\bibfield  {journal} {\bibinfo  {journal} {arXiv preprint arXiv:2011.01927}\
  } (\bibinfo {year} {2020})}\BibitemShut {NoStop}%
\bibitem [{\citenamefont {Dam}\ \emph {et~al.}(2017)\citenamefont {Dam},
  \citenamefont {Br{\o}ns}, \citenamefont {Juul~Rasmussen}, \citenamefont
  {Naulin},\ and\ \citenamefont {Hesthaven}}]{Dam2017pf}%
  \BibitemOpen
  \bibfield  {author} {\bibinfo {author} {\bibfnamefont {M.}~\bibnamefont
  {Dam}}, \bibinfo {author} {\bibfnamefont {M.}~\bibnamefont {Br{\o}ns}},
  \bibinfo {author} {\bibfnamefont {J.}~\bibnamefont {Juul~Rasmussen}},
  \bibinfo {author} {\bibfnamefont {V.}~\bibnamefont {Naulin}},\ and\ \bibinfo
  {author} {\bibfnamefont {J.~S.}\ \bibnamefont {Hesthaven}},\ }\bibfield
  {title} {\bibinfo {title} {Sparse identification of a predator-prey system
  from simulation data of a convection model},\ }\href@noop {} {\bibfield
  {journal} {\bibinfo  {journal} {Physics of Plasmas}\ }\textbf {\bibinfo
  {volume} {24}},\ \bibinfo {pages} {022310} (\bibinfo {year}
  {2017})}\BibitemShut {NoStop}%
\bibitem [{\citenamefont {Holmes}\ \emph {et~al.}(2012)\citenamefont {Holmes},
  \citenamefont {Lumley}, \citenamefont {Berkooz},\ and\ \citenamefont
  {Rowley}}]{holmes2012turbulence}%
  \BibitemOpen
  \bibfield  {author} {\bibinfo {author} {\bibfnamefont {P.}~\bibnamefont
  {Holmes}}, \bibinfo {author} {\bibfnamefont {J.~L.}\ \bibnamefont {Lumley}},
  \bibinfo {author} {\bibfnamefont {G.}~\bibnamefont {Berkooz}},\ and\ \bibinfo
  {author} {\bibfnamefont {C.~W.}\ \bibnamefont {Rowley}},\ }\href@noop {}
  {\emph {\bibinfo {title} {Turbulence, coherent structures, dynamical systems
  and symmetry}}}\ (\bibinfo  {publisher} {Cambridge university press},\
  \bibinfo {year} {2012})\BibitemShut {NoStop}%
\bibitem [{\citenamefont {Willcox}\ and\ \citenamefont
  {Peraire}(2002)}]{willcox2002balanced}%
  \BibitemOpen
  \bibfield  {author} {\bibinfo {author} {\bibfnamefont {K.}~\bibnamefont
  {Willcox}}\ and\ \bibinfo {author} {\bibfnamefont {J.}~\bibnamefont
  {Peraire}},\ }\bibfield  {title} {\bibinfo {title} {Balanced model reduction
  via the proper orthogonal decomposition},\ }\href@noop {} {\bibfield
  {journal} {\bibinfo  {journal} {AIAA journal}\ }\textbf {\bibinfo {volume}
  {40}},\ \bibinfo {pages} {2323} (\bibinfo {year} {2002})}\BibitemShut
  {NoStop}%
\bibitem [{\citenamefont {Benner}\ \emph {et~al.}(2015)\citenamefont {Benner},
  \citenamefont {Gugercin},\ and\ \citenamefont {Willcox}}]{benner2015survey}%
  \BibitemOpen
  \bibfield  {author} {\bibinfo {author} {\bibfnamefont {P.}~\bibnamefont
  {Benner}}, \bibinfo {author} {\bibfnamefont {S.}~\bibnamefont {Gugercin}},\
  and\ \bibinfo {author} {\bibfnamefont {K.}~\bibnamefont {Willcox}},\
  }\bibfield  {title} {\bibinfo {title} {A survey of projection-based model
  reduction methods for parametric dynamical systems},\ }\href@noop {}
  {\bibfield  {journal} {\bibinfo  {journal} {SIAM review}\ }\textbf {\bibinfo
  {volume} {57}},\ \bibinfo {pages} {483} (\bibinfo {year} {2015})}\BibitemShut
  {NoStop}%
\bibitem [{\citenamefont {Rowley}(2005)}]{rowley2005model}%
  \BibitemOpen
  \bibfield  {author} {\bibinfo {author} {\bibfnamefont {C.~W.}\ \bibnamefont
  {Rowley}},\ }\bibfield  {title} {\bibinfo {title} {Model reduction for
  fluids, using balanced proper orthogonal decomposition},\ }\href@noop {}
  {\bibfield  {journal} {\bibinfo  {journal} {International Journal of
  Bifurcation and Chaos}\ }\textbf {\bibinfo {volume} {15}},\ \bibinfo {pages}
  {997} (\bibinfo {year} {2005})}\BibitemShut {NoStop}%
\bibitem [{\citenamefont {Towne}\ \emph {et~al.}(2018)\citenamefont {Towne},
  \citenamefont {Schmidt},\ and\ \citenamefont {Colonius}}]{Towne2018jfm}%
  \BibitemOpen
  \bibfield  {author} {\bibinfo {author} {\bibfnamefont {A.}~\bibnamefont
  {Towne}}, \bibinfo {author} {\bibfnamefont {O.~T.}\ \bibnamefont {Schmidt}},\
  and\ \bibinfo {author} {\bibfnamefont {T.}~\bibnamefont {Colonius}},\
  }\bibfield  {title} {\bibinfo {title} {Spectral proper orthogonal
  decomposition and its relationship to dynamic mode decomposition and
  resolvent analysis},\ }\href@noop {} {\bibfield  {journal} {\bibinfo
  {journal} {Journal of Fluid Mechanics}\ }\textbf {\bibinfo {volume} {847}},\
  \bibinfo {pages} {821} (\bibinfo {year} {2018})}\BibitemShut {NoStop}%
\bibitem [{\citenamefont {Schmid}(2010)}]{schmid_dynamic_2010}%
  \BibitemOpen
  \bibfield  {author} {\bibinfo {author} {\bibfnamefont {P.~J.}\ \bibnamefont
  {Schmid}},\ }\bibfield  {title} {\bibinfo {title} {Dynamic mode decomposition
  of numerical and experimental data},\ }\href@noop {} {\bibfield  {journal}
  {\bibinfo  {journal} {Journal of Fluid Mechanics}\ }\textbf {\bibinfo
  {volume} {656}},\ \bibinfo {pages} {5} (\bibinfo {year} {2010})}\BibitemShut
  {NoStop}%
\bibitem [{\citenamefont {Rowley}\ \emph {et~al.}(2009)\citenamefont {Rowley},
  \citenamefont {Mezi\'c}, \citenamefont {Bagheri}, \citenamefont {Schlatter},\
  and\ \citenamefont {Henningson}}]{Rowley2009jfm}%
  \BibitemOpen
  \bibfield  {author} {\bibinfo {author} {\bibfnamefont {C.~W.}\ \bibnamefont
  {Rowley}}, \bibinfo {author} {\bibfnamefont {I.}~\bibnamefont {Mezi\'c}},
  \bibinfo {author} {\bibfnamefont {S.}~\bibnamefont {Bagheri}}, \bibinfo
  {author} {\bibfnamefont {P.}~\bibnamefont {Schlatter}},\ and\ \bibinfo
  {author} {\bibfnamefont {D.}~\bibnamefont {Henningson}},\ }\bibfield  {title}
  {\bibinfo {title} {Spectral analysis of nonlinear flows},\ }\href@noop {}
  {\bibfield  {journal} {\bibinfo  {journal} {J.\ Fluid Mech.}\ }\textbf
  {\bibinfo {volume} {645}},\ \bibinfo {pages} {115} (\bibinfo {year}
  {2009})}\BibitemShut {NoStop}%
\bibitem [{\citenamefont {Tu}\ \emph {et~al.}(2014)\citenamefont {Tu},
  \citenamefont {Rowley}, \citenamefont {Luchtenburg}, \citenamefont
  {Brunton},\ and\ \citenamefont {Kutz}}]{Tu2014jcd}%
  \BibitemOpen
  \bibfield  {author} {\bibinfo {author} {\bibfnamefont {J.~H.}\ \bibnamefont
  {Tu}}, \bibinfo {author} {\bibfnamefont {C.~W.}\ \bibnamefont {Rowley}},
  \bibinfo {author} {\bibfnamefont {D.~M.}\ \bibnamefont {Luchtenburg}},
  \bibinfo {author} {\bibfnamefont {S.~L.}\ \bibnamefont {Brunton}},\ and\
  \bibinfo {author} {\bibfnamefont {J.~N.}\ \bibnamefont {Kutz}},\ }\bibfield
  {title} {\bibinfo {title} {On dynamic mode decomposition: theory and
  applications},\ }\href@noop {} {\bibfield  {journal} {\bibinfo  {journal}
  {Journal of Computational Dynamics}\ }\textbf {\bibinfo {volume} {1}},\
  \bibinfo {pages} {391} (\bibinfo {year} {2014})}\BibitemShut {NoStop}%
\bibitem [{\citenamefont {Koopman}(1931)}]{koopman_hamiltonian_1931}%
  \BibitemOpen
  \bibfield  {author} {\bibinfo {author} {\bibfnamefont {B.~O.}\ \bibnamefont
  {Koopman}},\ }\bibfield  {title} {\bibinfo {title} {Hamiltonian systems and
  transformation in {Hilbert} space},\ }\href@noop {} {\bibfield  {journal}
  {\bibinfo  {journal} {Proceedings of the National Academy of Sciences}\
  }\textbf {\bibinfo {volume} {17}},\ \bibinfo {pages} {315} (\bibinfo {year}
  {1931})}\BibitemShut {NoStop}%
\bibitem [{\citenamefont {Mezic}(2013)}]{mezic_analysis_2013}%
  \BibitemOpen
  \bibfield  {author} {\bibinfo {author} {\bibfnamefont {I.}~\bibnamefont
  {Mezic}},\ }\bibfield  {title} {\bibinfo {title} {Analysis of fluid flows via
  spectral properties of the {Koopman} operator},\ }\href@noop {} {\bibfield
  {journal} {\bibinfo  {journal} {Annual Review of Fluid Mechanics}\ }\textbf
  {\bibinfo {volume} {45}},\ \bibinfo {pages} {357} (\bibinfo {year}
  {2013})}\BibitemShut {NoStop}%
\bibitem [{\citenamefont {Brunton}\ \emph {et~al.}(2021)\citenamefont
  {Brunton}, \citenamefont {Budi{\v{s}}i{\'c}}, \citenamefont {Kaiser},\ and\
  \citenamefont {Kutz}}]{brunton2021modern}%
  \BibitemOpen
  \bibfield  {author} {\bibinfo {author} {\bibfnamefont {S.~L.}\ \bibnamefont
  {Brunton}}, \bibinfo {author} {\bibfnamefont {M.}~\bibnamefont
  {Budi{\v{s}}i{\'c}}}, \bibinfo {author} {\bibfnamefont {E.}~\bibnamefont
  {Kaiser}},\ and\ \bibinfo {author} {\bibfnamefont {J.~N.}\ \bibnamefont
  {Kutz}},\ }\bibfield  {title} {\bibinfo {title} {Modern {K}oopman theory for
  dynamical systems},\ }\href@noop {} {\bibfield  {journal} {\bibinfo
  {journal} {arXiv preprint arXiv:2102.12086}\ } (\bibinfo {year}
  {2021})}\BibitemShut {NoStop}%
\bibitem [{\citenamefont {McKeon}\ and\ \citenamefont
  {Sharma}(2010)}]{mckeon2010critical}%
  \BibitemOpen
  \bibfield  {author} {\bibinfo {author} {\bibfnamefont {B.}~\bibnamefont
  {McKeon}}\ and\ \bibinfo {author} {\bibfnamefont {A.}~\bibnamefont
  {Sharma}},\ }\bibfield  {title} {\bibinfo {title} {A critical-layer framework
  for turbulent pipe flow},\ }\href@noop {} {\bibfield  {journal} {\bibinfo
  {journal} {Journal of Fluid Mechanics}\ }\textbf {\bibinfo {volume} {658}},\
  \bibinfo {pages} {336} (\bibinfo {year} {2010})}\BibitemShut {NoStop}%
\bibitem [{\citenamefont {Luhar}\ \emph {et~al.}(2014)\citenamefont {Luhar},
  \citenamefont {Sharma},\ and\ \citenamefont {McKeon}}]{luhar2014opposition}%
  \BibitemOpen
  \bibfield  {author} {\bibinfo {author} {\bibfnamefont {M.}~\bibnamefont
  {Luhar}}, \bibinfo {author} {\bibfnamefont {A.~S.}\ \bibnamefont {Sharma}},\
  and\ \bibinfo {author} {\bibfnamefont {B.~J.}\ \bibnamefont {McKeon}},\
  }\bibfield  {title} {\bibinfo {title} {Opposition control within the
  resolvent analysis framework},\ }\href@noop {} {\bibfield  {journal}
  {\bibinfo  {journal} {Journal of Fluid Mechanics}\ }\textbf {\bibinfo
  {volume} {749}},\ \bibinfo {pages} {597} (\bibinfo {year}
  {2014})}\BibitemShut {NoStop}%
\bibitem [{\citenamefont {Lusch}\ \emph {et~al.}(2018)\citenamefont {Lusch},
  \citenamefont {Kutz},\ and\ \citenamefont {Brunton}}]{lusch2018deep}%
  \BibitemOpen
  \bibfield  {author} {\bibinfo {author} {\bibfnamefont {B.}~\bibnamefont
  {Lusch}}, \bibinfo {author} {\bibfnamefont {J.~N.}\ \bibnamefont {Kutz}},\
  and\ \bibinfo {author} {\bibfnamefont {S.~L.}\ \bibnamefont {Brunton}},\
  }\bibfield  {title} {\bibinfo {title} {Deep learning for universal linear
  embeddings of nonlinear dynamics},\ }\href@noop {} {\bibfield  {journal}
  {\bibinfo  {journal} {Nature communications}\ }\textbf {\bibinfo {volume}
  {9}},\ \bibinfo {pages} {4950} (\bibinfo {year} {2018})}\BibitemShut
  {NoStop}%
\bibitem [{\citenamefont {Champion}\ \emph {et~al.}(2019)\citenamefont
  {Champion}, \citenamefont {Lusch}, \citenamefont {Kutz},\ and\ \citenamefont
  {Brunton}}]{champion2019data}%
  \BibitemOpen
  \bibfield  {author} {\bibinfo {author} {\bibfnamefont {K.}~\bibnamefont
  {Champion}}, \bibinfo {author} {\bibfnamefont {B.}~\bibnamefont {Lusch}},
  \bibinfo {author} {\bibfnamefont {J.~N.}\ \bibnamefont {Kutz}},\ and\
  \bibinfo {author} {\bibfnamefont {S.~L.}\ \bibnamefont {Brunton}},\
  }\bibfield  {title} {\bibinfo {title} {Data-driven discovery of coordinates
  and governing equations},\ }\href@noop {} {\bibfield  {journal} {\bibinfo
  {journal} {Proceedings of the National Academy of Sciences}\ }\textbf
  {\bibinfo {volume} {116}},\ \bibinfo {pages} {22445} (\bibinfo {year}
  {2019})}\BibitemShut {NoStop}%
\bibitem [{\citenamefont {Lee}\ and\ \citenamefont
  {Carlberg}(2020)}]{lee2020model}%
  \BibitemOpen
  \bibfield  {author} {\bibinfo {author} {\bibfnamefont {K.}~\bibnamefont
  {Lee}}\ and\ \bibinfo {author} {\bibfnamefont {K.~T.}\ \bibnamefont
  {Carlberg}},\ }\bibfield  {title} {\bibinfo {title} {Model reduction of
  dynamical systems on nonlinear manifolds using deep convolutional
  autoencoders},\ }\href@noop {} {\bibfield  {journal} {\bibinfo  {journal}
  {Journal of Computational Physics}\ }\textbf {\bibinfo {volume} {404}},\
  \bibinfo {pages} {108973} (\bibinfo {year} {2020})}\BibitemShut {NoStop}%
\bibitem [{\citenamefont {Bongard}\ and\ \citenamefont
  {Lipson}(2007)}]{bongard_automated_2007}%
  \BibitemOpen
  \bibfield  {author} {\bibinfo {author} {\bibfnamefont {J.}~\bibnamefont
  {Bongard}}\ and\ \bibinfo {author} {\bibfnamefont {H.}~\bibnamefont
  {Lipson}},\ }\bibfield  {title} {\bibinfo {title} {Automated reverse
  engineering of nonlinear dynamical systems},\ }\href@noop {} {\bibfield
  {journal} {\bibinfo  {journal} {Proceedings of the National Academy of
  Sciences}\ }\textbf {\bibinfo {volume} {104}},\ \bibinfo {pages} {9943}
  (\bibinfo {year} {2007})}\BibitemShut {NoStop}%
\bibitem [{\citenamefont {Schmidt}\ and\ \citenamefont
  {Lipson}(2009)}]{schmidt_distilling_2009}%
  \BibitemOpen
  \bibfield  {author} {\bibinfo {author} {\bibfnamefont {M.}~\bibnamefont
  {Schmidt}}\ and\ \bibinfo {author} {\bibfnamefont {H.}~\bibnamefont
  {Lipson}},\ }\bibfield  {title} {\bibinfo {title} {Distilling free-form
  natural laws from experimental data},\ }\href@noop {} {\bibfield  {journal}
  {\bibinfo  {journal} {Science}\ }\textbf {\bibinfo {volume} {324}},\ \bibinfo
  {pages} {81} (\bibinfo {year} {2009})}\BibitemShut {NoStop}%
\bibitem [{\citenamefont {Brunton}\ \emph {et~al.}(2016)\citenamefont
  {Brunton}, \citenamefont {Proctor},\ and\ \citenamefont
  {Kutz}}]{Brunton2016pnas}%
  \BibitemOpen
  \bibfield  {author} {\bibinfo {author} {\bibfnamefont {S.~L.}\ \bibnamefont
  {Brunton}}, \bibinfo {author} {\bibfnamefont {J.~L.}\ \bibnamefont
  {Proctor}},\ and\ \bibinfo {author} {\bibfnamefont {J.~N.}\ \bibnamefont
  {Kutz}},\ }\bibfield  {title} {\bibinfo {title} {Discovering governing
  equations from data by sparse identification of nonlinear dynamical
  systems},\ }\href@noop {} {\bibfield  {journal} {\bibinfo  {journal}
  {Proceedings of the National Academy of Sciences}\ }\textbf {\bibinfo
  {volume} {113}},\ \bibinfo {pages} {3932} (\bibinfo {year}
  {2016})}\BibitemShut {NoStop}%
\bibitem [{\citenamefont {Raissi}\ and\ \citenamefont
  {Karniadakis}(2017)}]{raissi2017hidden}%
  \BibitemOpen
  \bibfield  {author} {\bibinfo {author} {\bibfnamefont {M.}~\bibnamefont
  {Raissi}}\ and\ \bibinfo {author} {\bibfnamefont {G.~E.}\ \bibnamefont
  {Karniadakis}},\ }\bibfield  {title} {\bibinfo {title} {Hidden physics
  models: Machine learning of nonlinear partial differential equations},\
  }\href@noop {} {\bibfield  {journal} {\bibinfo  {journal} {arXiv preprint
  arXiv:1708.00588}\ } (\bibinfo {year} {2017})}\BibitemShut {NoStop}%
\bibitem [{\citenamefont {Raissi}\ \emph
  {et~al.}(2017{\natexlab{a}})\citenamefont {Raissi}, \citenamefont
  {Perdikaris},\ and\ \citenamefont {Karniadakis}}]{raissi2017physics1}%
  \BibitemOpen
  \bibfield  {author} {\bibinfo {author} {\bibfnamefont {M.}~\bibnamefont
  {Raissi}}, \bibinfo {author} {\bibfnamefont {P.}~\bibnamefont {Perdikaris}},\
  and\ \bibinfo {author} {\bibfnamefont {G.~E.}\ \bibnamefont {Karniadakis}},\
  }\bibfield  {title} {\bibinfo {title} {Physics informed deep learning (part
  i): Data-driven solutions of nonlinear partial differential equations},\
  }\href@noop {} {\bibfield  {journal} {\bibinfo  {journal} {arXiv preprint
  arXiv:1711.10561}\ } (\bibinfo {year} {2017}{\natexlab{a}})}\BibitemShut
  {NoStop}%
\bibitem [{\citenamefont {Yair}\ \emph {et~al.}(2017)\citenamefont {Yair},
  \citenamefont {Talmon}, \citenamefont {Coifman},\ and\ \citenamefont
  {Kevrekidis}}]{Yair2017pnas}%
  \BibitemOpen
  \bibfield  {author} {\bibinfo {author} {\bibfnamefont {O.}~\bibnamefont
  {Yair}}, \bibinfo {author} {\bibfnamefont {R.}~\bibnamefont {Talmon}},
  \bibinfo {author} {\bibfnamefont {R.~R.}\ \bibnamefont {Coifman}},\ and\
  \bibinfo {author} {\bibfnamefont {I.~G.}\ \bibnamefont {Kevrekidis}},\
  }\bibfield  {title} {\bibinfo {title} {Reconstruction of normal forms by
  learning informed observation geometries from data},\ }\href@noop {}
  {\bibfield  {journal} {\bibinfo  {journal} {Proceedings of the National
  Academy of Sciences}\ ,\ \bibinfo {pages} {201620045}} (\bibinfo {year}
  {2017})}\BibitemShut {NoStop}%
\bibitem [{\citenamefont {Klus}\ \emph {et~al.}(2018)\citenamefont {Klus},
  \citenamefont {N{\"u}ske}, \citenamefont {Koltai}, \citenamefont {Wu},
  \citenamefont {Kevrekidis}, \citenamefont {Sch{\"u}tte},\ and\ \citenamefont
  {No{\'e}}}]{klus2017data}%
  \BibitemOpen
  \bibfield  {author} {\bibinfo {author} {\bibfnamefont {S.}~\bibnamefont
  {Klus}}, \bibinfo {author} {\bibfnamefont {F.}~\bibnamefont {N{\"u}ske}},
  \bibinfo {author} {\bibfnamefont {P.}~\bibnamefont {Koltai}}, \bibinfo
  {author} {\bibfnamefont {H.}~\bibnamefont {Wu}}, \bibinfo {author}
  {\bibfnamefont {I.}~\bibnamefont {Kevrekidis}}, \bibinfo {author}
  {\bibfnamefont {C.}~\bibnamefont {Sch{\"u}tte}},\ and\ \bibinfo {author}
  {\bibfnamefont {F.}~\bibnamefont {No{\'e}}},\ }\bibfield  {title} {\bibinfo
  {title} {Data-driven model reduction and transfer operator approximation},\
  }\href@noop {} {\bibfield  {journal} {\bibinfo  {journal} {Journal of
  Nonlinear Science}\ } (\bibinfo {year} {2018})}\BibitemShut {NoStop}%
\bibitem [{\citenamefont {Wehmeyer}\ and\ \citenamefont
  {No{\'e}}(2018)}]{Wehmeyer2018jcp}%
  \BibitemOpen
  \bibfield  {author} {\bibinfo {author} {\bibfnamefont {C.}~\bibnamefont
  {Wehmeyer}}\ and\ \bibinfo {author} {\bibfnamefont {F.}~\bibnamefont
  {No{\'e}}},\ }\bibfield  {title} {\bibinfo {title} {Time-lagged autoencoders:
  Deep learning of slow collective variables for molecular kinetics},\
  }\href@noop {} {\bibfield  {journal} {\bibinfo  {journal} {The Journal of
  Chemical Physics}\ }\textbf {\bibinfo {volume} {148}},\ \bibinfo {pages} {1}
  (\bibinfo {year} {2018})}\BibitemShut {NoStop}%
\bibitem [{\citenamefont {Mardt}\ \emph {et~al.}(2018)\citenamefont {Mardt},
  \citenamefont {Pasquali}, \citenamefont {Wu},\ and\ \citenamefont
  {No{\'e}}}]{Mardt2018natcomm}%
  \BibitemOpen
  \bibfield  {author} {\bibinfo {author} {\bibfnamefont {A.}~\bibnamefont
  {Mardt}}, \bibinfo {author} {\bibfnamefont {L.}~\bibnamefont {Pasquali}},
  \bibinfo {author} {\bibfnamefont {H.}~\bibnamefont {Wu}},\ and\ \bibinfo
  {author} {\bibfnamefont {F.}~\bibnamefont {No{\'e}}},\ }\bibfield  {title}
  {\bibinfo {title} {{VAMP}nets: Deep learning of molecular kinetics},\
  }\href@noop {} {\bibfield  {journal} {\bibinfo  {journal} {Nature
  Communications}\ }\textbf {\bibinfo {volume} {9}} (\bibinfo {year}
  {2018})}\BibitemShut {NoStop}%
\bibitem [{\citenamefont {Duraisamy}\ \emph {et~al.}(2019)\citenamefont
  {Duraisamy}, \citenamefont {Iaccarino},\ and\ \citenamefont
  {Xiao}}]{Duraisamy2019arfm}%
  \BibitemOpen
  \bibfield  {author} {\bibinfo {author} {\bibfnamefont {K.}~\bibnamefont
  {Duraisamy}}, \bibinfo {author} {\bibfnamefont {G.}~\bibnamefont
  {Iaccarino}},\ and\ \bibinfo {author} {\bibfnamefont {H.}~\bibnamefont
  {Xiao}},\ }\bibfield  {title} {\bibinfo {title} {Turbulence modeling in the
  age of data},\ }\href@noop {} {\bibfield  {journal} {\bibinfo  {journal}
  {Annual Reviews of Fluid Mechanics}\ }\textbf {\bibinfo {volume} {51}},\
  \bibinfo {pages} {357} (\bibinfo {year} {2019})}\BibitemShut {NoStop}%
\bibitem [{\citenamefont {Pathak}\ \emph {et~al.}(2018)\citenamefont {Pathak},
  \citenamefont {Hunt}, \citenamefont {Girvan}, \citenamefont {Lu},\ and\
  \citenamefont {Ott}}]{pathak2018model}%
  \BibitemOpen
  \bibfield  {author} {\bibinfo {author} {\bibfnamefont {J.}~\bibnamefont
  {Pathak}}, \bibinfo {author} {\bibfnamefont {B.}~\bibnamefont {Hunt}},
  \bibinfo {author} {\bibfnamefont {M.}~\bibnamefont {Girvan}}, \bibinfo
  {author} {\bibfnamefont {Z.}~\bibnamefont {Lu}},\ and\ \bibinfo {author}
  {\bibfnamefont {E.}~\bibnamefont {Ott}},\ }\bibfield  {title} {\bibinfo
  {title} {Model-free prediction of large spatiotemporally chaotic systems from
  data: a reservoir computing approach},\ }\href@noop {} {\bibfield  {journal}
  {\bibinfo  {journal} {Physical review letters}\ }\textbf {\bibinfo {volume}
  {120}},\ \bibinfo {pages} {024102} (\bibinfo {year} {2018})}\BibitemShut
  {NoStop}%
\bibitem [{\citenamefont {No{\'e}}\ \emph {et~al.}(2019)\citenamefont
  {No{\'e}}, \citenamefont {Olsson}, \citenamefont {K{\"o}hler},\ and\
  \citenamefont {Wu}}]{Noe2019science}%
  \BibitemOpen
  \bibfield  {author} {\bibinfo {author} {\bibfnamefont {F.}~\bibnamefont
  {No{\'e}}}, \bibinfo {author} {\bibfnamefont {S.}~\bibnamefont {Olsson}},
  \bibinfo {author} {\bibfnamefont {J.}~\bibnamefont {K{\"o}hler}},\ and\
  \bibinfo {author} {\bibfnamefont {H.}~\bibnamefont {Wu}},\ }\bibfield
  {title} {\bibinfo {title} {Boltzmann generators: Sampling equilibrium states
  of many-body systems with deep learning},\ }\href@noop {} {\bibfield
  {journal} {\bibinfo  {journal} {Science}\ }\textbf {\bibinfo {volume}
  {365}},\ \bibinfo {pages} {eaaw1147} (\bibinfo {year} {2019})}\BibitemShut
  {NoStop}%
\bibitem [{\citenamefont {Bar-Sinai}\ \emph {et~al.}(2019)\citenamefont
  {Bar-Sinai}, \citenamefont {Hoyer}, \citenamefont {Hickey},\ and\
  \citenamefont {Brenner}}]{bar2019learning}%
  \BibitemOpen
  \bibfield  {author} {\bibinfo {author} {\bibfnamefont {Y.}~\bibnamefont
  {Bar-Sinai}}, \bibinfo {author} {\bibfnamefont {S.}~\bibnamefont {Hoyer}},
  \bibinfo {author} {\bibfnamefont {J.}~\bibnamefont {Hickey}},\ and\ \bibinfo
  {author} {\bibfnamefont {M.~P.}\ \bibnamefont {Brenner}},\ }\bibfield
  {title} {\bibinfo {title} {Learning data-driven discretizations for partial
  differential equations},\ }\href@noop {} {\bibfield  {journal} {\bibinfo
  {journal} {Proceedings of the National Academy of Sciences}\ }\textbf
  {\bibinfo {volume} {116}},\ \bibinfo {pages} {15344} (\bibinfo {year}
  {2019})}\BibitemShut {NoStop}%
\bibitem [{\citenamefont {Kochkov}\ \emph {et~al.}(2021)\citenamefont
  {Kochkov}, \citenamefont {Smith}, \citenamefont {Alieva}, \citenamefont
  {Wang}, \citenamefont {Brenner},\ and\ \citenamefont
  {Hoyer}}]{kochkov2021machine}%
  \BibitemOpen
  \bibfield  {author} {\bibinfo {author} {\bibfnamefont {D.}~\bibnamefont
  {Kochkov}}, \bibinfo {author} {\bibfnamefont {J.~A.}\ \bibnamefont {Smith}},
  \bibinfo {author} {\bibfnamefont {A.}~\bibnamefont {Alieva}}, \bibinfo
  {author} {\bibfnamefont {Q.}~\bibnamefont {Wang}}, \bibinfo {author}
  {\bibfnamefont {M.~P.}\ \bibnamefont {Brenner}},\ and\ \bibinfo {author}
  {\bibfnamefont {S.}~\bibnamefont {Hoyer}},\ }\bibfield  {title} {\bibinfo
  {title} {Machine learning--accelerated computational fluid dynamics},\
  }\href@noop {} {\bibfield  {journal} {\bibinfo  {journal} {Proceedings of the
  National Academy of Sciences}\ }\textbf {\bibinfo {volume} {118}} (\bibinfo
  {year} {2021})}\BibitemShut {NoStop}%
\bibitem [{\citenamefont {Raissi}\ \emph
  {et~al.}(2017{\natexlab{b}})\citenamefont {Raissi}, \citenamefont
  {Perdikaris},\ and\ \citenamefont {Karniadakis}}]{raissi2017physics2}%
  \BibitemOpen
  \bibfield  {author} {\bibinfo {author} {\bibfnamefont {M.}~\bibnamefont
  {Raissi}}, \bibinfo {author} {\bibfnamefont {P.}~\bibnamefont {Perdikaris}},\
  and\ \bibinfo {author} {\bibfnamefont {G.~E.}\ \bibnamefont {Karniadakis}},\
  }\bibfield  {title} {\bibinfo {title} {Physics informed deep learning (part
  ii): Data-driven discovery of nonlinear partial differential equations},\
  }\href@noop {} {\bibfield  {journal} {\bibinfo  {journal} {arXiv preprint
  arXiv:1711.10566}\ } (\bibinfo {year} {2017}{\natexlab{b}})}\BibitemShut
  {NoStop}%
\bibitem [{\citenamefont {Battaglia}\ \emph {et~al.}(2018)\citenamefont
  {Battaglia}, \citenamefont {Hamrick}, \citenamefont {Bapst}, \citenamefont
  {Sanchez-Gonzalez}, \citenamefont {Zambaldi}, \citenamefont {Malinowski},
  \citenamefont {Tacchetti}, \citenamefont {Raposo}, \citenamefont {Santoro},
  \citenamefont {Faulkner} \emph {et~al.}}]{battaglia2018relational}%
  \BibitemOpen
  \bibfield  {author} {\bibinfo {author} {\bibfnamefont {P.~W.}\ \bibnamefont
  {Battaglia}}, \bibinfo {author} {\bibfnamefont {J.~B.}\ \bibnamefont
  {Hamrick}}, \bibinfo {author} {\bibfnamefont {V.}~\bibnamefont {Bapst}},
  \bibinfo {author} {\bibfnamefont {A.}~\bibnamefont {Sanchez-Gonzalez}},
  \bibinfo {author} {\bibfnamefont {V.}~\bibnamefont {Zambaldi}}, \bibinfo
  {author} {\bibfnamefont {M.}~\bibnamefont {Malinowski}}, \bibinfo {author}
  {\bibfnamefont {A.}~\bibnamefont {Tacchetti}}, \bibinfo {author}
  {\bibfnamefont {D.}~\bibnamefont {Raposo}}, \bibinfo {author} {\bibfnamefont
  {A.}~\bibnamefont {Santoro}}, \bibinfo {author} {\bibfnamefont
  {R.}~\bibnamefont {Faulkner}}, \emph {et~al.},\ }\bibfield  {title} {\bibinfo
  {title} {Relational inductive biases, deep learning, and graph networks},\
  }\href@noop {} {\bibfield  {journal} {\bibinfo  {journal} {arXiv preprint
  arXiv:1806.01261}\ } (\bibinfo {year} {2018})}\BibitemShut {NoStop}%
\bibitem [{\citenamefont {Cranmer}\ \emph {et~al.}(2019)\citenamefont
  {Cranmer}, \citenamefont {Xu}, \citenamefont {Battaglia},\ and\ \citenamefont
  {Ho}}]{cranmer2019learning}%
  \BibitemOpen
  \bibfield  {author} {\bibinfo {author} {\bibfnamefont {M.~D.}\ \bibnamefont
  {Cranmer}}, \bibinfo {author} {\bibfnamefont {R.}~\bibnamefont {Xu}},
  \bibinfo {author} {\bibfnamefont {P.}~\bibnamefont {Battaglia}},\ and\
  \bibinfo {author} {\bibfnamefont {S.}~\bibnamefont {Ho}},\ }\bibfield
  {title} {\bibinfo {title} {Learning symbolic physics with graph networks},\
  }\href@noop {} {\bibfield  {journal} {\bibinfo  {journal} {arXiv preprint
  arXiv:1909.05862}\ } (\bibinfo {year} {2019})}\BibitemShut {NoStop}%
\bibitem [{\citenamefont {Mohebujjaman}\ \emph {et~al.}(2019)\citenamefont
  {Mohebujjaman}, \citenamefont {Rebholz},\ and\ \citenamefont
  {Iliescu}}]{mohebujjaman2019physically}%
  \BibitemOpen
  \bibfield  {author} {\bibinfo {author} {\bibfnamefont {M.}~\bibnamefont
  {Mohebujjaman}}, \bibinfo {author} {\bibfnamefont {L.~G.}\ \bibnamefont
  {Rebholz}},\ and\ \bibinfo {author} {\bibfnamefont {T.}~\bibnamefont
  {Iliescu}},\ }\bibfield  {title} {\bibinfo {title} {Physically constrained
  data-driven correction for reduced-order modeling of fluid flows},\
  }\href@noop {} {\bibfield  {journal} {\bibinfo  {journal} {International
  Journal for Numerical Methods in Fluids}\ }\textbf {\bibinfo {volume} {89}},\
  \bibinfo {pages} {103} (\bibinfo {year} {2019})}\BibitemShut {NoStop}%
\bibitem [{\citenamefont {Cranmer}\ \emph {et~al.}(2020)\citenamefont
  {Cranmer}, \citenamefont {Greydanus}, \citenamefont {Hoyer}, \citenamefont
  {Battaglia}, \citenamefont {Spergel},\ and\ \citenamefont
  {Ho}}]{cranmer2020lagrangian}%
  \BibitemOpen
  \bibfield  {author} {\bibinfo {author} {\bibfnamefont {M.}~\bibnamefont
  {Cranmer}}, \bibinfo {author} {\bibfnamefont {S.}~\bibnamefont {Greydanus}},
  \bibinfo {author} {\bibfnamefont {S.}~\bibnamefont {Hoyer}}, \bibinfo
  {author} {\bibfnamefont {P.}~\bibnamefont {Battaglia}}, \bibinfo {author}
  {\bibfnamefont {D.}~\bibnamefont {Spergel}},\ and\ \bibinfo {author}
  {\bibfnamefont {S.}~\bibnamefont {Ho}},\ }\bibfield  {title} {\bibinfo
  {title} {Lagrangian neural networks},\ }\href@noop {} {\bibfield  {journal}
  {\bibinfo  {journal} {arXiv preprint arXiv:2003.04630}\ } (\bibinfo {year}
  {2020})}\BibitemShut {NoStop}%
\bibitem [{\citenamefont {Loiseau}\ \emph {et~al.}(2018)\citenamefont
  {Loiseau}, \citenamefont {Noack},\ and\ \citenamefont
  {Brunton}}]{loiseau2018sparse}%
  \BibitemOpen
  \bibfield  {author} {\bibinfo {author} {\bibfnamefont {J.-C.}\ \bibnamefont
  {Loiseau}}, \bibinfo {author} {\bibfnamefont {B.~R.}\ \bibnamefont {Noack}},\
  and\ \bibinfo {author} {\bibfnamefont {S.~L.}\ \bibnamefont {Brunton}},\
  }\bibfield  {title} {\bibinfo {title} {Sparse reduced-order modeling:
  sensor-based dynamics to full-state estimation},\ }\href@noop {} {\bibfield
  {journal} {\bibinfo  {journal} {Journal of Fluid Mechanics}\ }\textbf
  {\bibinfo {volume} {844}},\ \bibinfo {pages} {459} (\bibinfo {year}
  {2018})}\BibitemShut {NoStop}%
\bibitem [{\citenamefont {Pouquet}\ \emph {et~al.}(2019)\citenamefont
  {Pouquet}, \citenamefont {Rosenberg}, \citenamefont {Stawarz},\ and\
  \citenamefont {Marino}}]{pouquet2019helicity}%
  \BibitemOpen
  \bibfield  {author} {\bibinfo {author} {\bibfnamefont {A.}~\bibnamefont
  {Pouquet}}, \bibinfo {author} {\bibfnamefont {D.}~\bibnamefont {Rosenberg}},
  \bibinfo {author} {\bibfnamefont {J.~E.}\ \bibnamefont {Stawarz}},\ and\
  \bibinfo {author} {\bibfnamefont {R.}~\bibnamefont {Marino}},\ }\bibfield
  {title} {\bibinfo {title} {Helicity dynamics, inverse, and bidirectional
  cascades in fluid and magnetohydrodynamic turbulence: a brief review},\
  }\href@noop {} {\bibfield  {journal} {\bibinfo  {journal} {Earth and Space
  Science}\ }\textbf {\bibinfo {volume} {6}},\ \bibinfo {pages} {351} (\bibinfo
  {year} {2019})}\BibitemShut {NoStop}%
\bibitem [{\citenamefont {Rowley}\ \emph {et~al.}(2004)\citenamefont {Rowley},
  \citenamefont {Colonius},\ and\ \citenamefont {Murray}}]{rowley2004model}%
  \BibitemOpen
  \bibfield  {author} {\bibinfo {author} {\bibfnamefont {C.~W.}\ \bibnamefont
  {Rowley}}, \bibinfo {author} {\bibfnamefont {T.}~\bibnamefont {Colonius}},\
  and\ \bibinfo {author} {\bibfnamefont {R.~M.}\ \bibnamefont {Murray}},\
  }\bibfield  {title} {\bibinfo {title} {Model reduction for compressible flows
  using {POD} and {G}alerkin projection},\ }\href@noop {} {\bibfield  {journal}
  {\bibinfo  {journal} {Physica D}\ }\textbf {\bibinfo {volume} {189}},\
  \bibinfo {pages} {115} (\bibinfo {year} {2004})}\BibitemShut {NoStop}%
\bibitem [{\citenamefont {Noack}\ \emph {et~al.}(2011)\citenamefont {Noack},
  \citenamefont {Schlegel}, \citenamefont {Morzynski},\ and\ \citenamefont
  {Tadmor}}]{noack2011galerkin}%
  \BibitemOpen
  \bibfield  {author} {\bibinfo {author} {\bibfnamefont {B.~R.}\ \bibnamefont
  {Noack}}, \bibinfo {author} {\bibfnamefont {M.}~\bibnamefont {Schlegel}},
  \bibinfo {author} {\bibfnamefont {M.}~\bibnamefont {Morzynski}},\ and\
  \bibinfo {author} {\bibfnamefont {G.}~\bibnamefont {Tadmor}},\ }\href@noop {}
  {\emph {\bibinfo {title} {Galerkin method for nonlinear dynamics}}}\
  (\bibinfo  {publisher} {Springer},\ \bibinfo {year} {2011})\BibitemShut
  {NoStop}%
\bibitem [{\citenamefont {Balajewicz}\ \emph {et~al.}(2013)\citenamefont
  {Balajewicz}, \citenamefont {Dowell},\ and\ \citenamefont
  {Noack}}]{balajewicz2013low}%
  \BibitemOpen
  \bibfield  {author} {\bibinfo {author} {\bibfnamefont {M.~J.}\ \bibnamefont
  {Balajewicz}}, \bibinfo {author} {\bibfnamefont {E.~H.}\ \bibnamefont
  {Dowell}},\ and\ \bibinfo {author} {\bibfnamefont {B.~R.}\ \bibnamefont
  {Noack}},\ }\bibfield  {title} {\bibinfo {title} {Low-dimensional modelling
  of high-{R}eynolds-number shear flows incorporating constraints from the
  {N}avier--{S}tokes equation},\ }\href@noop {} {\bibfield  {journal} {\bibinfo
   {journal} {Journal of Fluid Mechanics}\ }\textbf {\bibinfo {volume} {729}},\
  \bibinfo {pages} {285} (\bibinfo {year} {2013})}\BibitemShut {NoStop}%
\bibitem [{\citenamefont {Schlegel}\ and\ \citenamefont
  {Noack}(2015)}]{Schlegel2015jfm}%
  \BibitemOpen
  \bibfield  {author} {\bibinfo {author} {\bibfnamefont {M.}~\bibnamefont
  {Schlegel}}\ and\ \bibinfo {author} {\bibfnamefont {B.~R.}\ \bibnamefont
  {Noack}},\ }\bibfield  {title} {\bibinfo {title} {On long-term boundedness of
  {G}alerkin models},\ }\href@noop {} {\bibfield  {journal} {\bibinfo
  {journal} {Journal of Fluid Mechanics}\ }\textbf {\bibinfo {volume} {765}},\
  \bibinfo {pages} {325} (\bibinfo {year} {2015})}\BibitemShut {NoStop}%
\bibitem [{\citenamefont {Jarboe}\ \emph {et~al.}(2006)\citenamefont {Jarboe},
  \citenamefont {Hamp}, \citenamefont {Marklin}, \citenamefont {Nelson},
  \citenamefont {O’Neill}, \citenamefont {Redd}, \citenamefont {Sieck},
  \citenamefont {Smith},\ and\ \citenamefont {Wrobel}}]{jarboe2006spheromak}%
  \BibitemOpen
  \bibfield  {author} {\bibinfo {author} {\bibfnamefont {T.}~\bibnamefont
  {Jarboe}}, \bibinfo {author} {\bibfnamefont {W.}~\bibnamefont {Hamp}},
  \bibinfo {author} {\bibfnamefont {G.}~\bibnamefont {Marklin}}, \bibinfo
  {author} {\bibfnamefont {B.}~\bibnamefont {Nelson}}, \bibinfo {author}
  {\bibfnamefont {R.}~\bibnamefont {O’Neill}}, \bibinfo {author}
  {\bibfnamefont {A.}~\bibnamefont {Redd}}, \bibinfo {author} {\bibfnamefont
  {P.}~\bibnamefont {Sieck}}, \bibinfo {author} {\bibfnamefont
  {R.}~\bibnamefont {Smith}},\ and\ \bibinfo {author} {\bibfnamefont
  {J.}~\bibnamefont {Wrobel}},\ }\bibfield  {title} {\bibinfo {title}
  {Spheromak formation by steady inductive helicity injection},\ }\href@noop {}
  {\bibfield  {journal} {\bibinfo  {journal} {Physical review letters}\
  }\textbf {\bibinfo {volume} {97}},\ \bibinfo {pages} {115003} (\bibinfo
  {year} {2006})}\BibitemShut {NoStop}%
\bibitem [{\citenamefont {Schnack}\ \emph {et~al.}(2006)\citenamefont
  {Schnack}, \citenamefont {Barnes}, \citenamefont {Brennan}, \citenamefont
  {Hegna}, \citenamefont {Held}, \citenamefont {Kim}, \citenamefont {Kruger},
  \citenamefont {Pankin},\ and\ \citenamefont {Sovinec}}]{Schnack2006}%
  \BibitemOpen
  \bibfield  {author} {\bibinfo {author} {\bibfnamefont {D.~D.}\ \bibnamefont
  {Schnack}}, \bibinfo {author} {\bibfnamefont {D.~C.}\ \bibnamefont {Barnes}},
  \bibinfo {author} {\bibfnamefont {D.~P.}\ \bibnamefont {Brennan}}, \bibinfo
  {author} {\bibfnamefont {C.~C.}\ \bibnamefont {Hegna}}, \bibinfo {author}
  {\bibfnamefont {E.}~\bibnamefont {Held}}, \bibinfo {author} {\bibfnamefont
  {C.~C.}\ \bibnamefont {Kim}}, \bibinfo {author} {\bibfnamefont {S.~E.}\
  \bibnamefont {Kruger}}, \bibinfo {author} {\bibfnamefont {A.~Y.}\
  \bibnamefont {Pankin}},\ and\ \bibinfo {author} {\bibfnamefont {C.~R.}\
  \bibnamefont {Sovinec}},\ }\bibfield  {title} {\bibinfo {title}
  {Computational modeling of fully ionized magnetized plasmas using the fluid
  approximation},\ }\href@noop {} {\bibfield  {journal} {\bibinfo  {journal}
  {Physics of Plasmas}\ }\textbf {\bibinfo {volume} {13}},\ \bibinfo {pages}
  {058103} (\bibinfo {year} {2006})}\BibitemShut {NoStop}%
\bibitem [{\citenamefont {Ma}\ and\ \citenamefont
  {Bhattacharjee}(2001)}]{ma2001hall}%
  \BibitemOpen
  \bibfield  {author} {\bibinfo {author} {\bibfnamefont {Z.}~\bibnamefont
  {Ma}}\ and\ \bibinfo {author} {\bibfnamefont {A.}~\bibnamefont
  {Bhattacharjee}},\ }\bibfield  {title} {\bibinfo {title} {Hall
  magnetohydrodynamic reconnection: The geospace environment modeling
  challenge},\ }\href@noop {} {\bibfield  {journal} {\bibinfo  {journal}
  {Journal of Geophysical Research: Space Physics}\ }\textbf {\bibinfo {volume}
  {106}},\ \bibinfo {pages} {3773} (\bibinfo {year} {2001})}\BibitemShut
  {NoStop}%
\bibitem [{\citenamefont {Krishan}\ and\ \citenamefont
  {Mahajan}(2004)}]{krishan2004magnetic}%
  \BibitemOpen
  \bibfield  {author} {\bibinfo {author} {\bibfnamefont {V.}~\bibnamefont
  {Krishan}}\ and\ \bibinfo {author} {\bibfnamefont {S.}~\bibnamefont
  {Mahajan}},\ }\bibfield  {title} {\bibinfo {title} {Magnetic fluctuations and
  {H}all magnetohydrodynamic turbulence in the solar wind},\ }\href@noop {}
  {\bibfield  {journal} {\bibinfo  {journal} {Journal of Geophysical Research:
  Space Physics}\ }\textbf {\bibinfo {volume} {109}} (\bibinfo {year}
  {2004})}\BibitemShut {NoStop}%
\bibitem [{\citenamefont {Ebrahimi}\ \emph {et~al.}(2011)\citenamefont
  {Ebrahimi}, \citenamefont {Lefebvre}, \citenamefont {Forest},\ and\
  \citenamefont {Bhattacharjee}}]{Ebrahimi2011}%
  \BibitemOpen
  \bibfield  {author} {\bibinfo {author} {\bibfnamefont {F.}~\bibnamefont
  {Ebrahimi}}, \bibinfo {author} {\bibfnamefont {B.}~\bibnamefont {Lefebvre}},
  \bibinfo {author} {\bibfnamefont {C.~B.}\ \bibnamefont {Forest}},\ and\
  \bibinfo {author} {\bibfnamefont {A.}~\bibnamefont {Bhattacharjee}},\
  }\bibfield  {title} {\bibinfo {title} {Global {H}all-{MHD} simulations of
  magnetorotational instability in a plasma {C}ouette flow experiment},\
  }\href@noop {} {\bibfield  {journal} {\bibinfo  {journal} {Physics of
  Plasmas}\ }\textbf {\bibinfo {volume} {18}},\ \bibinfo {pages} {062904}
  (\bibinfo {year} {2011})}\BibitemShut {NoStop}%
\bibitem [{\citenamefont {Ferraro}(2012)}]{Ferraro2012}%
  \BibitemOpen
  \bibfield  {author} {\bibinfo {author} {\bibfnamefont {N.~M.}\ \bibnamefont
  {Ferraro}},\ }\bibfield  {title} {\bibinfo {title} {Calculations of two-fluid
  linear response to non-axisymmetric fields in tokamaks},\ }\href@noop {}
  {\bibfield  {journal} {\bibinfo  {journal} {Physics of Plasmas}\ }\textbf
  {\bibinfo {volume} {19}},\ \bibinfo {pages} {056105} (\bibinfo {year}
  {2012})}\BibitemShut {NoStop}%
\bibitem [{\citenamefont {Kaptanoglu}\ \emph
  {et~al.}(2020{\natexlab{b}})\citenamefont {Kaptanoglu}, \citenamefont
  {Benedett}, \citenamefont {Morgan}, \citenamefont {Hansen},\ and\
  \citenamefont {Jarboe}}]{kaptanoglu2020two}%
  \BibitemOpen
  \bibfield  {author} {\bibinfo {author} {\bibfnamefont {A.~A.}\ \bibnamefont
  {Kaptanoglu}}, \bibinfo {author} {\bibfnamefont {T.~E.}\ \bibnamefont
  {Benedett}}, \bibinfo {author} {\bibfnamefont {K.~D.}\ \bibnamefont
  {Morgan}}, \bibinfo {author} {\bibfnamefont {C.~J.}\ \bibnamefont {Hansen}},\
  and\ \bibinfo {author} {\bibfnamefont {T.~R.}\ \bibnamefont {Jarboe}},\
  }\bibfield  {title} {\bibinfo {title} {Two-temperature effects in {Hall-MHD}
  simulations of the {HIT-SI} experiment},\ }\href@noop {} {\bibfield
  {journal} {\bibinfo  {journal} {Physics of Plasmas}\ }\textbf {\bibinfo
  {volume} {27}},\ \bibinfo {pages} {072505} (\bibinfo {year}
  {2020}{\natexlab{b}})}\BibitemShut {NoStop}%
\bibitem [{\citenamefont {Dudok~de Wit}\ \emph {et~al.}(1994)\citenamefont
  {Dudok~de Wit}, \citenamefont {Pecquet}, \citenamefont {Vallet},\ and\
  \citenamefont {Lima}}]{dudok1994biorthogonal}%
  \BibitemOpen
  \bibfield  {author} {\bibinfo {author} {\bibfnamefont {T.}~\bibnamefont
  {Dudok~de Wit}}, \bibinfo {author} {\bibfnamefont {A.-L.}\ \bibnamefont
  {Pecquet}}, \bibinfo {author} {\bibfnamefont {J.-C.}\ \bibnamefont
  {Vallet}},\ and\ \bibinfo {author} {\bibfnamefont {R.}~\bibnamefont {Lima}},\
  }\bibfield  {title} {\bibinfo {title} {The biorthogonal decomposition as a
  tool for investigating fluctuations in plasmas},\ }\href@noop {} {\bibfield
  {journal} {\bibinfo  {journal} {Physics of Plasmas}\ }\textbf {\bibinfo
  {volume} {1}},\ \bibinfo {pages} {3288} (\bibinfo {year} {1994})}\BibitemShut
  {NoStop}%
\bibitem [{\citenamefont {Galperti}\ \emph {et~al.}(2014)\citenamefont
  {Galperti}, \citenamefont {Marchetto}, \citenamefont {Alessi}, \citenamefont
  {Minelli}, \citenamefont {Mosconi}, \citenamefont {Belli}, \citenamefont
  {Boncagni}, \citenamefont {Botrugno}, \citenamefont {Buratti}, \citenamefont
  {Esposito} \emph {et~al.}}]{galperti2014development}%
  \BibitemOpen
  \bibfield  {author} {\bibinfo {author} {\bibfnamefont {C.}~\bibnamefont
  {Galperti}}, \bibinfo {author} {\bibfnamefont {C.}~\bibnamefont {Marchetto}},
  \bibinfo {author} {\bibfnamefont {E.}~\bibnamefont {Alessi}}, \bibinfo
  {author} {\bibfnamefont {D.}~\bibnamefont {Minelli}}, \bibinfo {author}
  {\bibfnamefont {M.}~\bibnamefont {Mosconi}}, \bibinfo {author} {\bibfnamefont
  {F.}~\bibnamefont {Belli}}, \bibinfo {author} {\bibfnamefont
  {L.}~\bibnamefont {Boncagni}}, \bibinfo {author} {\bibfnamefont
  {A.}~\bibnamefont {Botrugno}}, \bibinfo {author} {\bibfnamefont
  {P.}~\bibnamefont {Buratti}}, \bibinfo {author} {\bibfnamefont
  {B.}~\bibnamefont {Esposito}}, \emph {et~al.},\ }\bibfield  {title} {\bibinfo
  {title} {Development of real-time {MHD} markers based on biorthogonal
  decomposition of signals from {M}irnov coils},\ }\href@noop {} {\bibfield
  {journal} {\bibinfo  {journal} {Plasma Physics and Controlled Fusion}\
  }\textbf {\bibinfo {volume} {56}},\ \bibinfo {pages} {114012} (\bibinfo
  {year} {2014})}\BibitemShut {NoStop}%
\bibitem [{\citenamefont {Van~Milligen}\ \emph {et~al.}(2014)\citenamefont
  {Van~Milligen}, \citenamefont {S{\'a}nchez}, \citenamefont {Alonso},
  \citenamefont {Pedrosa}, \citenamefont {Hidalgo}, \citenamefont
  {De~Aguilera},\ and\ \citenamefont {Fraguas}}]{van2014use}%
  \BibitemOpen
  \bibfield  {author} {\bibinfo {author} {\bibfnamefont {B.~P.}\ \bibnamefont
  {Van~Milligen}}, \bibinfo {author} {\bibfnamefont {E.}~\bibnamefont
  {S{\'a}nchez}}, \bibinfo {author} {\bibfnamefont {A.}~\bibnamefont {Alonso}},
  \bibinfo {author} {\bibfnamefont {M.}~\bibnamefont {Pedrosa}}, \bibinfo
  {author} {\bibfnamefont {C.}~\bibnamefont {Hidalgo}}, \bibinfo {author}
  {\bibfnamefont {A.~M.}\ \bibnamefont {De~Aguilera}},\ and\ \bibinfo {author}
  {\bibfnamefont {A.~L.}\ \bibnamefont {Fraguas}},\ }\bibfield  {title}
  {\bibinfo {title} {The use of the biorthogonal decomposition for the
  identification of zonal flows at {TJ-II}},\ }\href@noop {} {\bibfield
  {journal} {\bibinfo  {journal} {Plasma Physics and Controlled Fusion}\
  }\textbf {\bibinfo {volume} {57}},\ \bibinfo {pages} {025005} (\bibinfo
  {year} {2014})}\BibitemShut {NoStop}%
\bibitem [{\citenamefont {Hansen}\ \emph {et~al.}(2015)\citenamefont {Hansen},
  \citenamefont {Victor}, \citenamefont {Morgan}, \citenamefont {Jarboe},
  \citenamefont {Hossack}, \citenamefont {Marklin}, \citenamefont {Nelson},\
  and\ \citenamefont {Sutherland}}]{hansen2015numerical}%
  \BibitemOpen
  \bibfield  {author} {\bibinfo {author} {\bibfnamefont {C.}~\bibnamefont
  {Hansen}}, \bibinfo {author} {\bibfnamefont {B.}~\bibnamefont {Victor}},
  \bibinfo {author} {\bibfnamefont {K.}~\bibnamefont {Morgan}}, \bibinfo
  {author} {\bibfnamefont {T.}~\bibnamefont {Jarboe}}, \bibinfo {author}
  {\bibfnamefont {A.}~\bibnamefont {Hossack}}, \bibinfo {author} {\bibfnamefont
  {G.}~\bibnamefont {Marklin}}, \bibinfo {author} {\bibfnamefont
  {B.}~\bibnamefont {Nelson}},\ and\ \bibinfo {author} {\bibfnamefont
  {D.}~\bibnamefont {Sutherland}},\ }\bibfield  {title} {\bibinfo {title}
  {Numerical studies and metric development for validation of
  magnetohydrodynamic models on the {HIT-SI} experiment},\ }\href@noop {}
  {\bibfield  {journal} {\bibinfo  {journal} {Physics of Plasmas}\ }\textbf
  {\bibinfo {volume} {22}},\ \bibinfo {pages} {056105} (\bibinfo {year}
  {2015})}\BibitemShut {NoStop}%
\bibitem [{\citenamefont {Golub}\ \emph {et~al.}(1996)\citenamefont {Golub}
  \emph {et~al.}}]{golub1996cf}%
  \BibitemOpen
  \bibfield  {author} {\bibinfo {author} {\bibfnamefont {G.~H.}\ \bibnamefont
  {Golub}} \emph {et~al.},\ }\bibfield  {title} {\bibinfo {title} {Matrix
  computations},\ }\href@noop {} {\bibfield  {journal} {\bibinfo  {journal}
  {The Johns Hopkins}\ } (\bibinfo {year} {1996})}\BibitemShut {NoStop}%
\bibitem [{\citenamefont {Frieze}\ \emph {et~al.}(2004)\citenamefont {Frieze},
  \citenamefont {Kannan},\ and\ \citenamefont {Vempala}}]{frieze2004fast}%
  \BibitemOpen
  \bibfield  {author} {\bibinfo {author} {\bibfnamefont {A.}~\bibnamefont
  {Frieze}}, \bibinfo {author} {\bibfnamefont {R.}~\bibnamefont {Kannan}},\
  and\ \bibinfo {author} {\bibfnamefont {S.}~\bibnamefont {Vempala}},\
  }\bibfield  {title} {\bibinfo {title} {Fast {M}onte-{C}arlo algorithms for
  finding low-rank approximations},\ }\href@noop {} {\bibfield  {journal}
  {\bibinfo  {journal} {Journal of the ACM (JACM)}\ }\textbf {\bibinfo {volume}
  {51}},\ \bibinfo {pages} {1025} (\bibinfo {year} {2004})}\BibitemShut
  {NoStop}%
\bibitem [{\citenamefont {Liberty}\ \emph {et~al.}(2007)\citenamefont
  {Liberty}, \citenamefont {Woolfe}, \citenamefont {Martinsson}, \citenamefont
  {Rokhlin},\ and\ \citenamefont {Tygert}}]{liberty2007randomized}%
  \BibitemOpen
  \bibfield  {author} {\bibinfo {author} {\bibfnamefont {E.}~\bibnamefont
  {Liberty}}, \bibinfo {author} {\bibfnamefont {F.}~\bibnamefont {Woolfe}},
  \bibinfo {author} {\bibfnamefont {P.-G.}\ \bibnamefont {Martinsson}},
  \bibinfo {author} {\bibfnamefont {V.}~\bibnamefont {Rokhlin}},\ and\ \bibinfo
  {author} {\bibfnamefont {M.}~\bibnamefont {Tygert}},\ }\bibfield  {title}
  {\bibinfo {title} {Randomized algorithms for the low-rank approximation of
  matrices},\ }\href@noop {} {\bibfield  {journal} {\bibinfo  {journal}
  {Proceedings of the National Academy of Sciences}\ }\textbf {\bibinfo
  {volume} {104}},\ \bibinfo {pages} {20167} (\bibinfo {year}
  {2007})}\BibitemShut {NoStop}%
\bibitem [{\citenamefont {Woolfe}\ \emph {et~al.}(2008)\citenamefont {Woolfe},
  \citenamefont {Liberty}, \citenamefont {Rokhlin},\ and\ \citenamefont
  {Tygert}}]{woolfe2008fast}%
  \BibitemOpen
  \bibfield  {author} {\bibinfo {author} {\bibfnamefont {F.}~\bibnamefont
  {Woolfe}}, \bibinfo {author} {\bibfnamefont {E.}~\bibnamefont {Liberty}},
  \bibinfo {author} {\bibfnamefont {V.}~\bibnamefont {Rokhlin}},\ and\ \bibinfo
  {author} {\bibfnamefont {M.}~\bibnamefont {Tygert}},\ }\bibfield  {title}
  {\bibinfo {title} {A fast randomized algorithm for the approximation of
  matrices},\ }\href@noop {} {\bibfield  {journal} {\bibinfo  {journal}
  {Applied and Computational Harmonic Analysis}\ }\textbf {\bibinfo {volume}
  {25}},\ \bibinfo {pages} {335} (\bibinfo {year} {2008})}\BibitemShut
  {NoStop}%
\bibitem [{\citenamefont {Rempfer}\ and\ \citenamefont
  {Fasel}(1994)}]{rempfer1994dynamics}%
  \BibitemOpen
  \bibfield  {author} {\bibinfo {author} {\bibfnamefont {D.}~\bibnamefont
  {Rempfer}}\ and\ \bibinfo {author} {\bibfnamefont {H.~F.}\ \bibnamefont
  {Fasel}},\ }\bibfield  {title} {\bibinfo {title} {Dynamics of
  three-dimensional coherent structures in a flat-plate boundary layer},\
  }\href@noop {} {\bibfield  {journal} {\bibinfo  {journal} {Journal of Fluid
  Mechanics}\ }\textbf {\bibinfo {volume} {275}},\ \bibinfo {pages} {257}
  (\bibinfo {year} {1994})}\BibitemShut {NoStop}%
\bibitem [{\citenamefont {Spitzer}(2006)}]{spitzer2006physics}%
  \BibitemOpen
  \bibfield  {author} {\bibinfo {author} {\bibfnamefont {L.}~\bibnamefont
  {Spitzer}},\ }\href@noop {} {\emph {\bibinfo {title} {Physics of fully
  ionized gases}}}\ (\bibinfo  {publisher} {Courier Corporation},\ \bibinfo
  {year} {2006})\BibitemShut {NoStop}%
\bibitem [{\citenamefont {Braginskii}\ and\ \citenamefont
  {Leontovich}(1965)}]{braginskii1965reviews}%
  \BibitemOpen
  \bibfield  {author} {\bibinfo {author} {\bibfnamefont {S.}~\bibnamefont
  {Braginskii}}\ and\ \bibinfo {author} {\bibfnamefont {M.}~\bibnamefont
  {Leontovich}},\ }\href@noop {} {\bibinfo {title} {Reviews of plasma physics}}
  (\bibinfo {year} {1965})\BibitemShut {NoStop}%
\bibitem [{\citenamefont {Plunian}\ \emph {et~al.}(2013)\citenamefont
  {Plunian}, \citenamefont {Stepanov},\ and\ \citenamefont
  {Frick}}]{plunian2013shell}%
  \BibitemOpen
  \bibfield  {author} {\bibinfo {author} {\bibfnamefont {F.}~\bibnamefont
  {Plunian}}, \bibinfo {author} {\bibfnamefont {R.}~\bibnamefont {Stepanov}},\
  and\ \bibinfo {author} {\bibfnamefont {P.}~\bibnamefont {Frick}},\ }\bibfield
   {title} {\bibinfo {title} {Shell models of magnetohydrodynamic turbulence},\
  }\href@noop {} {\bibfield  {journal} {\bibinfo  {journal} {Physics Reports}\
  }\textbf {\bibinfo {volume} {523}},\ \bibinfo {pages} {1} (\bibinfo {year}
  {2013})}\BibitemShut {NoStop}%
\bibitem [{\citenamefont {Biskamp}(1994)}]{biskamp1994cascade}%
  \BibitemOpen
  \bibfield  {author} {\bibinfo {author} {\bibfnamefont {D.}~\bibnamefont
  {Biskamp}},\ }\bibfield  {title} {\bibinfo {title} {Cascade models for
  magnetohydrodynamic turbulence},\ }\href@noop {} {\bibfield  {journal}
  {\bibinfo  {journal} {Physical Review E}\ }\textbf {\bibinfo {volume} {50}},\
  \bibinfo {pages} {2702} (\bibinfo {year} {1994})}\BibitemShut {NoStop}%
\bibitem [{\citenamefont {Couplet}\ \emph {et~al.}(2003)\citenamefont
  {Couplet}, \citenamefont {Sagaut},\ and\ \citenamefont
  {Basdevant}}]{couplet2003intermodal}%
  \BibitemOpen
  \bibfield  {author} {\bibinfo {author} {\bibfnamefont {M.}~\bibnamefont
  {Couplet}}, \bibinfo {author} {\bibfnamefont {P.}~\bibnamefont {Sagaut}},\
  and\ \bibinfo {author} {\bibfnamefont {C.}~\bibnamefont {Basdevant}},\
  }\bibfield  {title} {\bibinfo {title} {Intermodal energy transfers in a
  proper orthogonal decomposition-{G}alerkin representation of a turbulent
  separated flow},\ }\href@noop {} {\bibfield  {journal} {\bibinfo  {journal}
  {Journal of Fluid Mechanics}\ }\textbf {\bibinfo {volume} {491}},\ \bibinfo
  {pages} {275} (\bibinfo {year} {2003})}\BibitemShut {NoStop}%
\bibitem [{\citenamefont {Kaptanoglu}\ \emph {et~al.}(2021)\citenamefont
  {Kaptanoglu}, \citenamefont {Callaham}, \citenamefont {Hansen}, \citenamefont
  {Aravkin},\ and\ \citenamefont {Brunton}}]{kaptanoglu2021promoting}%
  \BibitemOpen
  \bibfield  {author} {\bibinfo {author} {\bibfnamefont {A.~A.}\ \bibnamefont
  {Kaptanoglu}}, \bibinfo {author} {\bibfnamefont {J.~L.}\ \bibnamefont
  {Callaham}}, \bibinfo {author} {\bibfnamefont {C.~J.}\ \bibnamefont
  {Hansen}}, \bibinfo {author} {\bibfnamefont {A.}~\bibnamefont {Aravkin}},\
  and\ \bibinfo {author} {\bibfnamefont {S.~L.}\ \bibnamefont {Brunton}},\
  }\bibfield  {title} {\bibinfo {title} {Promoting global stability in
  data-driven models of quadratic nonlinear dynamics},\ }\href@noop {}
  {\bibfield  {journal} {\bibinfo  {journal} {arXiv preprint arXiv:2105.01843}\
  } (\bibinfo {year} {2021})}\BibitemShut {NoStop}%
\bibitem [{\citenamefont {Galtier}(2016)}]{galtier2016introduction}%
  \BibitemOpen
  \bibfield  {author} {\bibinfo {author} {\bibfnamefont {S.}~\bibnamefont
  {Galtier}},\ }\href@noop {} {\emph {\bibinfo {title} {Introduction to modern
  magnetohydrodynamics}}}\ (\bibinfo  {publisher} {Cambridge University
  Press},\ \bibinfo {year} {2016})\BibitemShut {NoStop}%
\bibitem [{\citenamefont {Chaturantabut}\ and\ \citenamefont
  {Sorensen}(2009)}]{chaturantabut2009discrete}%
  \BibitemOpen
  \bibfield  {author} {\bibinfo {author} {\bibfnamefont {S.}~\bibnamefont
  {Chaturantabut}}\ and\ \bibinfo {author} {\bibfnamefont {D.~C.}\ \bibnamefont
  {Sorensen}},\ }\bibfield  {title} {\bibinfo {title} {Discrete empirical
  interpolation for nonlinear model reduction},\ }in\ \href@noop {} {\emph
  {\bibinfo {booktitle} {Proceedings of the 48th IEEE Conference on Decision
  and Control (CDC) held jointly with 2009 28th Chinese Control Conference}}}\
  (\bibinfo {organization} {IEEE},\ \bibinfo {year} {2009})\ pp.\ \bibinfo
  {pages} {4316--4321}\BibitemShut {NoStop}%
\bibitem [{\citenamefont {Drmac}\ and\ \citenamefont
  {Gugercin}(2016)}]{drmac2016new}%
  \BibitemOpen
  \bibfield  {author} {\bibinfo {author} {\bibfnamefont {Z.}~\bibnamefont
  {Drmac}}\ and\ \bibinfo {author} {\bibfnamefont {S.}~\bibnamefont
  {Gugercin}},\ }\bibfield  {title} {\bibinfo {title} {A new selection operator
  for the discrete empirical interpolation method---improved a priori error
  bound and extensions},\ }\href@noop {} {\bibfield  {journal} {\bibinfo
  {journal} {SIAM Journal on Scientific Computing}\ }\textbf {\bibinfo {volume}
  {38}},\ \bibinfo {pages} {A631} (\bibinfo {year} {2016})}\BibitemShut
  {NoStop}%
\bibitem [{\citenamefont {Astrid}\ \emph {et~al.}(2008)\citenamefont {Astrid},
  \citenamefont {Weiland}, \citenamefont {Willcox},\ and\ \citenamefont
  {Backx}}]{astrid2008missing}%
  \BibitemOpen
  \bibfield  {author} {\bibinfo {author} {\bibfnamefont {P.}~\bibnamefont
  {Astrid}}, \bibinfo {author} {\bibfnamefont {S.}~\bibnamefont {Weiland}},
  \bibinfo {author} {\bibfnamefont {K.}~\bibnamefont {Willcox}},\ and\ \bibinfo
  {author} {\bibfnamefont {T.}~\bibnamefont {Backx}},\ }\bibfield  {title}
  {\bibinfo {title} {Missing point estimation in models described by proper
  orthogonal decomposition},\ }\href@noop {} {\bibfield  {journal} {\bibinfo
  {journal} {IEEE Transactions on Automatic Control}\ }\textbf {\bibinfo
  {volume} {53}},\ \bibinfo {pages} {2237} (\bibinfo {year}
  {2008})}\BibitemShut {NoStop}%
\bibitem [{\citenamefont {Willcox}(2006)}]{willcox2006unsteady}%
  \BibitemOpen
  \bibfield  {author} {\bibinfo {author} {\bibfnamefont {K.}~\bibnamefont
  {Willcox}},\ }\bibfield  {title} {\bibinfo {title} {Unsteady flow sensing and
  estimation via the gappy proper orthogonal decomposition},\ }\href@noop {}
  {\bibfield  {journal} {\bibinfo  {journal} {Computers \& Fluids}\ }\textbf
  {\bibinfo {volume} {35}},\ \bibinfo {pages} {208} (\bibinfo {year}
  {2006})}\BibitemShut {NoStop}%
\bibitem [{\citenamefont {Carlberg}\ \emph {et~al.}(2013)\citenamefont
  {Carlberg}, \citenamefont {Farhat}, \citenamefont {Cortial},\ and\
  \citenamefont {Amsallem}}]{carlberg2013gnat}%
  \BibitemOpen
  \bibfield  {author} {\bibinfo {author} {\bibfnamefont {K.}~\bibnamefont
  {Carlberg}}, \bibinfo {author} {\bibfnamefont {C.}~\bibnamefont {Farhat}},
  \bibinfo {author} {\bibfnamefont {J.}~\bibnamefont {Cortial}},\ and\ \bibinfo
  {author} {\bibfnamefont {D.}~\bibnamefont {Amsallem}},\ }\bibfield  {title}
  {\bibinfo {title} {The {GNAT} method for nonlinear model reduction: effective
  implementation and application to computational fluid dynamics and turbulent
  flows},\ }\href@noop {} {\bibfield  {journal} {\bibinfo  {journal} {Journal
  of Computational Physics}\ }\textbf {\bibinfo {volume} {242}},\ \bibinfo
  {pages} {623} (\bibinfo {year} {2013})}\BibitemShut {NoStop}%
\bibitem [{\citenamefont {de~Silva}\ \emph {et~al.}(2020)\citenamefont
  {de~Silva}, \citenamefont {Champion}, \citenamefont {Quade}, \citenamefont
  {Loiseau}, \citenamefont {Kutz},\ and\ \citenamefont
  {Brunton}}]{silva2020pysindy}%
  \BibitemOpen
  \bibfield  {author} {\bibinfo {author} {\bibfnamefont {B.~M.}\ \bibnamefont
  {de~Silva}}, \bibinfo {author} {\bibfnamefont {K.}~\bibnamefont {Champion}},
  \bibinfo {author} {\bibfnamefont {M.}~\bibnamefont {Quade}}, \bibinfo
  {author} {\bibfnamefont {J.-C.}\ \bibnamefont {Loiseau}}, \bibinfo {author}
  {\bibfnamefont {J.~N.}\ \bibnamefont {Kutz}},\ and\ \bibinfo {author}
  {\bibfnamefont {S.~L.}\ \bibnamefont {Brunton}},\ }\href@noop {} {\bibinfo
  {title} {{P}y{SIND}y: A python package for the sparse identification of
  nonlinear dynamics from data}} (\bibinfo {year} {2020}),\ \Eprint
  {https://arxiv.org/abs/2004.08424} {arXiv:2004.08424 [math.DS]} \BibitemShut
  {NoStop}%
\bibitem [{\citenamefont {Tibshirani}(1996)}]{Tibshirani1996lasso}%
  \BibitemOpen
  \bibfield  {author} {\bibinfo {author} {\bibfnamefont {R.}~\bibnamefont
  {Tibshirani}},\ }\bibfield  {title} {\bibinfo {title} {Regression shrinkage
  and selection via the lasso},\ }\href@noop {} {\bibfield  {journal} {\bibinfo
   {journal} {Journal of the Royal Statistical Society. Series B
  (Methodological)}\ ,\ \bibinfo {pages} {267}} (\bibinfo {year}
  {1996})}\BibitemShut {NoStop}%
\bibitem [{\citenamefont {Zheng}\ \emph {et~al.}(2019)\citenamefont {Zheng},
  \citenamefont {Askham}, \citenamefont {Brunton}, \citenamefont {Kutz},\ and\
  \citenamefont {Aravkin}}]{zheng2019unified}%
  \BibitemOpen
  \bibfield  {author} {\bibinfo {author} {\bibfnamefont {P.}~\bibnamefont
  {Zheng}}, \bibinfo {author} {\bibfnamefont {T.}~\bibnamefont {Askham}},
  \bibinfo {author} {\bibfnamefont {S.~L.}\ \bibnamefont {Brunton}}, \bibinfo
  {author} {\bibfnamefont {J.~N.}\ \bibnamefont {Kutz}},\ and\ \bibinfo
  {author} {\bibfnamefont {A.~Y.}\ \bibnamefont {Aravkin}},\ }\bibfield
  {title} {\bibinfo {title} {A unified framework for sparse relaxed regularized
  regression: {SR}3},\ }\href@noop {} {\bibfield  {journal} {\bibinfo
  {journal} {IEEE Access}\ }\textbf {\bibinfo {volume} {7}},\ \bibinfo {pages}
  {1404} (\bibinfo {year} {2019})}\BibitemShut {NoStop}%
\bibitem [{\citenamefont {Gel{\ss}}\ \emph {et~al.}(2019)\citenamefont
  {Gel{\ss}}, \citenamefont {Klus}, \citenamefont {Eisert},\ and\ \citenamefont
  {Sch{\"u}tte}}]{gelss2019multidimensional}%
  \BibitemOpen
  \bibfield  {author} {\bibinfo {author} {\bibfnamefont {P.}~\bibnamefont
  {Gel{\ss}}}, \bibinfo {author} {\bibfnamefont {S.}~\bibnamefont {Klus}},
  \bibinfo {author} {\bibfnamefont {J.}~\bibnamefont {Eisert}},\ and\ \bibinfo
  {author} {\bibfnamefont {C.}~\bibnamefont {Sch{\"u}tte}},\ }\bibfield
  {title} {\bibinfo {title} {Multidimensional approximation of nonlinear
  dynamical systems},\ }\href@noop {} {\bibfield  {journal} {\bibinfo
  {journal} {Journal of Computational and Nonlinear Dynamics}\ }\textbf
  {\bibinfo {volume} {14}} (\bibinfo {year} {2019})}\BibitemShut {NoStop}%
\bibitem [{\citenamefont {Callaham}\ \emph {et~al.}(2020)\citenamefont
  {Callaham}, \citenamefont {Loiseau}, \citenamefont {Rigas},\ and\
  \citenamefont {Brunton}}]{callaham2020nonlinear}%
  \BibitemOpen
  \bibfield  {author} {\bibinfo {author} {\bibfnamefont {J.~L.}\ \bibnamefont
  {Callaham}}, \bibinfo {author} {\bibfnamefont {J.-C.}\ \bibnamefont
  {Loiseau}}, \bibinfo {author} {\bibfnamefont {G.}~\bibnamefont {Rigas}},\
  and\ \bibinfo {author} {\bibfnamefont {S.~L.}\ \bibnamefont {Brunton}},\
  }\bibfield  {title} {\bibinfo {title} {Nonlinear stochastic modeling with
  {L}angevin regression},\ }\href@noop {} {\bibfield  {journal} {\bibinfo
  {journal} {arXiv preprint arXiv:2009.01006}\ } (\bibinfo {year}
  {2020})}\BibitemShut {NoStop}%
\bibitem [{\citenamefont {Bramburger}\ \emph {et~al.}(2020)\citenamefont
  {Bramburger}, \citenamefont {Dylewsky},\ and\ \citenamefont
  {Kutz}}]{bramburger2020sparse}%
  \BibitemOpen
  \bibfield  {author} {\bibinfo {author} {\bibfnamefont {J.~J.}\ \bibnamefont
  {Bramburger}}, \bibinfo {author} {\bibfnamefont {D.}~\bibnamefont
  {Dylewsky}},\ and\ \bibinfo {author} {\bibfnamefont {J.~N.}\ \bibnamefont
  {Kutz}},\ }\bibfield  {title} {\bibinfo {title} {Sparse identification of
  slow timescale dynamics},\ }\href@noop {} {\bibfield  {journal} {\bibinfo
  {journal} {arXiv preprint arXiv:2006.00940}\ } (\bibinfo {year}
  {2020})}\BibitemShut {NoStop}%
\bibitem [{\citenamefont {Kaheman}\ \emph {et~al.}(2020)\citenamefont
  {Kaheman}, \citenamefont {Kutz},\ and\ \citenamefont
  {Brunton}}]{kaheman2020sindy}%
  \BibitemOpen
  \bibfield  {author} {\bibinfo {author} {\bibfnamefont {K.}~\bibnamefont
  {Kaheman}}, \bibinfo {author} {\bibfnamefont {J.~N.}\ \bibnamefont {Kutz}},\
  and\ \bibinfo {author} {\bibfnamefont {S.~L.}\ \bibnamefont {Brunton}},\
  }\bibfield  {title} {\bibinfo {title} {{SIND}y-{PI}: a robust algorithm for
  parallel implicit sparse identification of nonlinear dynamics},\ }\href@noop
  {} {\bibfield  {journal} {\bibinfo  {journal} {Proceedings of the Royal
  Society A}\ }\textbf {\bibinfo {volume} {476}},\ \bibinfo {pages} {20200279}
  (\bibinfo {year} {2020})}\BibitemShut {NoStop}%
\bibitem [{\citenamefont {Cortiella}\ \emph {et~al.}(2021)\citenamefont
  {Cortiella}, \citenamefont {Park},\ and\ \citenamefont
  {Doostan}}]{cortiella2021sparse}%
  \BibitemOpen
  \bibfield  {author} {\bibinfo {author} {\bibfnamefont {A.}~\bibnamefont
  {Cortiella}}, \bibinfo {author} {\bibfnamefont {K.-C.}\ \bibnamefont
  {Park}},\ and\ \bibinfo {author} {\bibfnamefont {A.}~\bibnamefont
  {Doostan}},\ }\bibfield  {title} {\bibinfo {title} {Sparse identification of
  nonlinear dynamical systems via reweighted $l_1$-regularized least squares},\
  }\href@noop {} {\bibfield  {journal} {\bibinfo  {journal} {Computer Methods
  in Applied Mechanics and Engineering}\ }\textbf {\bibinfo {volume} {376}},\
  \bibinfo {pages} {113620} (\bibinfo {year} {2021})}\BibitemShut {NoStop}%
\bibitem [{\citenamefont {Br{\"u}ckner}\ \emph {et~al.}(2020)\citenamefont
  {Br{\"u}ckner}, \citenamefont {Ronceray},\ and\ \citenamefont
  {Broedersz}}]{bruckner2020inferring}%
  \BibitemOpen
  \bibfield  {author} {\bibinfo {author} {\bibfnamefont {D.~B.}\ \bibnamefont
  {Br{\"u}ckner}}, \bibinfo {author} {\bibfnamefont {P.}~\bibnamefont
  {Ronceray}},\ and\ \bibinfo {author} {\bibfnamefont {C.~P.}\ \bibnamefont
  {Broedersz}},\ }\bibfield  {title} {\bibinfo {title} {Inferring the dynamics
  of underdamped stochastic systems},\ }\href@noop {} {\bibfield  {journal}
  {\bibinfo  {journal} {Physical review letters}\ }\textbf {\bibinfo {volume}
  {125}},\ \bibinfo {pages} {058103} (\bibinfo {year} {2020})}\BibitemShut
  {NoStop}%
\bibitem [{\citenamefont {Beetham}\ and\ \citenamefont
  {Capecelatro}(2020)}]{beetham2020formulating}%
  \BibitemOpen
  \bibfield  {author} {\bibinfo {author} {\bibfnamefont {S.}~\bibnamefont
  {Beetham}}\ and\ \bibinfo {author} {\bibfnamefont {J.}~\bibnamefont
  {Capecelatro}},\ }\bibfield  {title} {\bibinfo {title} {Formulating
  turbulence closures using sparse regression with embedded form invariance},\
  }\href@noop {} {\bibfield  {journal} {\bibinfo  {journal} {Physical Review
  Fluids}\ }\textbf {\bibinfo {volume} {5}},\ \bibinfo {pages} {084611}
  (\bibinfo {year} {2020})}\BibitemShut {NoStop}%
\bibitem [{\citenamefont {Kolter}\ and\ \citenamefont
  {Manek}(2019)}]{kolter2019learning}%
  \BibitemOpen
  \bibfield  {author} {\bibinfo {author} {\bibfnamefont {J.~Z.}\ \bibnamefont
  {Kolter}}\ and\ \bibinfo {author} {\bibfnamefont {G.}~\bibnamefont {Manek}},\
  }\bibfield  {title} {\bibinfo {title} {Learning stable deep dynamics
  models},\ }in\ \href@noop {} {\emph {\bibinfo {booktitle} {Advances in Neural
  Information Processing Systems}}},\ Vol.~\bibinfo {volume} {32}\ (\bibinfo
  {year} {2019})\ pp.\ \bibinfo {pages} {11128--11136}\BibitemShut {NoStop}%
\bibitem [{\citenamefont {Pan}\ and\ \citenamefont
  {Duraisamy}(2020)}]{pan2020physics}%
  \BibitemOpen
  \bibfield  {author} {\bibinfo {author} {\bibfnamefont {S.}~\bibnamefont
  {Pan}}\ and\ \bibinfo {author} {\bibfnamefont {K.}~\bibnamefont
  {Duraisamy}},\ }\bibfield  {title} {\bibinfo {title} {Physics-informed
  probabilistic learning of linear embeddings of nonlinear dynamics with
  guaranteed stability},\ }\href@noop {} {\bibfield  {journal} {\bibinfo
  {journal} {SIAM Journal on Applied Dynamical Systems}\ }\textbf {\bibinfo
  {volume} {19}},\ \bibinfo {pages} {480} (\bibinfo {year} {2020})}\BibitemShut
  {NoStop}%
\bibitem [{\citenamefont {Boninsegna}\ \emph {et~al.}(2018)\citenamefont
  {Boninsegna}, \citenamefont {N{\"u}ske},\ and\ \citenamefont
  {Clementi}}]{boninsegna2018sparse}%
  \BibitemOpen
  \bibfield  {author} {\bibinfo {author} {\bibfnamefont {L.}~\bibnamefont
  {Boninsegna}}, \bibinfo {author} {\bibfnamefont {F.}~\bibnamefont
  {N{\"u}ske}},\ and\ \bibinfo {author} {\bibfnamefont {C.}~\bibnamefont
  {Clementi}},\ }\bibfield  {title} {\bibinfo {title} {Sparse learning of
  stochastic dynamical equations},\ }\href@noop {} {\bibfield  {journal}
  {\bibinfo  {journal} {The Journal of Chemical Physics}\ }\textbf {\bibinfo
  {volume} {148}},\ \bibinfo {pages} {241723} (\bibinfo {year}
  {2018})}\BibitemShut {NoStop}%
\bibitem [{\citenamefont {Zucatti}\ and\ \citenamefont
  {Wolf}(2021)}]{zucatti2021data}%
  \BibitemOpen
  \bibfield  {author} {\bibinfo {author} {\bibfnamefont {V.}~\bibnamefont
  {Zucatti}}\ and\ \bibinfo {author} {\bibfnamefont {W.}~\bibnamefont {Wolf}},\
  }\bibfield  {title} {\bibinfo {title} {Data-driven closure of
  projection-based reduced order models for unsteady compressible flows},\
  }\href@noop {} {\bibfield  {journal} {\bibinfo  {journal} {arXiv preprint
  arXiv:2103.12727}\ } (\bibinfo {year} {2021})}\BibitemShut {NoStop}%
\bibitem [{\citenamefont {Wrobel}(2011)}]{wrobel2011study}%
  \BibitemOpen
  \bibfield  {author} {\bibinfo {author} {\bibfnamefont {J.~S.}\ \bibnamefont
  {Wrobel}},\ }\href@noop {} {\emph {\bibinfo {title} {A study of {HIT-SI}
  plasma dynamics using surface magnetic field measurements}}}\ (\bibinfo
  {publisher} {University of Washington},\ \bibinfo {year} {2011})\BibitemShut
  {NoStop}%
\bibitem [{\citenamefont {Victor}\ \emph {et~al.}(2014)\citenamefont {Victor},
  \citenamefont {Jarboe}, \citenamefont {Hansen}, \citenamefont {Akcay},
  \citenamefont {Morgan}, \citenamefont {Hossack},\ and\ \citenamefont
  {Nelson}}]{victor2014sustained}%
  \BibitemOpen
  \bibfield  {author} {\bibinfo {author} {\bibfnamefont {B.}~\bibnamefont
  {Victor}}, \bibinfo {author} {\bibfnamefont {T.}~\bibnamefont {Jarboe}},
  \bibinfo {author} {\bibfnamefont {C.}~\bibnamefont {Hansen}}, \bibinfo
  {author} {\bibfnamefont {C.}~\bibnamefont {Akcay}}, \bibinfo {author}
  {\bibfnamefont {K.}~\bibnamefont {Morgan}}, \bibinfo {author} {\bibfnamefont
  {A.}~\bibnamefont {Hossack}},\ and\ \bibinfo {author} {\bibfnamefont
  {B.}~\bibnamefont {Nelson}},\ }\bibfield  {title} {\bibinfo {title}
  {Sustained spheromaks with ideal n=1 kink stability and pressure
  confinement},\ }\href@noop {} {\bibfield  {journal} {\bibinfo  {journal}
  {Physics of Plasmas}\ }\textbf {\bibinfo {volume} {21}},\ \bibinfo {pages}
  {082504} (\bibinfo {year} {2014})}\BibitemShut {NoStop}%
\bibitem [{\citenamefont {Sovinec}\ \emph {et~al.}(2004)\citenamefont
  {Sovinec}, \citenamefont {Glasser}, \citenamefont {Gianakon}, \citenamefont
  {Barnes}, \citenamefont {Nebel}, \citenamefont {Kruger}, \citenamefont
  {Schnack}, \citenamefont {Plimpton}, \citenamefont {Tarditi}, \citenamefont
  {Chu} \emph {et~al.}}]{sovinec2004nonlinear}%
  \BibitemOpen
  \bibfield  {author} {\bibinfo {author} {\bibfnamefont {C.}~\bibnamefont
  {Sovinec}}, \bibinfo {author} {\bibfnamefont {A.}~\bibnamefont {Glasser}},
  \bibinfo {author} {\bibfnamefont {T.}~\bibnamefont {Gianakon}}, \bibinfo
  {author} {\bibfnamefont {D.}~\bibnamefont {Barnes}}, \bibinfo {author}
  {\bibfnamefont {R.}~\bibnamefont {Nebel}}, \bibinfo {author} {\bibfnamefont
  {S.}~\bibnamefont {Kruger}}, \bibinfo {author} {\bibfnamefont
  {D.}~\bibnamefont {Schnack}}, \bibinfo {author} {\bibfnamefont
  {S.}~\bibnamefont {Plimpton}}, \bibinfo {author} {\bibfnamefont
  {A.}~\bibnamefont {Tarditi}}, \bibinfo {author} {\bibfnamefont {M.-S.}\
  \bibnamefont {Chu}}, \emph {et~al.},\ }\bibfield  {title} {\bibinfo {title}
  {Nonlinear magnetohydrodynamics simulation using high-order finite
  elements},\ }\href@noop {} {\bibfield  {journal} {\bibinfo  {journal}
  {Journal of Computational Physics}\ }\textbf {\bibinfo {volume} {195}},\
  \bibinfo {pages} {355} (\bibinfo {year} {2004})}\BibitemShut {NoStop}%
\bibitem [{\citenamefont {Akcay}(2013)}]{akcay2013extended}%
  \BibitemOpen
  \bibfield  {author} {\bibinfo {author} {\bibfnamefont {C.}~\bibnamefont
  {Akcay}},\ }\emph {\bibinfo {title} {Extended magnetohydrodynamic simulations
  of the helicity injected torus ({HIT-SI}) spheromak experiment with the
  {NIMROD} code}},\ \href@noop {} {Ph.D. thesis},\ \bibinfo  {school}
  {{U}niversity of {W}ashington, {S}eattle} (\bibinfo {year}
  {2013})\BibitemShut {NoStop}%
\bibitem [{\citenamefont {Akcay}\ \emph {et~al.}(2013)\citenamefont {Akcay},
  \citenamefont {Kim}, \citenamefont {Victor},\ and\ \citenamefont
  {Jarboe}}]{akcay2013}%
  \BibitemOpen
  \bibfield  {author} {\bibinfo {author} {\bibfnamefont {C.}~\bibnamefont
  {Akcay}}, \bibinfo {author} {\bibfnamefont {C.~C.}\ \bibnamefont {Kim}},
  \bibinfo {author} {\bibfnamefont {B.~S.}\ \bibnamefont {Victor}},\ and\
  \bibinfo {author} {\bibfnamefont {T.~R.}\ \bibnamefont {Jarboe}},\ }\bibfield
   {title} {\bibinfo {title} {Validation of single-fluid and two-fluid
  magnetohydrodynamic models of the helicity injected torus spheromak
  experiment with the {NIMROD} code},\ }\href@noop {} {\bibfield  {journal}
  {\bibinfo  {journal} {Physics of Plasmas}\ }\textbf {\bibinfo {volume}
  {20}},\ \bibinfo {pages} {082512} (\bibinfo {year} {2013})}\BibitemShut
  {NoStop}%
\bibitem [{\citenamefont {Morgan}\ \emph {et~al.}(2017)\citenamefont {Morgan},
  \citenamefont {Jarboe}, \citenamefont {Hossack}, \citenamefont {Chandra},\
  and\ \citenamefont {Everson}}]{morgan2017validation}%
  \BibitemOpen
  \bibfield  {author} {\bibinfo {author} {\bibfnamefont {K.}~\bibnamefont
  {Morgan}}, \bibinfo {author} {\bibfnamefont {T.}~\bibnamefont {Jarboe}},
  \bibinfo {author} {\bibfnamefont {A.}~\bibnamefont {Hossack}}, \bibinfo
  {author} {\bibfnamefont {R.}~\bibnamefont {Chandra}},\ and\ \bibinfo {author}
  {\bibfnamefont {C.}~\bibnamefont {Everson}},\ }\bibfield  {title} {\bibinfo
  {title} {Validation of extended magnetohydrodynamic simulations of the
  {HIT-SI}3 experiment using the {NIMROD} code},\ }\href@noop {} {\bibfield
  {journal} {\bibinfo  {journal} {Physics of Plasmas}\ }\textbf {\bibinfo
  {volume} {24}},\ \bibinfo {pages} {122510} (\bibinfo {year}
  {2017})}\BibitemShut {NoStop}%
\bibitem [{\citenamefont {Morrison}\ and\ \citenamefont
  {Greene}(1980)}]{morrison1980noncanonical}%
  \BibitemOpen
  \bibfield  {author} {\bibinfo {author} {\bibfnamefont {P.~J.}\ \bibnamefont
  {Morrison}}\ and\ \bibinfo {author} {\bibfnamefont {J.~M.}\ \bibnamefont
  {Greene}},\ }\bibfield  {title} {\bibinfo {title} {Noncanonical {H}amiltonian
  density formulation of hydrodynamics and ideal magnetohydrodynamics},\
  }\href@noop {} {\bibfield  {journal} {\bibinfo  {journal} {Physical Review
  Letters}\ }\textbf {\bibinfo {volume} {45}},\ \bibinfo {pages} {790}
  (\bibinfo {year} {1980})}\BibitemShut {NoStop}%
\bibitem [{\citenamefont {Yoshida}\ and\ \citenamefont
  {Hameiri}(2013)}]{yoshida2013canonical}%
  \BibitemOpen
  \bibfield  {author} {\bibinfo {author} {\bibfnamefont {Z.}~\bibnamefont
  {Yoshida}}\ and\ \bibinfo {author} {\bibfnamefont {E.}~\bibnamefont
  {Hameiri}},\ }\bibfield  {title} {\bibinfo {title} {Canonical {H}amiltonian
  mechanics of {H}all magnetohydrodynamics and its limit to ideal
  magnetohydrodynamics},\ }\href@noop {} {\bibfield  {journal} {\bibinfo
  {journal} {Journal of Physics A: Mathematical and Theoretical}\ }\textbf
  {\bibinfo {volume} {46}},\ \bibinfo {pages} {335502} (\bibinfo {year}
  {2013})}\BibitemShut {NoStop}%
\bibitem [{\citenamefont {Abdelhamid}\ \emph {et~al.}(2015)\citenamefont
  {Abdelhamid}, \citenamefont {Kawazura},\ and\ \citenamefont
  {Yoshida}}]{abdelhamid2015hamiltonian}%
  \BibitemOpen
  \bibfield  {author} {\bibinfo {author} {\bibfnamefont {H.~M.}\ \bibnamefont
  {Abdelhamid}}, \bibinfo {author} {\bibfnamefont {Y.}~\bibnamefont
  {Kawazura}},\ and\ \bibinfo {author} {\bibfnamefont {Z.}~\bibnamefont
  {Yoshida}},\ }\bibfield  {title} {\bibinfo {title} {Hamiltonian formalism of
  extended magnetohydrodynamics},\ }\href@noop {} {\bibfield  {journal}
  {\bibinfo  {journal} {Journal of Physics A: Mathematical and Theoretical}\
  }\textbf {\bibinfo {volume} {48}},\ \bibinfo {pages} {235502} (\bibinfo
  {year} {2015})}\BibitemShut {NoStop}%
\bibitem [{\citenamefont {Chu}\ and\ \citenamefont
  {Hayashibe}(2020)}]{chu2020discovering}%
  \BibitemOpen
  \bibfield  {author} {\bibinfo {author} {\bibfnamefont {H.~K.}\ \bibnamefont
  {Chu}}\ and\ \bibinfo {author} {\bibfnamefont {M.}~\bibnamefont
  {Hayashibe}},\ }\bibfield  {title} {\bibinfo {title} {Discovering
  interpretable dynamics by sparsity promotion on energy and the
  {L}agrangian},\ }\href@noop {} {\bibfield  {journal} {\bibinfo  {journal}
  {IEEE Robotics and Automation Letters}\ }\textbf {\bibinfo {volume} {5}},\
  \bibinfo {pages} {2154} (\bibinfo {year} {2020})}\BibitemShut {NoStop}%
\end{thebibliography}%
\end{document}